# GWTC-1: A Gravitational-Wave Transient Catalog of Compact Binary Mergers Observed by LIGO and Virgo during the First and Second Observing Runs

B. P. Abbott *et al.*[*]
(LIGO Scientific Collaboration and Virgo Collaboration)

We present the results from three gravitational-wave searches for coalescing compact binaries with component masses above 1 $M_\odot$ during the first and second observing runs of the advanced gravitational-wave detector network. During the first observing run ($O1$), from September 12, 2015 to January 19, 2016, gravitational waves from three binary black hole mergers were detected. The second observing run ($O2$), which ran from November 30, 2016 to August 25, 2017, saw the first detection of gravitational waves from a binary neutron star inspiral, in addition to the observation of gravitational waves from a total of seven binary black hole mergers, four of which we report here for the first time: GW170729, GW170809, GW170818, and GW170823. For all significant gravitational-wave events, we provide estimates of the source properties. The detected binary black holes have total masses between $18.6^{+3.2}_{-0.7}\,M_\odot$ and $84.4^{+15.8}_{-11.1}\,M_\odot$ and range in distance between $320^{+120}_{-110}$ and $2840^{+1400}_{-1360}$ Mpc. No neutron star–black hole mergers were detected. In addition to highly significant gravitational-wave events, we also provide a list of marginal event candidates with an estimated false-alarm rate less than 1 per 30 days. From these results over the first two observing runs, which include approximately one gravitational-wave detection per 15 days of data searched, we infer merger rates at the 90% confidence intervals of $110 - 3840\ \mathrm{Gpc}^{-3}\,\mathrm{y}^{-1}$ for binary neutron stars and $9.7 - 101\ \mathrm{Gpc}^{-3}\,\mathrm{y}^{-1}$ for binary black holes assuming fixed population distributions and determine a neutron star–black hole merger rate 90% upper limit of $610\ \mathrm{Gpc}^{-3}\,\mathrm{y}^{-1}$.

## I. INTRODUCTION

The first observing run ($O1$) of Advanced LIGO, which took place from September 12, 2015 until January 19, 2016, saw the first detections of gravitational waves (GWs) from stellar-mass binary black holes (BBHs) [1–4]. After an upgrade and commissioning period, the second observing run ($O2$) of the Advanced LIGO detectors [5] commenced on November 30, 2016 and ended on August 25, 2017. On August 1, 2017, the Advanced Virgo detector [6] joined the observing run, enabling the first three-detector observations of GWs. This network of ground-based interferometric detectors is sensitive to GWs from the inspiral, merger, and ringdown of compact binary coalescences (CBCs), covering a frequency range from about 15 Hz up to a few kilohertz (see Fig. 1). In this catalog, we report 11 confident detections of GWs from compact binary mergers as well as a selection of less significant triggers from both observing runs. The observations reported here and future GW detections will shed light on binary formation channels, enable precision tests of general relativity (GR) in its strong-field regime, and open up new avenues of astronomy research.

The events presented here are obtained from a total of three searches: two matched-filter searches, PyCBC [7,8] and GstLAL [9,10], using relativistic models of GWs from CBCs, as well as one unmodeled search for short-duration transient signals or bursts, coherent WaveBurst (cWB) [11]. The two matched-filter searches target GWs from compact binaries with a redshifted total mass $M(1+z)$ of 2–500 $M_\odot$ for PyCBC and 2–400 $M_\odot$ for GstLAL, where $z$ is the cosmological redshift of the source binary [12], and with maximal dimensionless spins of 0.998 for black holes (BHs) and 0.05 for neutron stars (NSs). The results of a matched-filter search for sub-solar-mass compact objects in $O1$ can be found in Ref. [13]; the results for $O2$ will be discussed elsewhere. The burst search cWB does not use waveform models to compare against the data but instead identifies regions of excess power in the time-frequency representation of the gravitational strain. We report results from a cWB analysis that is optimized for the detection of compact binaries with a total mass less than 100 $M_\odot$. A different tuning of the cWB analysis is used for a search for intermediate-mass BBHs with total masses greater than 100 $M_\odot$; the results of that analysis are discussed elsewhere. The three searches reported here use different methodologies to identify GWs from compact binaries in an overlapping but not identical search space, thus providing three largely independent analyses that allow for important cross-checks and yield consistent results. All searches have undergone improvements since $O1$, making it scientifically valuable to reanalyze the $O1$ data in order to reevaluate the significance of previously identified GW events and to potentially discover new ones.

[*]Full author list given at the end of the article.

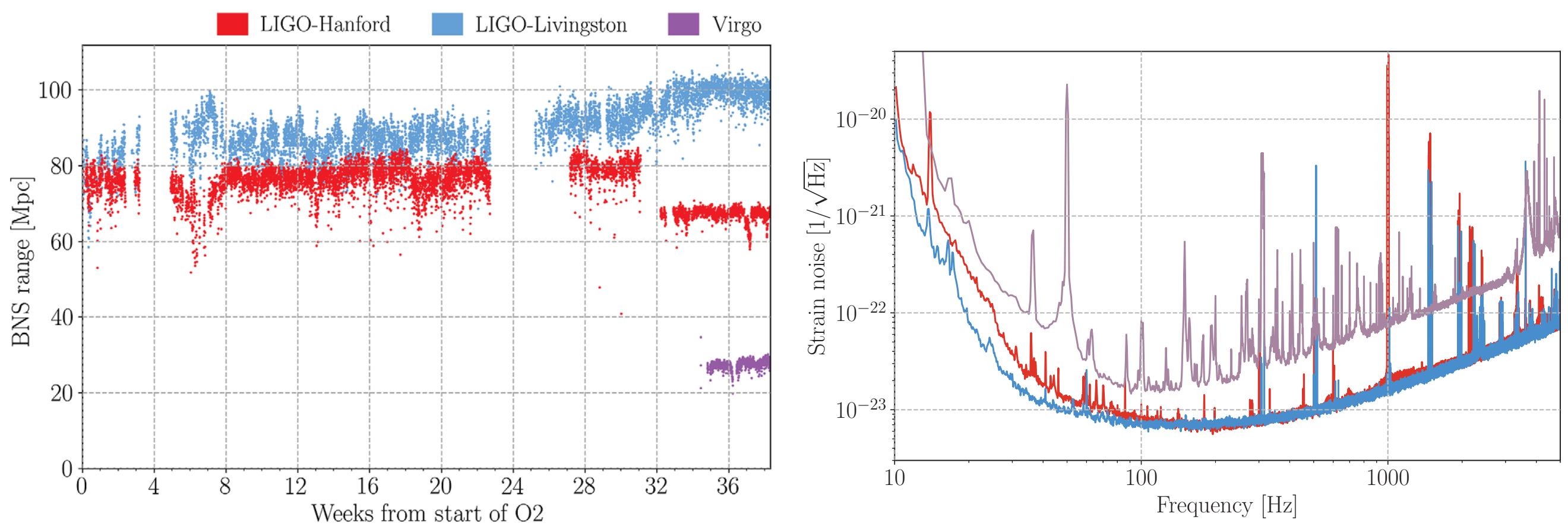

FIG. 1. Left: BNS range for each instrument during $O2$. The break at week 3 is for the 2016 end-of-year holidays. There is an additional break in the run at week 23 to make improvements to instrument sensitivity. The Montana earthquake's impact on the LHO instrument sensitivity can be seen at week 31. Virgo joins $O2$ in week 34. Right: Amplitude spectral density of the total strain noise of the Virgo, LHO, and LLO detectors. The curves are representative of the best performance of each detector during $O2$.

The searches identified a total of ten BBH mergers and one binary neutron star (BNS) signal. The GW events GW150914, GW151012 [14], GW151226, GW170104, GW170608, GW170814, and GW170817 have been reported previously [4,15–18]. In this catalog, we announce four previously unpublished BBH mergers observed during $O2$: GW170729, GW170809, GW170818, and GW170823. We estimate the total mass of GW170729 to be $84.4^{+15.8}_{-11.1}\ M_{\odot}$, making it the highest-mass BBH observed to date. GW170818 is the second BBH observed in triple coincidence between the two LIGO observatories and Virgo after GW170814 [16]. As the sky location is primarily determined by the differences in the times of arrival of the GW signal at the different detector sites, LIGO-Virgo coincident events have a vastly improved sky localization, which is crucial for electromagnetic follow-up campaigns [19–22]. The reanalysis of the $O1$ data did not result in the discovery of any new GW events, but GW151012 is now detected with increased significance. In addition, we list 14 GW candidate events that have an estimated false-alarm rate (FAR) less than 1 per 30 days in either of the two matched-filter analyses but whose astrophysical origin cannot be established nor excluded unambiguously (Sec. VII).

Gravitational waves from compact binaries carry information about the properties of the source such as the masses and spins. These can be extracted via Bayesian inference by using theoretical models of the GW signal that describe the inspiral, merger, and ringdown of the final object for BBH [23–30] and the inspiral (and merger) for BNS [31–33]. Such models are built by combining post-Newtonian calculations [34–38], the effective-one-body formalism [39–44], and numerical relativity [45–50]. Based on a variety of theoretical models, we provide key source properties of all confident GW detections. For previously reported detections, we provide updated parameter estimates which exploit refined instrumental calibration, noise subtraction (for $O2$ data) [51,52], and updated amplitude power spectral density estimates [53,54].

The observation of these GW events allows us to place constraints on the rates of stellar-mass BBH and BNS mergers in the Universe and probe their mass and spin distributions, putting them into astrophysical context. The nonobservation of GWs from a neutron star–black hole binary (NSBH) yields a stronger 90% upper limit on the rate. The details of the astrophysical implications of our observations are discussed in Ref. [55].

This paper is organized as follows: In Sec. II, we provide an overview of the operating detectors during $O2$, as well as the data used in the searches and parameter estimation. Section III briefly summarizes the three different searches, before we define the event selection criteria and present the results in Sec. IV. Tables I and II summarize some key search parameters for the clear GW detections and the marginal events. Details about the source properties of the GW events are given in Sec. V, and the values of some important parameters obtained from Bayesian inference are listed in Table III. We do not provide parameter estimation results for marginal events. An independent consistency analysis between the waveform-based results and the data is performed in Sec. VI. In Sec. VII, we describe how the probability of astrophysical origin is calculated and give its value for each significant and marginal event in Table IV. We provide an updated estimate of binary merger rates in this section before concluding in Sec. VIII. We also provide the Appendixes containing additional technical details.

TABLE I. Search results for the 11 GW events. We report a false-alarm rate for each search that found a given event; otherwise, we display ⋯. The network SNR for the two matched-filter searches is that of the template ranked highest by that search, which is not necessarily the template with the highest SNR. Moreover, the network SNR is the quadrature sum of the detectors coincident in the highest-ranked trigger; in some cases, only two detectors contribute, even if all three are operating nominally at the time of that event.

| | | FAR $[y^{-1}]$ | | | Network SNR | | |
|---|---|---|---|---|---|---|---|
| Event | UTC time | PyCBC | GstLAL | cWB | PyCBC | GstLAL | cWB |
| GW150914 | 09:50:45.4 | $<1.53\times10^{-5}$ | $<1.00\times10^{-7}$ | $<1.63\times10^{-4}$ | 23.6 | 24.4 | 25.2 |
| GW151012 | 09:54:43.4 | 0.17 | $7.92\times10^{-3}$ | ⋯ | 9.5 | 10.0 | ⋯ |
| GW151226 | 03:38:53.6 | $<1.69\times10^{-5}$ | $<1.00\times10^{-7}$ | 0.02 | 13.1 | 13.1 | 11.9 |
| GW170104 | 10:11:58.6 | $<1.37\times10^{-5}$ | $<1.00\times10^{-7}$ | $2.91\times10^{-4}$ | 13.0 | 13.0 | 13.0 |
| GW170608 | 02:01:16.5 | $<3.09\times10^{-4}$ | $<1.00\times10^{-7}$ | $1.44\times10^{-4}$ | 15.4 | 14.9 | 14.1 |
| GW170729 | 18:56:29.3 | 1.36 | 0.18 | 0.02 | 9.8 | 10.8 | 10.2 |
| GW170809 | 08:28:21.8 | $1.45\times10^{-4}$ | $<1.00\times10^{-7}$ | ⋯ | 12.2 | 12.4 | ⋯ |
| GW170814 | 10:30:43.5 | $<1.25\times10^{-5}$ | $<1.00\times10^{-7}$ | $<2.08\times10^{-4}$ | 16.3 | 15.9 | 17.2 |
| GW170817 | 12:41:04.4 | $<1.25\times10^{-5}$ | $<1.00\times10^{-7}$ | ⋯ | 30.9 | 33.0 | ⋯ |
| GW170818 | 02:25:09.1 | ⋯ | $4.20\times10^{-5}$ | ⋯ | ⋯ | 11.3 | ⋯ |
| GW170823 | 13:13:58.5 | $<3.29\times10^{-5}$ | $<1.00\times10^{-7}$ | $2.14\times10^{-3}$ | 11.1 | 11.5 | 10.8 |

TABLE II. Marginal triggers from the two matched-filter CBC searches. To distinguish events occurring on the same UTC day, we extend the YYMMDD label by decimal fractions of a day as needed, always rounding down (truncating) the decimal. The search that identifies each trigger is given, and the false alarm and network SNR. This network SNR is the quadrature sum of the individual detector SNRs for all detectors involved in the reported trigger; that can be fewer than the number of nominally operational detectors at the time, depending on the ranking algorithm of each pipeline. The detector chirp mass reported is that of the most significant template of the search. The concentration of our marginal triggers at low chirp masses is consistent with expectations for noise triggers, because search template waveforms are much more densely packed at low masses. The final column indicates whether there are any detector characterization concerns with the trigger; for an explanation and more details, see the text.

| Date | UTC | Search | FAR $[y^{-1}]$ | Network SNR | $\mathcal{M}^{\mathrm{det}}$ $[M_\odot]$ | Data quality |
|---|---|---|---|---|---|---|
| 151008 | 14:09:17.5 | PyCBC | 10.17 | 8.8 | 5.12 | No artifacts |
| 151012.2 | 06:30:45.2 | GstLAL | 8.56 | 9.6 | 2.01 | Artifacts present |
| 151116 | 22:41:48.7 | PyCBC | 4.77 | 9.0 | 1.24 | No artifacts |
| 161202 | 03:53:44.9 | GstLAL | 6.00 | 10.5 | 1.54 | Artifacts possibly caused |
| 161217 | 07:16:24.4 | GstLAL | 10.12 | 10.7 | 7.86 | Artifacts possibly caused |
| 170208 | 10:39:25.8 | GstLAL | 11.18 | 10.0 | 7.39 | Artifacts present |
| 170219 | 14:04:09.0 | GstLAL | 6.26 | 9.6 | 1.53 | No artifacts |
| 170405 | 11:04:52.7 | GstLAL | 4.55 | 9.3 | 1.44 | Artifacts present |
| 170412 | 15:56:39.0 | GstLAL | 8.22 | 9.7 | 4.36 | Artifacts possibly caused |
| 170423 | 12:10:45.0 | GstLAL | 6.47 | 8.9 | 1.17 | No artifacts |
| 170616 | 19:47:20.8 | PyCBC | 1.94 | 9.1 | 2.75 | Artifacts present |
| 170630 | 16:17:07.8 | GstLAL | 10.46 | 9.7 | 0.90 | Artifacts present |
| 170705 | 08:45:16.3 | GstLAL | 10.97 | 9.3 | 3.40 | No artifacts |
| 170720 | 22:44:31.8 | GstLAL | 10.75 | 13.0 | 5.96 | Artifacts possibly caused |

A variety of additional information on each event, data products including strain data and posterior samples, and postprocessing tools can be obtained from the accompanying data release [56] hosted by the Gravitational Wave Open Science Center [57].

## II. INSTRUMENTAL OVERVIEW AND DATA

### A. LIGO instruments

The Advanced LIGO detectors [58,59] began scientific operations in September, 2015 and almost immediately detected the first gravitational waves from the BBH merger GW150914.

Between $O1$ and $O2$, improvements were made to both LIGO instruments. At LIGO-Livingston (LLO), a malfunctioning temperature sensor [60] was replaced immediately after $O1$, contributing to an increase in the BNS range from approximately 60 Mpc to approximately 80 Mpc [61]. Other major changes included adding passive tuned mass dampers on the end test mass suspensions to reduce ringing up of mechanical modes, installing a new output Faraday isolator, adding a new in-vacuum array of photodiodes for stabilizing the laser intensity, installing higher quantum-efficiency photodiodes at the output port, and replacing the compensation plate on the input test mass suspension for the $Y$ arm. An attempt to upgrade the LLO laser to provide higher input power was not successful. During $O2$, improvements to the detector sensitivity continued, and sources of scattered light noise were mitigated. As a result, the sensitivity of the LLO instrument rose from a BNS range of 80 Mpc at the beginning of $O2$ to greater than 100 Mpc by the run's end.

The LIGO-Hanford (LHO) detector had a range of approximately 80 Mpc as $O1$ ended, and it was decided to concentrate on increasing the input laser power and forgo any incursions into the vacuum system. Increasing the input laser power to 50 W was successful, but since this increase did not result in an improvement in sensitivity, the LHO detector operated with 30 W input power during $O2$. It was eventually discovered that there was a point absorber on one of the input test mass optics, which we speculate led to increased coupling of input "jitter" noise from the laser table into the interferometer. By use of appropriate witness sensors, it was possible to perform an offline noise subtraction on the data, leading to an increase in the BNS range at LHO by an average of about 20% over all of $O2$ [51,52].

On July 6, 2017, LHO was severely affected by a 5.8 magnitude earthquake in Montana. Postearthquake, the sensitivity of the detector dropped by approximately 10 Mpc and remained in this condition until the end of the run on August 25, 2017.

### B. Virgo instrument

Advanced Virgo [6] aims to increase the sensitivity of the Virgo interferometer by one order of magnitude, and several upgrades were performed after the decommissioning of the first-generation detector in 2011. The main modifications include a new optical design, heavier mirrors, and suspended optical benches, including photodiodes in a vacuum. Special care was also taken to improve the decoupling of the instrument from environmental disturbances. One of the main limiting noise sources below 100 Hz is the thermal Brownian excitation of the wires used for suspending the mirrors. A first test performed on the Virgo configuration showed that silica fibers would reduce this contribution. A vacuum contamination issue, which has since been corrected, led to failures of these silica suspension fibers, so metal wires were used to avoid delaying Virgo's participation in $O2$. Unlike the LIGO instruments, Virgo has not yet implemented signal recycling, which will be installed in a later upgrade of the instrument.

After several months of commissioning, Virgo joined $O2$ on August 1, 2017 with a BNS range of approximately 25 Mpc. The performance experienced a temporary degradation on August 11 and 12, when the microseismic activity on site was highly elevated and it was difficult to keep the interferometer in its low-noise operating mode.

### C. Data

Figure 1 shows the BNS ranges of the LIGO and Virgo instruments over the course of $O2$ and the representative amplitude spectral density plots of the total strain noise for each detector.

We subtract several independent contributions to the instrumental noise from the data at both LIGO detectors [51]. For all of $O2$, the average increase in the BNS range from this noise subtraction process at LHO is approximately 20% [51]. At LLO, the noise-subtraction process targeted narrow line features, resulting in a negligible increase in the BNS range.

Calibrated strain data from each interferometer are produced online for use in low-latency searches. Following the run, a final frequency-dependent calibration is generated for each interferometer.

For the LIGO instruments, this final calibration benefits from the use of postrun measurements and the removal of instrumental lines. The calibration uncertainties are 3.8% in amplitude and 2.1° in phase for LLO and 2.6% in amplitude and 2.4° in phase for LHO. The results cited in this paper use the full frequency-dependent calibration uncertainties described in Refs. [64,65]. The LIGO timing uncertainty of $<1\ \mu$s [66] is included in the phase correction factor.

The calibration of strain data produced online by Virgo has large uncertainties due to the short time available for measurements. The data are reprocessed to reduce the errors by taking into account better calibration models obtained from postrun measurements and the subtraction of frequency noise. The reprocessing includes a time dependence for the noise subtraction and for the determination of the finesse of the cavities. The final uncertainties are 5.1% in amplitude and 2.3° in phase [67]. The Virgo calibration has an additional uncertainty of 20 $\mu$s originating from the time stamping of the data.

During $O2$, the individual LIGO detectors had duty factors of approximately 60% with a LIGO network duty factor of about 45%. Times with significant instrumental disturbances are flagged and removed, resulting in about 118 days of data suitable for coincident analysis [68]. Of these data, about 15 days are collected in coincident operation with Virgo, which after joining $O2$ operated with a duty factor of about 80%. Times with excess instrumental noise, which is not expected to render the data unusable, are also flagged [68]. Individual searches may then decide to include or not include such times in their final results.

## III. SEARCHES

The search results presented in the next section are obtained by two different, largely independent matched-filter searches, PyCBC and GstLAL, and the burst search cWB. Because of the sensitivity imbalance between the Advanced Virgo detector as compared to the two Advanced LIGO detectors, neither PyCBC nor cWB elect to analyze data from Virgo. GstLAL, however, includes Virgo into its search during the month of August. The two matched-filter searches assume sources that can be modeled by general relativity and, in particular, quasicircular binaries whose spin angular momenta are either aligned or antialigned with their orbital angular momenta. They are still capable, however, of detecting many systems that exhibit precession [69]. In contrast, the cWB search relies on no specific physical models of the source waveform, though in results presented here it does impose a restriction that signals are "chirping" in the time-frequency plane. We therefore refer to it as *weakly modeled*. In the remainder of this section, we present a brief description of each of these searches, summarizing both the parameter space searched and improvements made since their use in $O1$ [4].

### A. The PyCBC search

A pipeline to search for GWs from CBCs is constructed using the PyCBC software package [7,8]. This analysis performs direct matched filtering of the data against a bank of template waveforms to calculate the signal-to-noise ratio (SNR) for each combination of detector, template waveform, and coalescence time [70]. Whenever the local maximum of this SNR time series is larger than a threshold of 5.5, the pipeline produces a single-detector trigger associated with the detector, the parameters of the template, and the coalescence time. In order to suppress triggers caused by high-amplitude noise transients ("glitches"), two signal-based vetoes may be calculated [71,72]. Using the SNR, the results of these two vetoes, and a fitting and smoothing procedure designed to ensure that the rate of single-detector triggers is approximately constant across the search parameter space, a single-detector rank $\varrho$ is calculated for each single-detector trigger [73].

After generating triggers in the Hanford and Livingston detectors as described above, PyCBC finds two-detector coincidences by requiring a trigger from each detector associated with the same template and with coalescence times within 15 ms of each other. This time window accounts for the maximum light-travel time between LHO and LLO as well as the uncertainty in the inferred coalescence time at each detector. Coincident triggers are assigned a ranking statistic that approximates the relative likelihood of obtaining the event's measured trigger parameters in the presence of a GW signal versus in the presence of noise alone [73]. The detailed construction of this network statistic, as well as the single-detector rank $\varrho$, is improved from the corresponding statistics used in $O1$, partially motivating the reanalysis of $O1$ by this pipeline.

Finally, the statistical significances of coincident triggers are quantified by their inverse false-alarm rate (IFAR). This rate is estimated by applying the same coincidence procedure after repeatedly time shifting the triggers from one detector and using the resulting coincidences as a background sample. Each foreground coincident trigger is assigned a false-alarm rate (FAR) given by the number of background triggers with an equal or larger ranking, divided by the total time searched for time-shifted coincidences. For an event with a given IFAR observed in the data of duration $T$, the probability of obtaining one or more equally highly ranked events due to noise is

$$p = 1 - e^{-T/\mathrm{IFAR}}. \tag{1}$$

In the analysis of this paper, the data are divided into analysis periods that allow at least 5.2 days of coincident data between the two LIGO detectors [74]. Though previous publications performed time shifting across larger amounts of time [1,2,4], the results here consider only time shifts within a given analysis period, which is done because the noise characteristics of the detector vary significantly from the beginning of $O1$ through the end of $O2$, so this restriction more accurately reflects the variation in detector performance. This restriction means, however, that the minimum bound on the false-alarm rate of candidates that have a higher ranking statistic than any trigger in the background sample is larger than it would be if longer periods of data are used for the time-shift analysis.

For the PyCBC analysis presented here, the template bank described in Ref. [75] is used. This bank covers binary systems with a total mass between 2 and 500 $M_\odot$ and mass ratios down to 1/98. Components of the binary with a mass below 2 $M_\odot$ are assumed to be neutron stars and have a maximum spin magnitude of 0.05; otherwise, the maximum magnitude is 0.998. The high-mass boundary of the search space is determined by the requirement that the waveform duration be at least 0.15 s, which reduces the number of false-alarm triggers from short instrumental glitches. The waveform models used are a reduced-order-model (ROM) [29,76–78] of SEOBNRv4 [29] for systems with a total mass greater than 4 $M_\odot$ and TaylorF2 [38,80] otherwise.

### B. The GstLAL search

A largely independent matched-filter pipeline based on the GstLAL library [9,10] (henceforth GstLAL) also performs a matched-filter search for CBC signals. GstLAL produces triggers for each template waveform and each detector by maximizing the matched-filter SNR $\rho$ over one-second windows and requiring that it exceed a threshold of 4 for the two LIGO detectors and 3.5 for Virgo. The relatively lower Virgo SNR threshold is an *ad hoc* choice designed to improve the network sensitivity of the search given Virgo's smaller horizon distance. For the search described here, candidates are formed by requiring a temporal coincidence between triggers from the same template but from different detectors, with the coincidence window set by the light-travel time between detectors plus 5 ms [81]. GstLAL ranks candidates using the logarithm of the likelihood ratio, $\mathcal{L}$, a measure of how likely it is to observe that candidate if a signal is present compared to if only noise is present [9,82,83]. The noise model is constructed, in part, from single-detector triggers that are not found in coincidence, so as to minimize the possibility of contamination by real signals. In the search presented here, the likelihood ratio is a function of $\rho$, a signal-consistency test, the differences in time and phase between the coincident triggers, the detectors that contribute triggers to the candidate, the sensitivity of the detectors to signals at the time of the candidate, and the rate of triggers in each of the detectors at the time of the candidate [9]. This function is an expansion of the parameters used to model the likelihood ratio in earlier versions of GstLAL and improves the sensitivity of the pipeline used for this search over that used in $O1$.

The GstLAL search uses Monte Carlo methods and the likelihood ratio's noise model to determine the probability of observing a candidate with a log likelihood ratio greater than or equal to $\log\mathcal{L}$, $P(\log\mathcal{L}^* \geq \log\mathcal{L}|\text{noise})$. The expected number of candidates from noise with log likelihood ratios at least as high as $\log\mathcal{L}$ is then $NP(\log\mathcal{L}^* \geq \log\mathcal{L}|\text{noise})$, where $N$ is the number of observed candidates. The FAR is then the total number of expected candidates from noise divided by the live time of the experiment, $T$, and the $p$ value is obtained by assuming the noise is a Poisson process:

$$\text{FAR} = \frac{NP(\log\mathcal{L}^* \geq \log\mathcal{L}|\text{noise})}{T}, \tag{2}$$

$$p = 1 - e^{-NP(\log\mathcal{L}^* \geq \log\mathcal{L}|\text{noise})}. \tag{3}$$

For the analysis in this paper, GstLAL analyzes the same periods of data as PyCBC. However, FARs are assigned using the distribution of likelihood ratios in noise computed from marginalizing $P(\log\mathcal{L}^* \geq \log\mathcal{L}|\text{noise}, \text{period})$ over all analysis periods; thus, all of $O1$ and $O2$ are used to inform the noise model for FAR assignment. The only exception is GW170608. The analysis period used to estimate the significance of GW170608 is unique from the other ones [17], and thus its FAR is assigned using only its local background statistics.

For this search, GstLAL uses a bank of templates with a total mass between 2 and 400 $M_\odot$ and a mass ratio between 1/98 and 1. Components with a mass less than 2 $M_\odot$ have a maximum spin magnitude of 0.05 (as for PyCBC); otherwise, the spin magnitude is less than 0.999. The TaylorF2 waveform approximant is used to generate templates for systems with a chirp mass [see Eq. (5)] less than 1.73, and the reduced-order model of the SEOBNRv4 approximant is used elsewhere. More details on the bank construction can be found in Ref. [84].

### C. Coherent WaveBurst

Coherent WaveBurst (cWB) is an analysis algorithm used in searches for weakly modeled (or unmodeled) transient signals with networks of GW detectors. Designed to operate without a specific waveform model, cWB identifies coincident excess power in the multiresolution time-frequency representations of the detector strain data [85], for signal frequencies up to 1 kHz and durations up to a few seconds. The search identifies events that are coherent in multiple detectors and reconstructs the source sky location and signal waveforms by using the constrained maximum likelihood method [11]. The cWB detection statistic is based on the coherent energy $E_c$ obtained by cross-correlating the signal waveforms reconstructed in the two detectors. It is proportional to the coherent network SNR and used to rank each cWB candidate event. For an estimation of its statistical significance, each candidate event is ranked against a sample of background triggers obtained by repeating the analysis on time-shifted data, similar to the background estimation in the PyCBC search. To exclude astrophysical events from the background sample, the time shifts are selected to be much larger than the expected signal delay between the detectors. Each cWB event is assigned a FAR given by the rate of background triggers with a larger coherent network SNR.

To increase robustness against nonstationary detector noise, cWB uses signal-independent vetoes, which reduce the high rate of the initial excess power triggers. The primary veto cut is on the network correlation coefficient $c_c = E_c/(E_c + E_n)$, where $E_n$ is the residual noise energy estimated after the reconstructed signal is subtracted from the data. Typically, for a GW signal $c_c \approx 1$, and for instrumental glitches $c_c \ll 1$. Therefore, candidate events with $c_c < 0.7$ are rejected as potential glitches.

Finally, to improve the detection efficiency for a specific class of stellar-mass BBH sources and further reduce the number of false alarms, cWB selects a subset of detected events for which the frequency is increasing with time, i.e., events with a chirping time-frequency pattern. Such a time-frequency pattern captures the phenomenological behavior of most CBC sources. This flexibility allows cWB to potentially identify CBC sources with features such as higher-order modes, high mass ratios, misaligned spins, and eccentric orbits; it complements the existing templated algorithms by searching for new and possibly unexpected CBC populations.

For events that passed the signal-independent vetoes and chirp cut, the detection significance is characterized by a FAR computed as described above; otherwise, cWB provides only the reconstructed waveforms (see Sec. VI).

## IV. SEARCH RESULTS

### A. Selection criteria

In this section, we motivate and describe the selection of gravitational-wave events for presentation in this paper. We include *any* candidate event that can be identified with a nontrivial probability of association to an astrophysical binary merger event, as opposed to instrumental noise [86].

The matched-filter and cWB search pipelines produce large numbers of candidate events, but the majority of these are of very low significance and have a correspondingly low probability of being of astrophysical origin.

We desire to identify all events that are confidently astrophysical in origin and additionally provide a manageable set of marginal triggers that may include some true signals but certainly also includes noise triggers. To do this identification, we establish an initial threshold on estimated FAR of 1 per 30 days (about 12.2 per year), excluding any event that does not have a FAR less than this threshold in at least one of the two matched-filter analyses (see Sec. III). The cWB search results are not used in the event selection process. At this FAR threshold, if each pipeline produces independent noise events, we would expect on average two such noise events (false alarms) per month of analyzed coincident time. During these first two observing runs, we also empirically observe approximately two likely signal events per month of analyzed time. Thus, for *O*1 and *O*2, any sample of events all of whose measured FARs are *greater* than 1 per 30 days is expected to consist of at least 50% noise triggers. Individual triggers within such a sample are then considered to be of little astrophysical interest. Since the number of triggers with a FAR less than 1 per 30 days is manageable, restricting our attention to triggers with lower FAR captures all confident detections while also probing noise triggers.

Within the sample of triggers with a FAR less than the ceiling of 1 per 30 days in at least one of the matched-filter searches, we assign the "GW" designation to any event for which the probability of astrophysical origin from either matched-filter search is greater than 50% (for the exact definition and calculation of the astrophysical probability, see Sec. VII). We list these events in Table I.

For the remaining events in the sample that pass the initial FAR threshold, neither matched-filter search finds a greater than 50% probability of astrophysical origin. These are considered *marginal* events and are listed in Table II. The astrophysical probabilities of all events, confident and marginal, are given in Table IV.

### B. Gravitational-wave events

Results from the two matched-filter searches are shown in Fig. 2 and that of the unmodeled burst search in Fig. 3. In each plot, we show the observed distribution of events as a function of the inverse false-alarm rate, as well as the expected background for the analysis time, with Poisson uncertainty bands. The foreground distributions clearly stand out from the background, even though we show only rightward-pointing arrows for any event with a measured or bounded IFAR greater than 3000 y.

We present more quantitative details below on the 11 gravitational events, as selected by the criteria in Sec. IV A, in Table I. Of these 11 events, seven have been previously reported: the three gravitational-wave events from *O*1 [1–4] and, from *O*2, the binary neutron star merger GW170817 [18] and the binary black hole events GW170104 [15], GW170608 [17], and GW170814 [16]. The updated results we report here supersede those previously published. Four new gravitational-wave events are reported here for the first time: GW170729, GW170809, GW170818, and GW170823. All four are binary black hole events.

As noted in Sec. III, data from *O*1 are reanalyzed because of improvements in the search pipelines and the expansion of the parameter space searched. For the $O2$ events already published, our reanalysis is motivated by updates to the data itself. The noise subtraction procedure [52] that is available for parameter estimation of three of the published $O2$ events was not initially applied to the entire $O2$ dataset and, therefore, could not be used by searches. Following the procedures of Ref. [51], this noise subtraction is applied to all of $O2$ and is reflected in Table I for the four previously published $O2$ GW events, as well as the four events presented here for the first time.

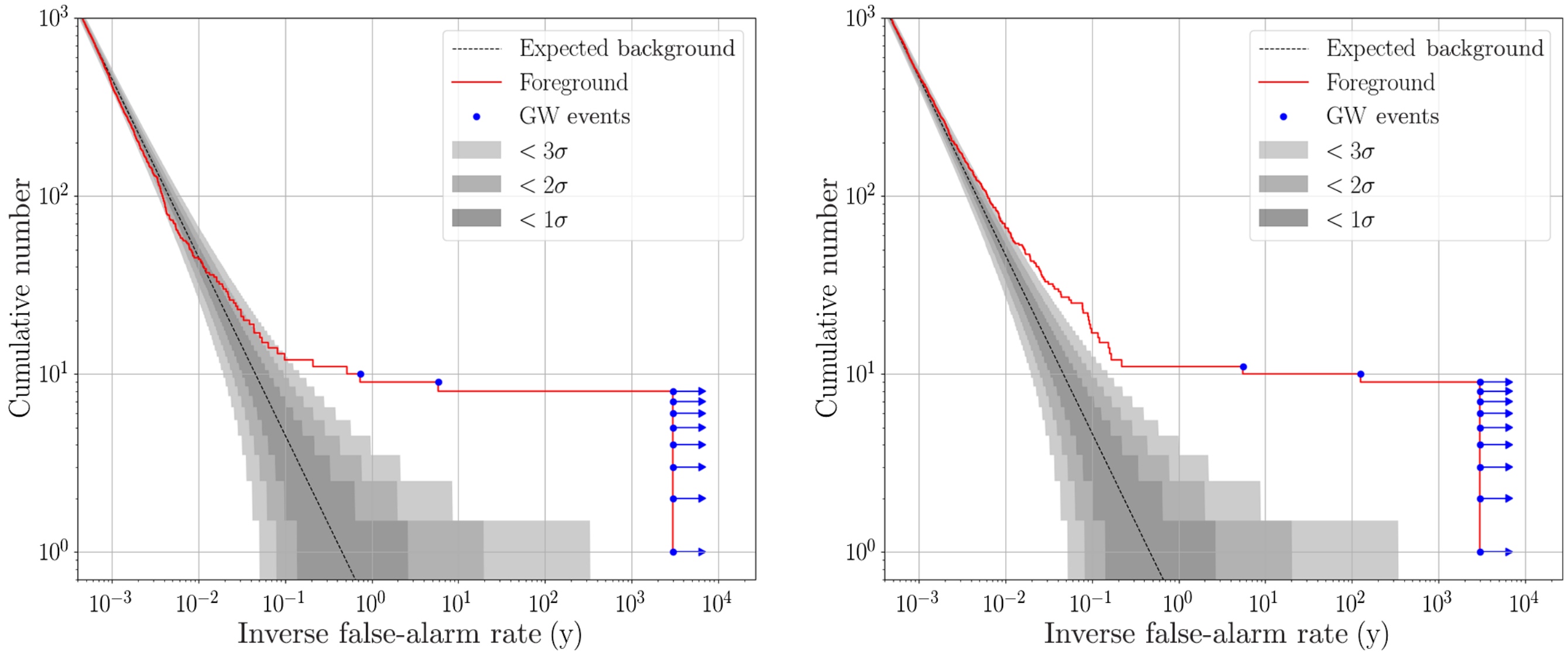

FIG. 2. Cumulative histograms of search results for the matched-filter searches, plotted versus inverse false-alarm rate. The dashed lines show the expected background, given the analysis time. Shaded regions denote sigma uncertainty bounds for Poisson uncertainty. The blue dots are the named gravitational-wave events found by each respective search. Any events with a measured or bounded inverse false-alarm rate greater than 3000 y are shown with an arrow pointing right. Left: PyCBC results. Right: GstLAL results.

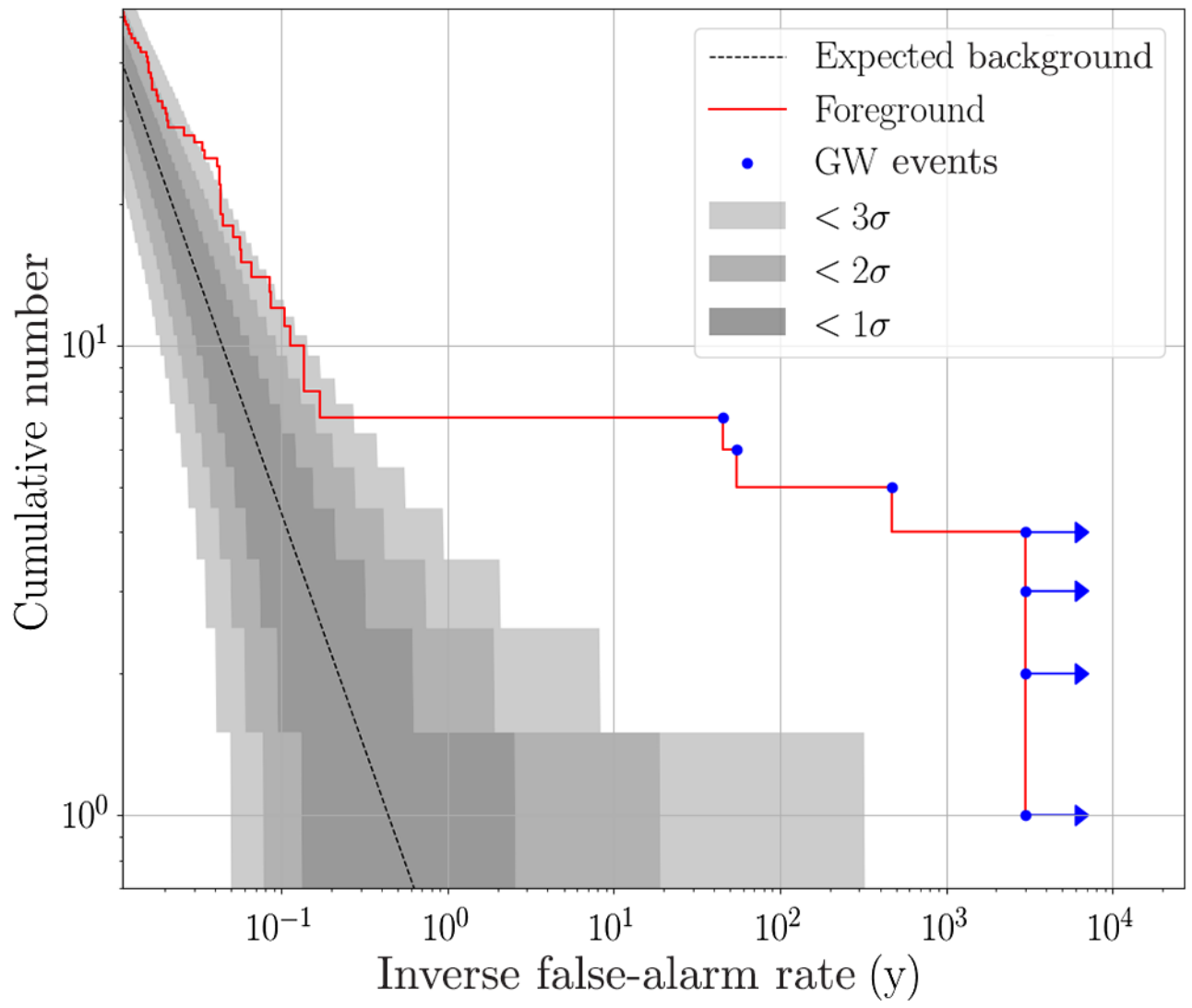

FIG. 3. Cumulative histograms of search results for the cWB search, plotted versus the inverse false-alarm rate. The dashed lines show the expected background, given the analysis time. Shaded regions denote sigma uncertainty bounds for Poisson uncertainty. The blue dots are the named gravitational-wave events found by each respective search. Any events with a measured or bounded inverse false-alarm rate greater than 3000 y are shown with an arrow pointing right.

For both PyCBC and cWB, the time-shift method of background estimation may result in only an upper bound on the false-alarm rate, if an event has a larger value of the ranking statistic than any trigger in the time-shifted background; this result is indicated in Table I. For GW150914 and GW151226, the bound that PyCBC places on the FAR in these updated results is in fact higher than that previously published [1,2,4], because, as noted in Sec. III A, this search elects to use shorter periods of time shifting to better capture the variation in the detectors' sensitivities. For GstLAL, the FAR is reported in Table I as an upper bound of $1.00 \times 10^{-7}$ whenever a smaller number is obtained, which reflects a more conservative noise hypothesis within the GstLAL analysis and follows the procedures and motivations detailed in Sec. IV in Ref. [3].

Five of the GW events reported here occurred during August 2017, which comprises approximately 10% of the total observation time. There are ten nonoverlapping periods of similar duration, with an average event rate of 1.1 per period. The probability that a Poisson process would produce five events or more in at least one of those periods is 5.3%. Thus, seeing five events in one month is statistically consistent with expectations. For more details, see Ref. [87].

For the remainder of this section, we briefly discuss each of the gravitational-wave events, highlighting interesting features from the perspectives of the three searches. A discussion of the properties of these sources may be found in Sec. V. Though the results presented are from the final, offline analysis of each search, for the four new GW events, we also indicate whether the event is found in a low-latency search and an alert sent to electromagnetic observing partners. Where this process did occur, we mention in this paper only the low-latency versions of the three searches with offline results presented here; in some cases, additional low-latency pipelines also found events. A more thorough discussion of all of the low-latency analyses and the electromagnetic follow-up of $O2$ events may be found in Ref. [22].

#### 1. GW150914, GW151012, and GW151226

During $O1$, two confident detections of binary black holes were made: GW150914 [1] and GW151226 [2]. Additionally, a third trigger was noted in the $O1$ catalog of binary black holes [3,4] and labeled LVT151012. That label is a consequence of the higher FAR of that trigger, though detector characterization studies show no instrumental or environmental artifact, and the results of parameter estimation are consistent with an astrophysical BBH source. Even with the significance that is measured with the $O1$ search pipelines [4], this event meets the criteria of Sec. IV A for a gravitational-wave event, and we henceforth relabel this event as GW151012.

The improved $O2$ pipelines substantially reduce the FAR assigned to GW151012: It is now $0.17\ \mathrm{y}^{-1}$ in the PyCBC search (previously, $0.37\ \mathrm{y}^{-1}$) and $7.92 \times 10^{-3}\ \mathrm{y}^{-1}$ in the GstLAL search (previously, $0.17\ \mathrm{y}^{-1}$). These improved FAR measurements for GW151012 are the most salient result of the reanalysis of $O1$ with the $O2$ pipelines; no new gravitational-wave events were discovered. The first binary black hole observation, GW150914, remains the highest SNR event in $O1$ and the second highest in the combined $O1$ and $O2$ datasets, behind only the binary neutron star inspiral GW170817.

Recently, Ref. [88] appeared. That catalog also presents search results from the PyCBC pipeline for $O1$ and also finds GW150914, GW151012, and GW151226 as the only confident gravitational-wave events in $O1$, with identical bounds on FAR to the PyCBC results in Table I for GW150914 and GW151226. The measured FAR for GW151012 is not identical but is consistent with the results we present in Table I.

#### 2. GW170104, GW170608, and GW170814

Three binary black hole events from $O2$ have already been published: GW170104 [15], GW170608 [17], and GW170814 [16]. Updated search results for these events are presented in Table I. As noted in the original publication for GW170608 [17], the Hanford detector was undergoing a procedure to stabilize angular noise at the time of the event; the Livingston detector was operating in a nominal configuration. For this reason, a specialized analysis time when both LIGO detectors were operating in that same configuration is identified, between June 7, 2017 and June 9, 2017. This period that was used to analyze GW170608 in the initial publication is again used for the results in Table I, though with the noise subtraction applied.

In the reanalysis of $O2$ data, GW170814 is identified as a double-coincident event between LLO and LHO by GstLAL. This results from the noise subtraction in the LIGO data and updated calibration of the Virgo data. Because of the noise subtraction in the LIGO data, under GstLAL's ranking of multiple triggers [9], a new template generates the highest ranked trigger as double coincident, with a Hanford SNR of 9.1 (the previous highest ranked trigger, a triple, had 7.3). Though this highest ranked event is a double-coincident trigger, the pipeline does identify other highly significant triggers, some double coincident and some triple coincident. As the search uses a discrete template bank, peaks from the SNR time series of the individual detectors, and clustering of several coincident triggers over the bank, it is difficult in this case to tell from the search results alone whether the event is truly a triple-coincident detection. For a definitive answer, we perform a fully Bayesian analysis with and without the Virgo data, similar to the results in Ref. [16]. Comparing the evidence, this Bayesian analysis—which enforces coherence and therefore more fully exploits consistency among detected amplitudes, phases, and times of arrival than the search pipelines—finds that a triple-coincident detection is strongly favored over a double-coincident detection, by a factor of approximately 60. Thus, the updated results are consistent with those that were previously published.

### 3. GW170817

Across the entirety of $O1$ and $O2$, the binary neutron star inspiral GW170817 remains the event with the highest network SNR and is accordingly assigned the most stringent possible bound on its FAR by PyCBC and the highest value of $\mathcal{L}$ (the logarithm of the likelihood ratio) of any event in the combined $O1$ and $O2$ dataset by GstLAL. As explained in detail in the original detection paper [18], a loud glitch occurs near the end of this signal in LLO. For the matched-filter searches, this glitch is excised via time-domain gating (and that gating is applied consistently to all such glitches throughout $O2$). Because the cWB pipeline is designed to detect short signals, it does not use that gating technique, and it rejects this event because of the glitch.

### 4. GW170729

We turn now to gravitational-wave events not previously announced. The first of these is GW170729, observed at 18:56:29.3 UTC on July 29, 2017. The PyCBC pipeline assigns it a FAR of 1.36 $\mathrm{y}^{-1}$, the GstLAL pipeline a FAR of 0.18 $\mathrm{y}^{-1}$, and the cWB pipeline a FAR of 0.02 $\mathrm{y}^{-1}$. As it is identified with the highest significance among all three search pipelines by the weakly modeled pipeline, it is worth investigating whether this event is unusual in some way, exhibiting effects (for instance, precession or higher-order modes) not adequately modeled by the templates used in the matched-filter searches. As a relatively simple way of investigating this question, a comparison study is done between the PyCBC pipeline and cWB, using software injections with parameters drawn from the SEOBNRv4 ROM parameter estimation of this event. That waveform does *not* incorporate precession or higher-order modes, but, by using these samples as inputs to both searches, we can probe how often we see comparable results. It is found that approximately 4% of these SEOBNRv4 ROM samples are recovered by both the PyCBC and cWB pipelines with FAR $\geq 1$ $\mathrm{y}^{-1}$ and FAR $\leq 0.02$ $\mathrm{y}^{-1}$, respectively. Thus, the observed difference in FARs between the two pipelines is not exceptionally unlikely and is consistent with a noise fluctuation which happens to decrease the significance of the event as seen by PyCBC and increase it for cWB. The detailed CBC parameter estimation studies in Sec. V also indicate no significant evidence for observationally important precession or higher-order modes. This event was identified only in the offline analyses, so no alert was sent to electromagnetic partners.

### 5. GW170809

GW170809 was observed on August 9, 2017 at 08:28:21.8 UTC with a FAR of $1.45 \times 10^{-4}$ $\mathrm{y}^{-1}$ by PyCBC and $<1.00 \times 10^{-7}$ $\mathrm{y}^{-1}$ by GstLAL. This event was identified in low latency by both the GstLAL and cWB pipelines, and an alert was sent to electromagnetic observing partners. In the final offline cWB analysis with updated calibration and noise subtracted from LIGO data, this event did not pass one of the signal-independent vetoes (Sec. III C) and was therefore not assigned a FAR.

### 6. GW170818

GW170818 was observed at 02:25:09.1 UTC on August 18, 2017, by GstLAL with a FAR of $4.20 \times 10^{-5}$ $\mathrm{y}^{-1}$; it was not observed by either the PyCBC or cWB pipelines. It is observed as a triple-coincident event by GstLAL, with an SNR in Virgo of 4.2, a Hanford SNR of 4.1, and a Livingston SNR of 9.7. In the PyCBC search, a trigger is seen in the Livingston detector with a comparable SNR and is noted as a "chirplike" single-detector trigger. When the Hanford and Virgo data are analyzed with modified settings around the time of that event, there are triggers with a similar SNR to those of GstLAL and, therefore, well below the threshold of 5.5 needed for a single-detector trigger in the PyCBC search to be considered further for possible coincidence. This event is initially identified in low latency by the GstLAL pipeline as a LLO-Virgo double-detector trigger. Online, the Virgo trigger is not included in significance estimation, and the LLO-only trigger does not pass the false-alarm threshold for the online search. Therefore, at that time no alerts was sent to electromagnetic observing partners.

#### 7. GW170823

On August 23, 2017, GW170823 was observed at 13:13:58.5 UTC. Its FAR is $<3.29\times10^{-5}\ \mathrm{y}^{-1}$ in the PyCBC pipeline, $<1.00\times10^{-7}\ \mathrm{y}^{-1}$ in the GstLAL pipeline, and $2.14\times10^{-3}\ \mathrm{y}^{-1}$ in the cWB pipeline. The online versions of each of these pipelines detect this event in low latency, and an alert is sent to electromagnetic observing partners.

### C. Marginal triggers and instrumental artifacts

In Table II, we present the remaining 14 triggers from $O1$ and $O2$ that pass the initial threshold of a FAR less than one per 30 days in at least one of the two matched-filter searches but are not assigned a probability of astrophysical origin of more than 50% by either pipeline (see Table IV). As noise triggers are generically a function of the details of the pipeline that identifies a trigger, we do not typically expect to see the same noise triggers in each pipeline. In Table II, we therefore present which of the two pipelines identified the trigger, as well as the FAR of that trigger, its SNR, and the chirp mass of the template generating the trigger; these chirp masses do *not* come from a detailed parameter estimation as is performed in Sec. V for the gravitational-wave events.

Before discussing the final column in Table II, we consider the reasonableness of the number of these triggers. The matched-filter pipelines analyze 0.46 y of coincident data, so, at a false-alarm threshold of once per 30 days, we would expect about six triggers purely from noise. We see from Table II that the PyCBC search observed three marginal triggers and the GstLAL search observed 11. Though the probability that 11 triggers could arise only from noise when six are expected is low, it is by itself not sufficiently low to confidently assert that some fraction of these triggers are astrophysical in origin. It is possible, however, that either search's marginal triggers could contain a population of real GW signals. In particular, the multicomponent population analysis [89,90] (see Sec. VII) explicitly considers the possibility of triggers (both confident detections and marginal triggers) arising from a combination of noise and distinct source populations. For GstLAL, the combined count of GWs and marginal triggers is 22, and the analysis in Sec. VII finds that to be within expectations at the 90% level. Although it may be the case that some of these marginal triggers are of astrophysical origin, we cannot then determine which ones.

Now we turn to a summary of the detector characterization information for each marginal trigger, briefly indicated in the final column in Table II. Following a subset of procedures used for previous gravitational-wave detections [91], we evaluate the possibility that artifacts from instrumental or environmental noise could have caused each of the marginal triggers. Using auxiliary sensors at each detector, as well as the gravitational-wave strain data, we evaluate the state of the detectors at the time of each marginal trigger, identify and investigate any artifacts in the data due to noise, and test whether any identified artifacts might explain the excess SNR observed in the analysis. Of the marginal triggers presented in this catalog, nine have excess power from known sources of noise occurring during times when the matched-filter template that yields the trigger has a GW frequency within the sensitive band of the detectors. For four of these cases, the observed instrumental artifact overlaps the signal region and possibly causes the marginal trigger.

Details on the physical couplings that create these instrumental artifacts and possible mitigation strategies useful for analysis of LIGO-Virgo data are discussed in Appendix A. For the remainder of this subsection, we describe how the different categories discussed in that Appendix apply to the marginal triggers in Table II.

To determine whether artifacts identified as noise "could account for" marginal triggers, we use two metrics: (i) whether the type of noise has been shown to produce an excess of triggers consistent with the properties of the trigger present and (ii) the noise artifact could account for the presence of the trigger as reported by that search, including SNR and time-frequency evolution, without the presence of an astrophysical signal.

In and of themselves, these classifications do not affect the probability that any particular marginal trigger is associated with a signal as measured by the searches but are statements about the evidence of transient noise in the detectors. It is expected that a substantial fraction of marginal events at the false-alarm rate values reported are caused by noise, given the estimated background of our searches and the expected rate of signals. See Sec. VII, Fig. 11, and Table IV below for a more detailed discussion of the probabilities of astrophysical or noise origin of such events.

#### 1. No noise artifacts present: 151008, 151116, 170219, 170423, and 170705

Investigations into this set of marginal triggers have identified no instrumental artifacts in time coincidence with the triggers.

#### 2. Light scattering possibly caused: 161217 and 170720

All marginal triggers in this class and the next are in time coincidence with artifacts from scattered light in one of the detectors. Scattered light leads to excess power at low frequencies that appear in time-frequency spectrograms as archlike shapes. In some cases, the frequencies affected are above the minimum frequency used in the analysis. When this happens, scattered light transients can create significant triggers in matched-filter searches [68,92,93].

The two marginal triggers 161217 and 170720 occur during periods of scattered light affecting frequencies up to 80 Hz with high-amplitude arches. In both cases, a significant overlap with the trigger template and the excess power from scattering is observed. Investigations into the status of the observatories at the times in question identified high-amplitude ground motion correlated with the scattering.

The marginal trigger 161217 occurs during a period of high-amplitude ground motion at Livingston caused by storm activity. During this storm activity, the Livingston detector is not able to maintain a stable interferometer for periods longer than 10 min. The presence of intense scattering artifacts contribute to the unstable state of the interferometer and could account for the SNR of the marginal trigger. Because of the short observing duration, this time period is not analyzed by the PyCBC search.

Within 20 s of trigger 170720, excess ground motion from earthquakes forced the Livingston detector to drop out of its nominal mode of operation. Before the detector dropped out of the observing state, the data are heavily polluted with scattering artifacts that could account for the SNR of the triggers. As the PyCBC search does not consider times near the edges of observing periods, this time period is also not analyzed by that search. Artifacts related to scattered light are also observed at Hanford at this time.

#### 3. Light scattering present: 151012.2, 170208, and 170616

In the case of trigger 151012.2, light scattering does not introduce significant power above 30 Hz prior to the reported trigger time. Investigations into the relationship between the trigger and the scattered light find no power overlap, suggesting that the artifacts could not account for the observed marginal trigger.

Investigations into triggers 170208 and 170616 find similar results. In the case of these triggers, a slight overlap with excess power from scattering is observed. Multiple efforts, including BayesWave [53] glitch subtraction and gating [8], are used to mitigate the scattered light artifacts. After subtraction of the noise artifacts, the data are reanalyzed to evaluate whether the excess power subtracted could have accounted for the trigger. In both cases, the marginal trigger remains with similar significance, suggesting that the observed scattering artifact could not have accounted for the SNR of the marginal trigger.

#### 4. 60–200 Hz nonstationarity possibly caused: 161202 and 170412

This class of marginal triggers occurs during periods of noise referred to as “60–200 Hz nonstationarity.” This nonstationarity appears in time-frequency spectrograms as excess power with slowly varying frequencies over time periods of multiple minutes.

Previous work [68] has shown that periods of 60–200 Hz nonstationarity can cause significant triggers in the searches, both impacting the ability of searches to accurately measure the noise spectrum of the data and contributing excess noise to matched-filter searches. Triggers 161202 and 170412 demonstrate a significant overlap with excess power from the nonstationarity noise. BayesWave [53] glitch subtraction is unable to completely mitigate the 60–200 Hz nonstationarity due to its long duration.

#### 5. Short-duration, high-amplitude artifacts present: 170405 and 170630

The marginal triggers in this class occur in time coincidence with short-duration, high-amplitude noise transients that are removed in the data-conditioning step of the search pipelines [8]. The times surrounding these transients do not demonstrate an elevated trigger rate after the transient has been removed. Trigger 170405 is in coincidence with this class of transient at Hanford, and trigger 170630 is in coincidence with this class of transient at Livingston. As triggers 170405 and 170630 are identified as significant after the removal of the short-duration transients, the presence of noise artifacts cannot account for the SNR of these marginal triggers.

## V. SOURCE PROPERTIES

Here, we present inferred source properties of gravitational-wave signals observed by the LIGO and Virgo detectors under the assumption that they originate from compact binary coalescences described by general relativity. We analyze all GW events described in Sec. IV. Full parameter estimation (PE) results for $O1$ events are provided for GW150914 in Refs. [4,95,96], for GW151226 in Refs. [2,4], and for GW151012 in Refs. [3,4]. PE results for four $O2$ events are provided for GW170104 in Ref. [15], for GW170608 in Ref. [17], for GW170814 in Ref. [16], and for GW170817 in Refs. [18,97]. Data from the three-detector LIGO-Virgo network are used to obtain parameter estimates for GW170729, GW170809, GW170814, GW170817, and GW170818. Virgo data for GW170729, from the commissioning phase before it officially joined, are included for PE analyses, because the calibration of the data and the sensitivity of the instrument are comparable to that in August. For the remaining events, the analysis uses data from the two LIGO detectors.

We perform a reanalysis of the data for the entirety of $O1$ and $O2$ including the published events. As discussed in Sec. II C, the $O2$ data are recalibrated and cleaned [52]. These improvements increase the sensitivity of the detector network and motivate a reanalysis also for already-published events found in $O2$ data. While the $O1$ data and calibration have not changed, a reanalysis is valuable for the following reasons: (i) Parameter estimation analyses use an improved method for estimating the power spectral density of the detector noise [53,54] and frequency-dependent calibration envelopes [98]; (ii) we use two waveform models that incorporate precession and combine their posteriors to mitigate model uncertainties.

TABLE III. Selected source parameters of the 11 confident detections. We report median values with 90% credible intervals that include statistical errors and systematic errors from averaging the results of two waveform models for BBHs. For GW170817, credible intervals and statistical errors are shown for IMRPhenomPv2NRT with a low spin prior, while the sky area is computed from TaylorF2 samples. The redshift for NGC 4993 from Ref. [94] and its associated uncertainties are used to calculate source-frame masses for GW170817. For BBH events, the redshift is calculated from the luminosity distance and assumed cosmology as discussed in Appendix B. The columns show source-frame component masses $m_i$ and chirp mass $\mathcal{M}$, dimensionless effective aligned spin $\chi_{\rm eff}$, final source-frame mass $M_f$, final spin $a_f$, radiated energy $E_{\rm rad}$, peak luminosity $l_{\rm peak}$, luminosity distance $d_L$, redshift $z$, and sky localization $\Delta\Omega$. The sky localization is the area of the 90% credible region. For GW170817, we give conservative bounds on parameters of the final remnant discussed in Sec. V E.

| Event | $m_1/M_\odot$ | $m_2/M_\odot$ | $\mathcal{M}/M_\odot$ | $\chi_{\rm eff}$ | $M_f/M_\odot$ | $a_f$ | $E_{\rm rad}/(M_\odot c^2)$ | $\ell_{\rm peak}/({\rm erg\,s^{-1}})$ | $d_L$/Mpc | $z$ | $\Delta\Omega/{\rm deg}^2$ |
|---|---|---|---|---|---|---|---|---|---|---|---|
| GW150914 | $35.6^{+4.7}_{-3.1}$ | $30.6^{+3.0}_{-4.4}$ | $28.6^{+1.7}_{-1.5}$ | $-0.01^{+0.12}_{-0.13}$ | $63.1^{+3.4}_{-3.0}$ | $0.69^{+0.05}_{-0.04}$ | $3.1^{+0.4}_{-0.4}$ | $3.6^{+0.4}_{-0.4}\times10^{56}$ | $440^{+150}_{-170}$ | $0.09^{+0.03}_{-0.03}$ | 182 |
| GW151012 | $23.2^{+14.9}_{-5.5}$ | $13.6^{+4.1}_{-4.8}$ | $15.2^{+2.1}_{-1.2}$ | $0.05^{+0.31}_{-0.20}$ | $35.6^{+10.8}_{-3.8}$ | $0.67^{+0.13}_{-0.11}$ | $1.6^{+0.6}_{-0.5}$ | $3.2^{+0.8}_{-1.7}\times10^{56}$ | $1080^{+550}_{-490}$ | $0.21^{+0.09}_{-0.09}$ | 1523 |
| GW151226 | $13.7^{+8.8}_{-3.2}$ | $7.7^{+2.2}_{-2.5}$ | $8.9^{+0.3}_{-0.3}$ | $0.18^{+0.20}_{-0.12}$ | $20.5^{+6.4}_{-1.5}$ | $0.74^{+0.07}_{-0.05}$ | $1.0^{+0.1}_{-0.2}$ | $3.4^{+0.7}_{-1.7}\times10^{56}$ | $450^{+180}_{-190}$ | $0.09^{+0.04}_{-0.04}$ | 1033 |
| GW170104 | $30.8^{+7.3}_{-5.6}$ | $20.0^{+4.9}_{-4.6}$ | $21.4^{+2.2}_{-1.8}$ | $-0.04^{+0.17}_{-0.21}$ | $48.9^{+5.1}_{-4.0}$ | $0.66^{+0.08}_{-0.11}$ | $2.2^{+0.5}_{-0.5}$ | $3.3^{+0.6}_{-1.0}\times10^{56}$ | $990^{+440}_{-430}$ | $0.20^{+0.08}_{-0.08}$ | 921 |
| GW170608 | $11.0^{+5.5}_{-1.7}$ | $7.6^{+1.4}_{-2.2}$ | $7.9^{+0.2}_{-0.2}$ | $0.03^{+0.19}_{-0.07}$ | $17.8^{+3.4}_{-0.7}$ | $0.69^{+0.04}_{-0.04}$ | $0.9^{+0.0}_{-0.1}$ | $3.5^{+0.4}_{-1.3}\times10^{56}$ | $320^{+120}_{-110}$ | $0.07^{+0.02}_{-0.02}$ | 392 |
| GW170729 | $50.2^{+16.2}_{-10.2}$ | $34.0^{+9.1}_{-10.1}$ | $35.4^{+6.5}_{-4.8}$ | $0.37^{+0.21}_{-0.25}$ | $79.5^{+14.7}_{-10.2}$ | $0.81^{+0.07}_{-0.13}$ | $4.8^{+1.7}_{-1.7}$ | $4.2^{+0.9}_{-1.5}\times10^{56}$ | $2840^{+1400}_{-1360}$ | $0.49^{+0.19}_{-0.21}$ | 1041 |
| GW170809 | $35.0^{+8.3}_{-5.9}$ | $23.8^{+5.1}_{-5.2}$ | $24.9^{+2.1}_{-1.7}$ | $0.08^{+0.17}_{-0.17}$ | $56.3^{+5.2}_{-3.8}$ | $0.70^{+0.08}_{-0.09}$ | $2.7^{+0.6}_{-0.6}$ | $3.5^{+0.6}_{-0.9}\times10^{56}$ | $1030^{+320}_{-390}$ | $0.20^{+0.05}_{-0.07}$ | 308 |
| GW170814 | $30.6^{+5.6}_{-3.0}$ | $25.2^{+2.8}_{-4.0}$ | $24.1^{+1.4}_{-1.1}$ | $0.07^{+0.12}_{-0.12}$ | $53.2^{+3.2}_{-2.4}$ | $0.72^{+0.07}_{-0.05}$ | $2.7^{+0.4}_{-0.3}$ | $3.7^{+0.4}_{-0.5}\times10^{56}$ | $600^{+150}_{-220}$ | $0.12^{+0.03}_{-0.04}$ | 87 |
| GW170817 | $1.46^{+0.12}_{-0.10}$ | $1.27^{+0.09}_{-0.09}$ | $1.186^{+0.001}_{-0.001}$ | $0.00^{+0.02}_{-0.01}$ | $\leq 2.8$ | $\leq 0.89$ | $\geq 0.04$ | $\geq 0.1\times10^{56}$ | $40^{+7}_{-15}$ | $0.01^{+0.00}_{-0.00}$ | 16 |
| GW170818 | $35.4^{+7.5}_{-4.7}$ | $26.7^{+4.3}_{-5.2}$ | $26.5^{+2.1}_{-1.7}$ | $-0.09^{+0.18}_{-0.21}$ | $59.4^{+4.9}_{-3.8}$ | $0.67^{+0.07}_{-0.08}$ | $2.7^{+0.5}_{-0.5}$ | $3.4^{+0.5}_{-0.7}\times10^{56}$ | $1060^{+420}_{-380}$ | $0.21^{+0.07}_{-0.07}$ | 39 |
| GW170823 | $39.5^{+11.2}_{-6.7}$ | $29.0^{+6.7}_{-7.8}$ | $29.2^{+4.6}_{-3.6}$ | $0.09^{+0.22}_{-0.26}$ | $65.4^{+10.1}_{-7.4}$ | $0.72^{+0.09}_{-0.12}$ | $3.3^{+1.0}_{-0.9}$ | $3.6^{+0.7}_{-1.1}\times10^{56}$ | $1940^{+970}_{-900}$ | $0.35^{+0.15}_{-0.15}$ | 1666 |

Key source parameters for the ten BBHs and one BNS are shown in Table III. We quote the median and symmetric 90% credible intervals for inferred quantities. For BBH coalescences, parameter uncertainties include statistical and systematic errors from averaging posterior probability distributions over the two waveform models, as well as calibration uncertainty. Apart from GW170817, all posterior distributions of GW events are consistent with originating from BBHs. Posterior distributions for all GW events are shown in Figs. 4–8. Mass and tidal deformability posteriors for GW170817 are shown in Fig. 9. For BBH coalescences, we present combined posterior distributions from an effective precessing spin waveform model (IMRPhenomPv2) [25,26,49] and a fully precessing model (SEOBNRv3) [27,28,30]. For the analysis of GW170817, we present results for three frequency-domain models IMRPhenomPv2NRT [25,26,32,49,99], SEOBNRv4NRT [29,32,77,99], and TaylorF2 [35,36,38,100–112] and two time-domain models SEOBNRv4T [31] and TEOBResumS [33,113]. Details on Bayesian parameter estimation methods, prior choices, and waveform models used for BBH and BNS systems are provided in Appendix B, B 1, and B 2, respectively. We discuss an analysis including higher harmonics in the waveform in Appendix B 3 and find results broadly consistent with the analysis presented below. The impact of prior choices on selected results is discussed in Appendix C.

### A. Source parameters

The GW signal emitted from a BBH coalescence depends on intrinsic parameters that directly characterize the binary's dynamics and emitted waveform, and extrinsic parameters that encode the relation of the source to the detector network. In general relativity, an isolated BH is uniquely described by its mass, spin, and electric charge [114–118]. For astrophysical BHs, we assume the electric charge to be negligible. A BBH undergoing quasicircular inspiral can be described by eight intrinsic parameters, the two masses $m_i$, and the two three-dimensional spin vectors $\vec{S}_i$ of its component BHs defined at a reference frequency. Seven additional extrinsic parameters are needed to describe a BH binary: the sky location (right ascension $\alpha$ and declination $\delta$), luminosity distance $d_L$, the orbital inclination $\iota$ and polarization angle $\psi$, the time $t_c$, and phase $\phi_c$ at coalescence.

Since the maximum spin a Kerr BH of mass $m$ can reach is $(Gm^2)/c$, we define dimensionless spin vectors $\vec{\chi}_i = c\vec{S}_i/(Gm_i^2)$ and spin magnitudes $a_i = c|\vec{S}_i|/(Gm_i^2)$. If the spins have a component in the orbital plane, then the binary's orbital angular momentum $\vec{L}$ and its spin vectors precess [119,120] around the total angular momentum $\vec{J} = \vec{L} + \vec{S}_1 + \vec{S}_2$.

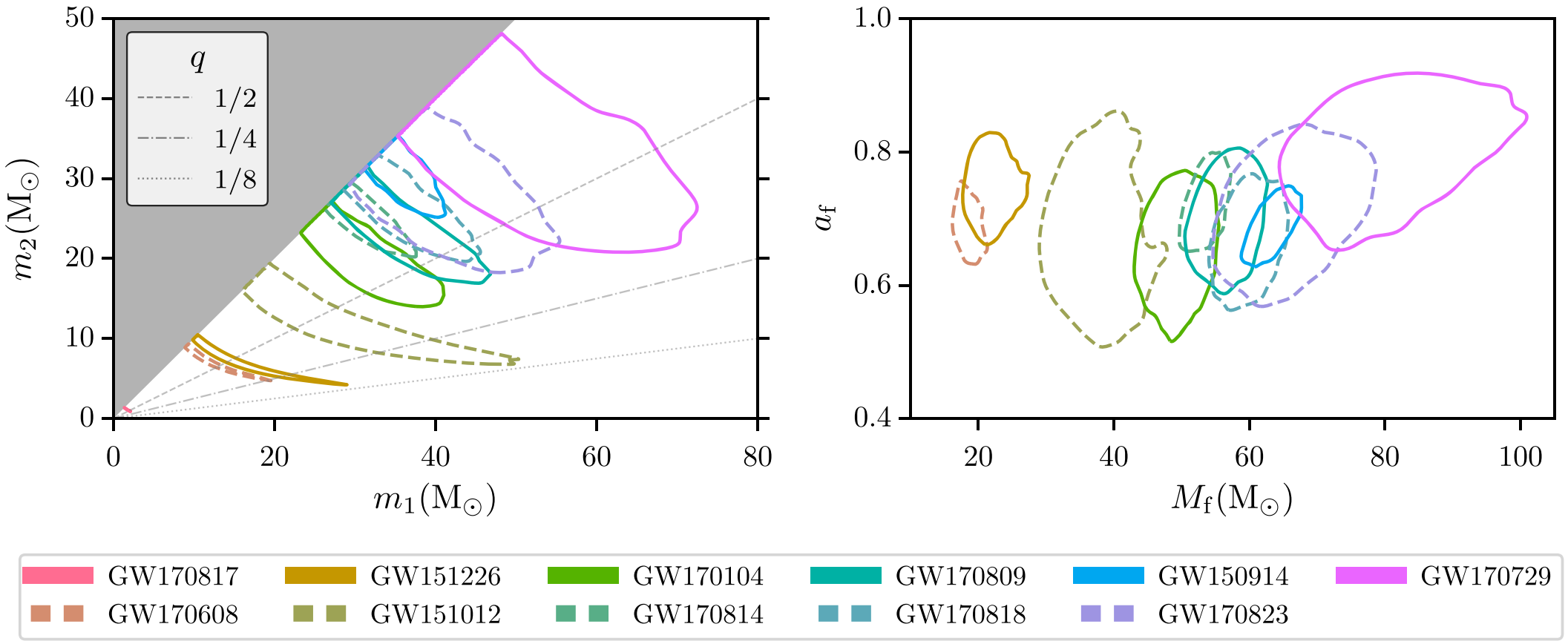

FIG. 4. Parameter estimation summary plots I. Posterior probability densities of the component masses and final masses and spins of the GW events. For the two-dimensional distributions, the contours show 90% credible regions. Left: Source-frame component masses $m_1$ and $m_2$. We use the convention that $m_1 \geq m_2$, which produces the sharp cut in the two-dimensional distribution. Lines of constant mass ratio $q = m_2/m_1$ are shown for $1/q = 2, 4, 8$. For low-mass events, the contours follow lines of constant chirp mass. Right: The mass $M_f$ and dimensionless spin magnitude $a_f$ of the final black holes. The colored event labels are ordered by source-frame chirp mass. The same color code and ordering (where appropriate) apply to Figs. 5–8.

We describe the dominant spin effects by introducing effective parameters. The effective aligned spin is defined as a simple mass-weighted linear combination of the spins [23,24,121] projected onto the Newtonian angular momentum $\hat{L}_N$, which is normal to the orbital plane ($\hat{L} = \hat{L}_N$ for aligned-spin binaries)

$$\chi_{\rm eff} = \frac{(m_1\vec{\chi}_1 + m_2\vec{\chi}_2)\cdot\hat{L}_N}{M}, \tag{4}$$

where $M = m_1 + m_2$ is the total mass of the binary and $m_1$ is defined to be the mass of the larger component of the binary, such that $m_1 \geq m_2$. Different parameterizations of spin effects are possible and can be motivated from their appearance in the GW phase or dynamics [122–124]. $\chi_{\rm eff}$ is approximately conserved throughout the inspiral [121]. To assess whether a binary is precessing, we use a single effective precession spin parameter $\chi_p$ [125] (see Appendix C).

During the inspiral, the phase evolution depends at leading order on the chirp mass [34,126,127]

$$\mathcal{M} = \frac{(m_1 m_2)^{3/5}}{M^{1/5}}, \tag{5}$$

which is also the best measured parameter for low-mass systems dominated by the inspiral [63,101,122,128]. The mass ratio

$$q = \frac{m_2}{m_1} \leq 1 \tag{6}$$

and effective aligned spin $\chi_{\rm eff}$ appear in the phasing at higher orders [101,121,123].

For precessing binaries, the orbital angular momentum vector $\vec{L}$ is not a stable direction, and it is preferable to describe the source inclination by the angle $\theta_{JN}$ between the total angular momentum $\vec{J}$ (which typically is approximately constant throughout the inspiral) and the line-of-sight vector $\vec{N}$ instead of the orbital inclination angle $\iota$ between $\vec{L}$ and $\vec{N}$ [119,129]. We quote frequency-dependent quantities such as spin vectors and derived quantities as $\chi_p$ at a GW reference frequency $f_{\rm ref} = 20$ Hz.

Binary neutron stars have additional degrees of freedom (d.o.f.) related to their response to a tidal field. The dominant quadrupolar ($\ell = 2$) tidal deformation is described by the dimensionless tidal deformability $\Lambda = (2/3)k_2[(c^2/G)(R/m)]^5$ of each neutron star (NS), where $k_2$ is the dimensionless $\ell = 2$ Love number and $R$ is the NS radius. The tidal deformabilities depend on the NS mass $m$ and the equation of state (EOS). The dominant tidal contribution to the GW phase evolution is encapsulated in an effective tidal deformability parameter [130,131]:

$$\tilde{\Lambda} = \frac{16}{13}\frac{(m_1 + 12m_2)m_1^4\Lambda_1 + (m_2 + 12m_1)m_2^4\Lambda_2}{M^5}. \tag{7}$$

### B. Masses

In the left panel in Fig. 4, we show the inferred component masses of the binaries in the source frame as contours in the $m_1$-$m_2$ plane. Because of the mass prior, we consider only systems with $m_1 \geq m_2$ and exclude the shaded region. The component masses of the detected BH binaries cover a wide range from about 5 $M_\odot$ to about 70 $M_\odot$ and lie within the range expected for stellar-mass BHs [132–134]. The posterior distribution of the heavier component in the heaviest BBH, GW170729, grazes the lower boundary of the possible mass gap expected from pulsational pair instability and pair instability supernovae at approximately 60–120 $M_\odot$ [135–138]. The lowest-mass BBH systems, GW151226 and GW170608, have 90% credible lower bounds on $m_2$ of 5.6 $M_\odot$ and 5.9 $M_\odot$, respectively, and therefore lie above the proposed BH mass gap region [139–142] of 2–5 $M_\odot$. The component masses of the BBHs show a strong degeneracy with each other. Lower-mass systems are dominated by the inspiral of the binary, and the component mass contours trace out a line of constant chirp mass Eq. (5) which is the best measured parameter in the inspiral [34,63,122]. Since higher-mass systems merge at a lower GW frequency, their GW signal is dominated by the merger of the binary. For high-mass binaries, the total mass can be measured with an accuracy comparable to that of the chirp mass [143–146].

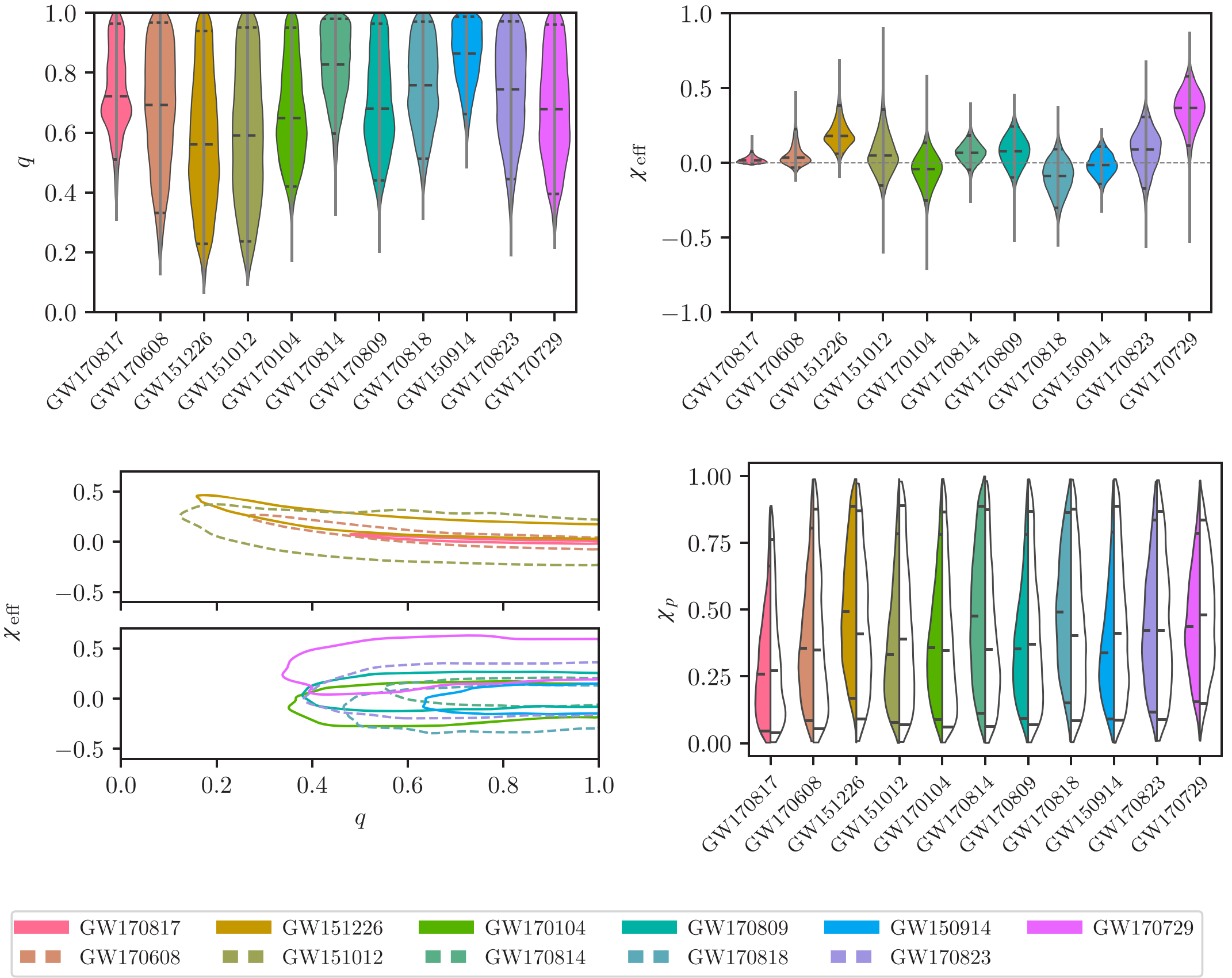

FIG. 5. Parameter estimation summary plots II. Posterior probability densities of the mass ratio and spin parameters of the GW events. The shaded probability distributions have equal maximum widths, and horizontal lines indicate the medians and 90% credible intervals of the distributions. For the two-dimensional distributions, the contours show 90% credible regions. Events are ordered by source-frame chirp mass. The colors correspond to the colors used in summary plots. For GW170817, we show results for the high-spin prior $a_i < 0.89$. Top left: The mass ratio $q = m_2/m_1$. Top right: The effective aligned spin magnitude $\chi_{\mathrm{eff}}$. Bottom left: Contours of 90% credible regions for the effective aligned spin and mass ratio of the binary components for low- (high-) mass binaries are shown in the upper (lower) panel. Bottom right: The effective precession spin posterior (colored) and its effective prior distribution (white) for BBH (BNS) events. The priors are conditioned on the $\chi_{\mathrm{eff}}$ posterior distributions.

We show posteriors for the ratio of the component masses Eq. (6) in the top left in Fig. 5. This parameter is much harder to constrain than the chirp mass. The width of the posteriors depends mostly on the SNR, and so the mass ratio is best measured for the loudest events, GW170817, GW150914, and GW170814. Even though GW170817 has the highest SNR of all events, its mass ratio is less well constrained, because the signal power comes predominantly from the inspiral, while the merger contributes little compared to the BBH [147]. GW151226 and GW151012 have posterior support for more unequal mass ratios than the other events, with lower bounds of 0.28 and 0.29, respectively, at 90% credible level.

The final mass, radiated energy, final spin, and peak luminosity of the BH remnant from a BBH coalescence are computed using averages of fits to numerical relativity (NR) results [15,148–153]. Posteriors for the mass and spin of the BH remnant for BBH coalescences are shown in the right in Fig. 4. Only a fraction (0.02–0.07) of the binary's total mass is radiated away in GWs. The amount of radiated energy scales with its total mass. The heaviest remnant BH found is GW170729, at $79.5^{+14.7}_{-10.2}\ M_\odot$ while the lightest remnant BH is GW170608, at $17.8^{+3.4}_{-0.7}\ M_\odot$.

GW mergers reach extraordinary values of peak luminosity which is independent of the total mass. While it depends on the mass ratio and spins, the posteriors overlap to a large degree for the observed BBH events. Because of its relatively high spin, GW170729 has the highest value of $\ell_{\rm peak} = 4.2^{+0.9}_{-1.5} \times 10^{56}$ erg s$^{-1}$.

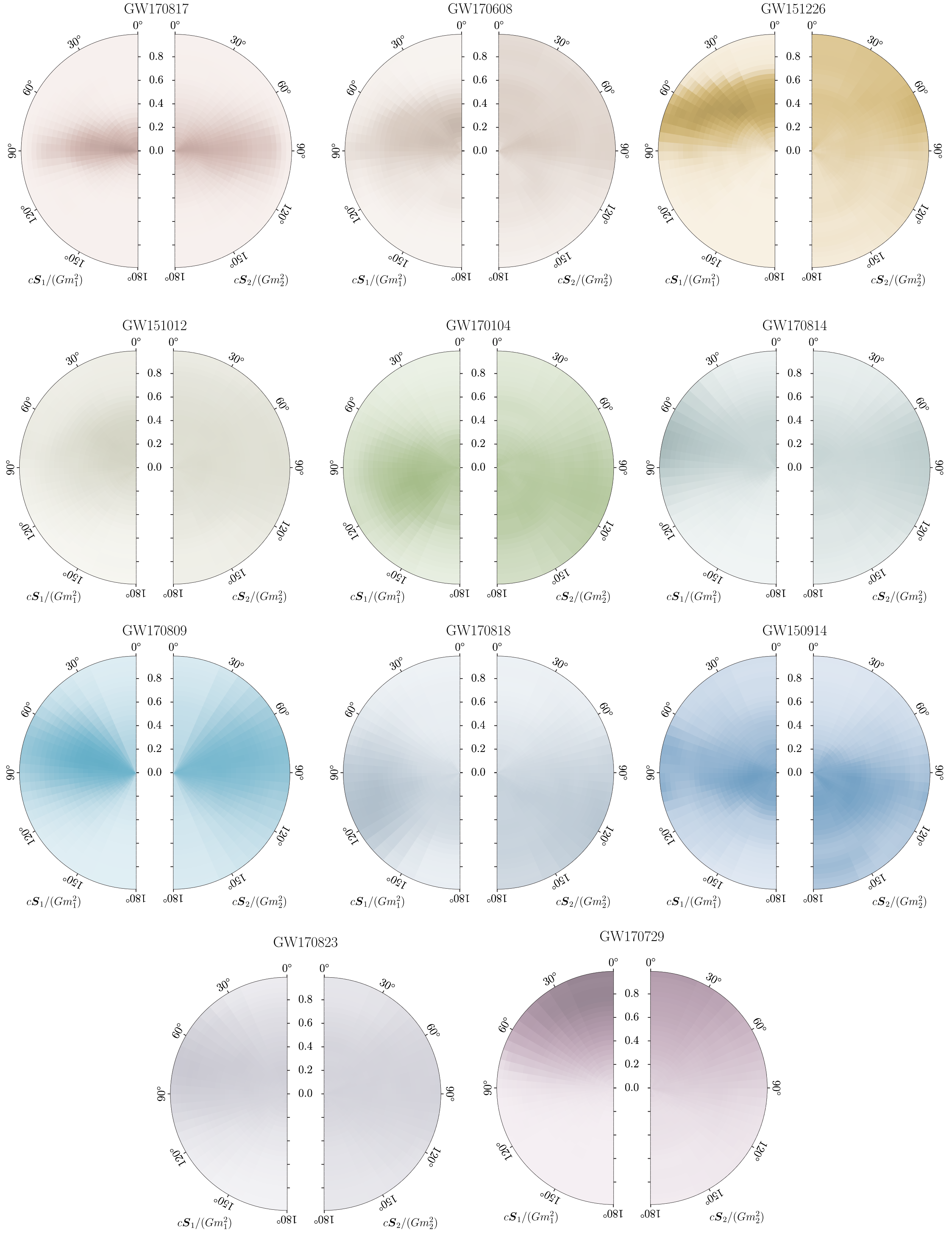

FIG. 6. Parameter estimation summary plots III. Posterior probability distributions for the dimensionless component spins $c\vec{S}_1/(Gm_1^2)$ and $c\vec{S}_2/(Gm_2^2)$ relative to the normal to the orbital plane $\vec{L}$, marginalized over the azimuthal angles. The bins are constructed linearly in spin magnitude and the cosine of the tilt angles and are assigned equal prior probability. Events are ordered by source-frame chirp mass. The colors correspond to the colors used in summary plots. For GW170817, we show results for the high-spin prior $a_i < 0.89$.

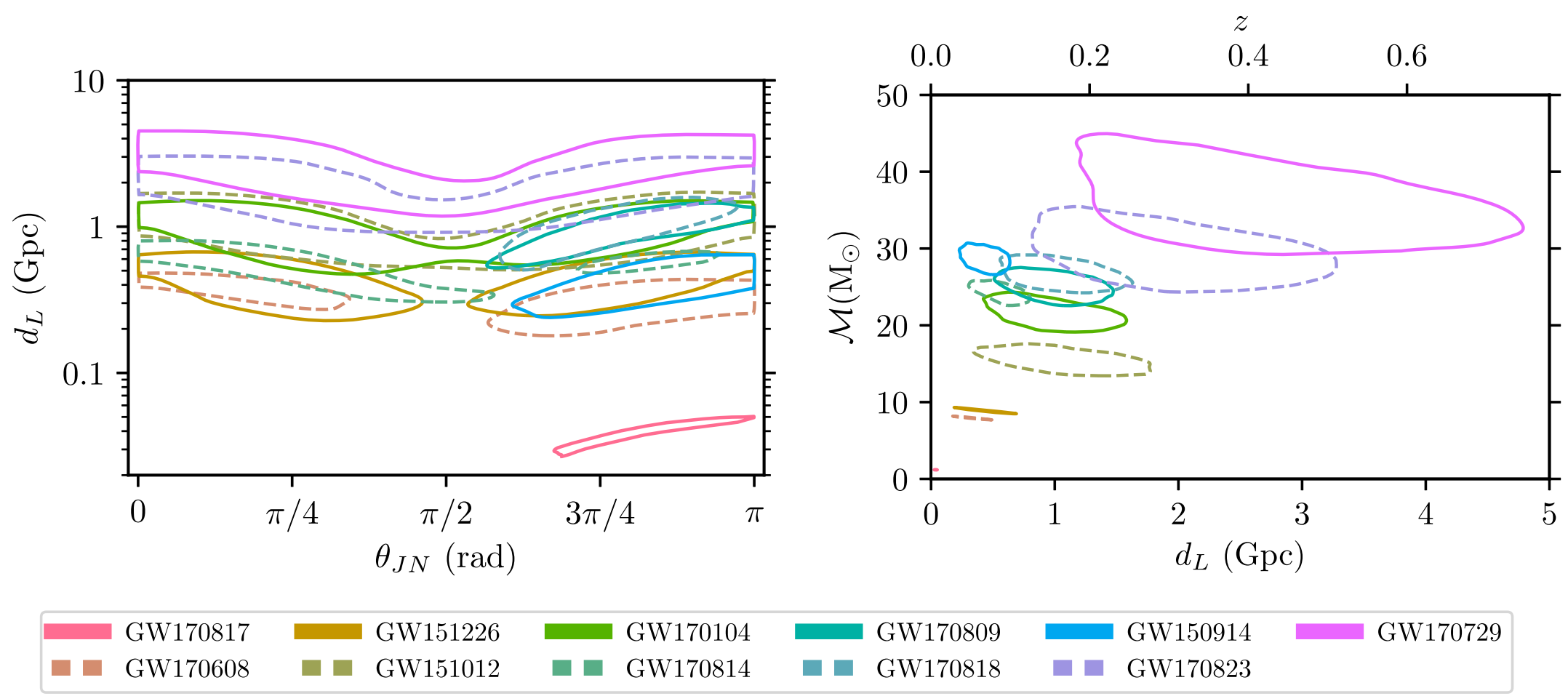

FIG. 7. Parameter estimation summary plots IV. Posterior probability densities of distance $d_L$, inclination angle $\theta_{JN}$, and chirp mass $\mathcal{M}$ of the GW events. For the two-dimensional distributions, the contours show 90% credible regions. For GW170817, we show results for the high-spin prior $a_i < 0.89$. Left: The inclination angle and luminosity distance of the binaries. Right: The luminosity distance (or redshift $z$) and source-frame chirp mass. The colored event labels are ordered by source-frame chirp mass.

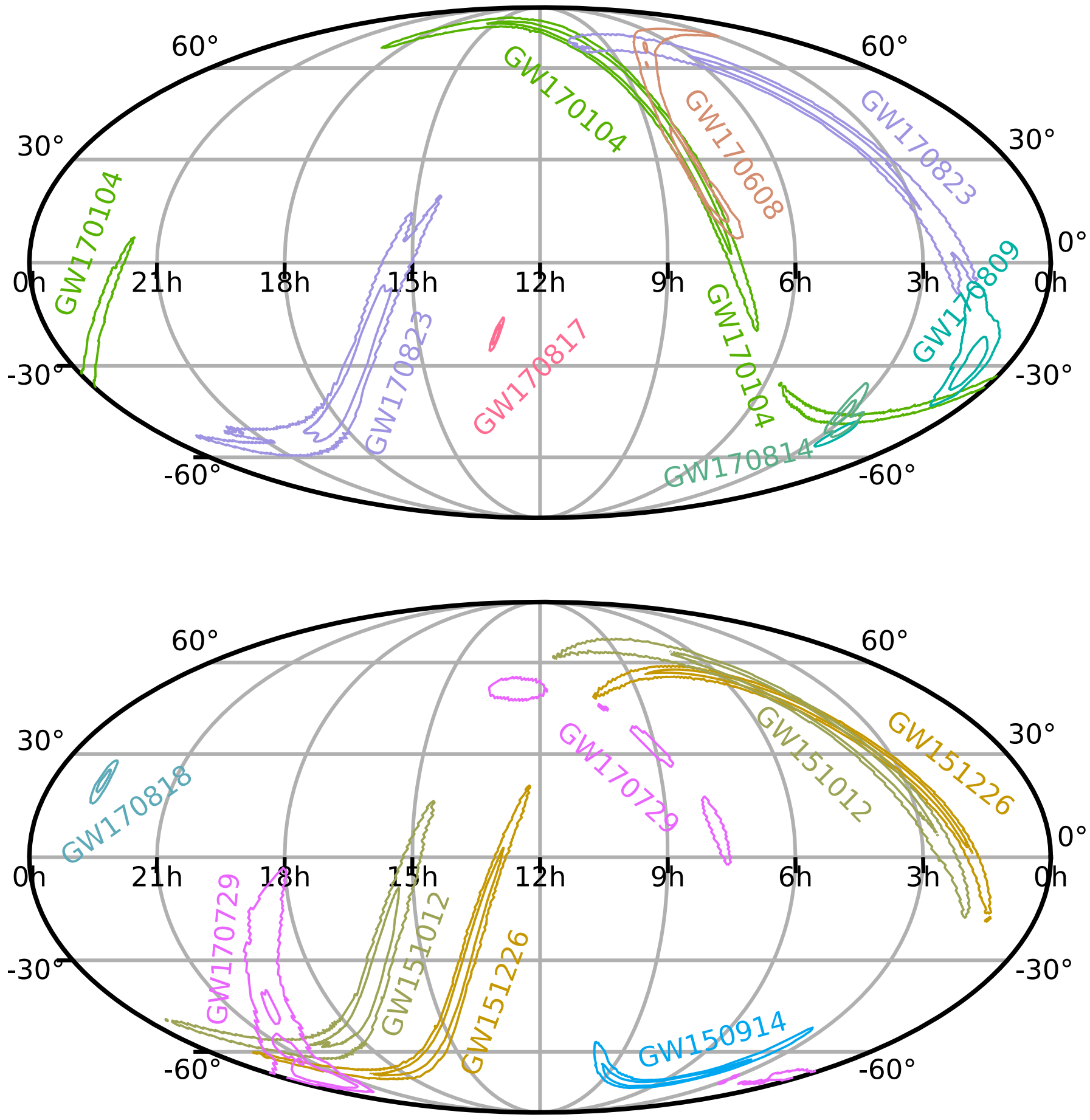

FIG. 8. Parameter estimation summary plots V. The contours show 90% and 50% credible regions for the sky locations of all GW events in a Mollweide projection. The probable position of the source is shown in equatorial coordinates (right ascension is measured in hours, and declination is measured in degrees). 50% and 90% credible regions of posterior probability sky areas for the GW events. Top: Confidently detected $O2$ GW events [22] (GW170817, GW170104, GW170823, GW170608, GW170809, and GW170814) for which alerts were sent to EM observers. Bottom: $O1$ events (GW150914, GW151226, and GW151012), along with $O2$ events (GW170729 and GW170818) not previously released to EM observers. Where applicable, the initial sky maps shared with EM partners in low latency are available from Ref. [185].

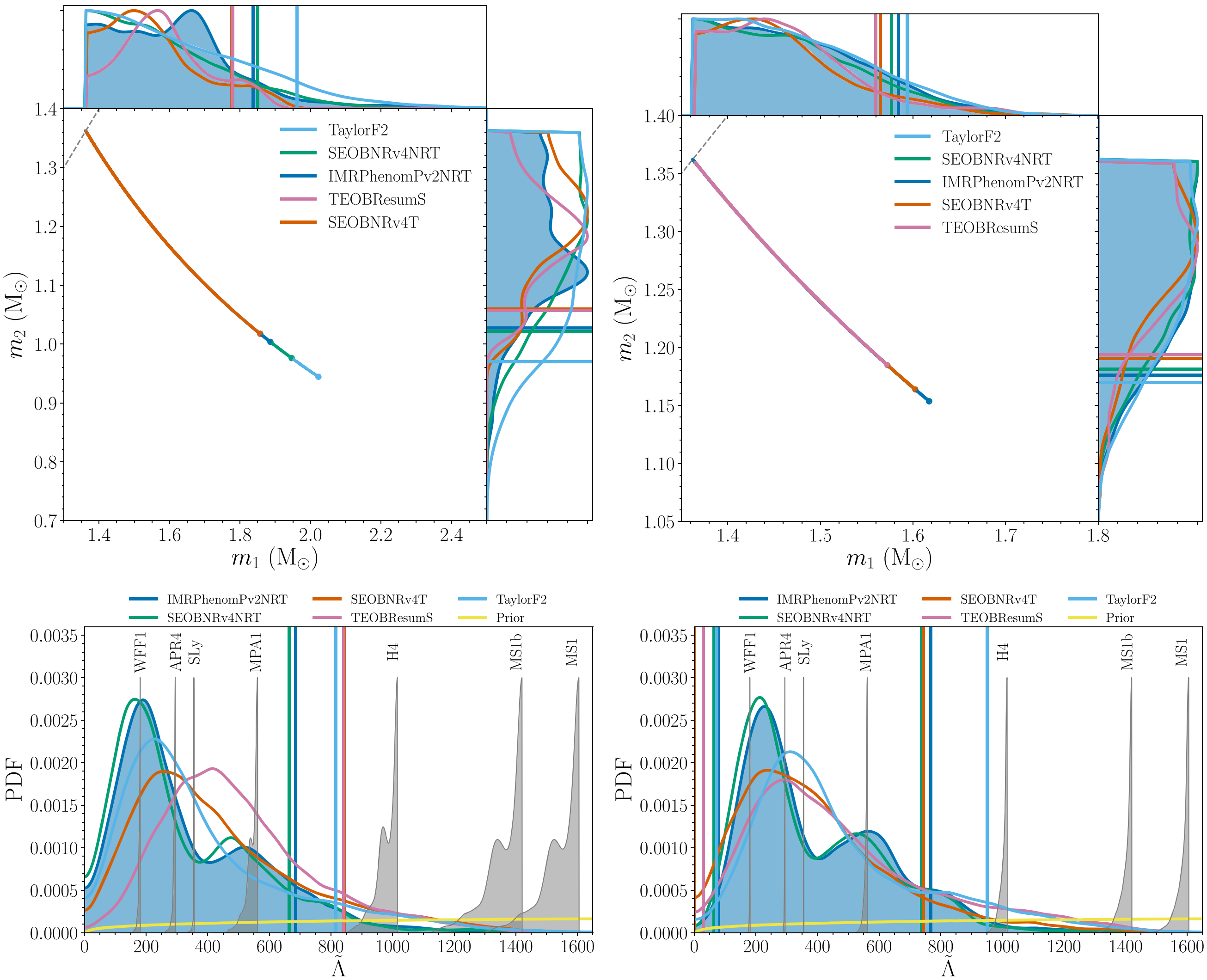

FIG. 9. Posterior distributions for component masses and tidal deformability for GW170817 for the waveform models: IMRPhenomPv2NRT, SEOBNRv4NRT, TaylorF2, SEOBNRv4T, and TEOBResumS. Top: 90% credible regions for the component masses for the high-spin prior $a_i < 0.89$ (left) and low-spin prior $a_i < 0.05$ (right). The edge of the 90% credible regions is marked by points; the uncertainty in the contour is smaller than the thickness shown because of the precise chirp mass determination. 1D marginal distributions are renormalized to have equal maxima, and the vertical and horizontal lines give the 90% upper and lower limits on $m_1$ and $m_2$, respectively. Bottom: Posterior distributions of the effective tidal deformability parameter $\tilde{\Lambda}$ for the high-spin (left) and low-spin (right) priors. These PDFs are reweighted to have a flat prior distribution. The original $\tilde{\Lambda}$ prior is shown in yellow. 90% upper bounds are represented by vertical lines for the high-spin prior (left). For the low-spin prior (right), 90% highest posterior density (HPD) credible intervals are shown instead. Gray PDFs indicate seven representative equations of state (EOSs) using masses estimated with the IMRPhenomPv2NRT model.

### C. Spins

The spin vectors of compact binaries can *a priori* point in any direction. Particular directions in the spin space are easier to constrain, and we focus on these first. An averaged projection of the spins parallel to the Newtonian orbital angular momentum of the binary can be measured best. This effective aligned spin $\chi_{\rm eff}$ is defined by Eq. (4). Positive (negative) values of $\chi_{\rm eff}$ increase (decrease) the number of orbits from any given separation to merger with respect to a nonspinning binary [38,154]. We show posterior distributions for this quantity in the top right in Fig. 5. Most posteriors peak around zero. The posteriors for GW170729 and GW151226 exclude $\chi_{\rm eff} = 0$ at $> 90\%$ confidence, but see Sec. V F. As can be seen from Table III, the 90% intervals are 0.11–0.58 for GW170729 and 0.06–0.38 for GW151226.

As shown in the bottom left in Fig. 5, the mass ratio and effective aligned spin parameters can be degenerate [122,128,155], which makes them difficult to measure individually. For lower-mass binaries, most of the waveform is in the inspiral regime, and the posterior has a shape that curves upwards towards larger values of $\chi_{\mathrm{eff}}$ and lower values of $q$, exhibiting a degeneracy between these parameters. This degeneracy is broken for high-mass binaries for which the signal is short and is dominated by the late inspiral and merger [147]. For all observed binaries, the posteriors reach up to the equal mass boundary ($q = 1$). With current detector sensitivity, it is difficult to measure the individual BH's spins [147,156–158], and, in contrast to $\chi_{\mathrm{eff}}$, the posteriors of an antisymmetric mass-weighted linear combination of $\chi_1$ and $\chi_2$ are rather wide.

The remaining spin d.o.f. are due to a misalignment of the spin vectors with the normal to the orbital plane and give rise to a precession of the orbital plane and spin vectors around the total angular momentum of the binary. The bottom right in Fig. 5 shows posterior and prior distributions for the quantity $\chi_p$, which encapsulates the dominant effective precession spin. The prior distribution for $\chi_p$ is induced by the spin prior assumptions (see Appendixes B 1 and C). Since $\chi_p$ and $\chi_{\mathrm{eff}}$ are correlated, we show prior distributions conditioned on the $\chi_{\mathrm{eff}}$ posteriors. The $\chi_p$ posteriors are broad, covering the entire domain from 0 to 1, and are overall similar to the conditioned priors. A more detailed representation of the spin distributions is given in Fig. 6, showing the probability of the spin magnitudes and tilt angles relative to the Newtonian orbital angular momentum. Deviations from uniformity in the shading indicate the strength of precession effects. Overall, it is easier to measure the spin of the heavier component in each binary [147,157]. None of the GW events exhibit clear precession. To quantify this result, we compute the Kullback-Leibler divergence $D_{\mathrm{KL}}$ [159] for the information gain from the $\chi_p$ prior to the posterior. Since $\chi_p$ and $\chi_{\mathrm{eff}}$ are correlated, we can condition the prior on the $\chi_{\mathrm{eff}}$ posterior before we compute $D_{\mathrm{KL}}^{\chi_p}$. We give $D_{\mathrm{KL}}^{\chi_p}$ values with and without conditioning in Table V in Appendix B. Among the BBH events, the highest values of $D_{\mathrm{KL}}^{\chi_p}$ are found for GW170814 ($0.13^{+0.03}_{-0.02}$ bits) and GW151226 ($0.12^{+0.05}_{-0.02}$ bits). In all cases, the information gain for $D_{\mathrm{KL}}^{\chi_p}$ is much less than a single bit. Conversely, we gain more than one bit of information in $\chi_{\mathrm{eff}}$ for several events (see Table V and also Table VI). For very well measured quantities such as the chirp mass for GW170817, we can gain approximately 10 bits of information and come close to the information entropy [160] in the posterior data. A clear imprint of precession could also help break the degeneracy between the mass ratio and effective aligned spin [161–164]. We discuss the influence of the choice of priors for spin parameters (and distance) in Appendix C.

As a weighted average of the mass and aligned spin of the binary, $\chi_{\mathrm{eff}}$ provides a convenient tool to test models of compact object binary spin properties via GW measurements [165–170]. Several authors suggest [171–179] how stellar binary evolutionary pathways leave imprints on the overall distribution of detected parameters such as masses and spins. By inferring the population properties of the events observed to date [55], we disfavor scenarios in which most black holes merge with large spins aligned with the binary's orbital angular momentum. With more detections, it will be possible to determine, for example, if the BH spin is preferentially aligned or isotropically distributed.

For comparable-mass binaries, the spin of a remnant black hole comes predominantly from the orbital angular momentum of the progenitor binary at merger. For nonspinning equal-mass binaries, the final spin of the remnant is expected to be approximately 0.7 [180–184]. The final spin posteriors are more precisely constrained than the component spins and also the effective aligned spin $\chi_{\mathrm{eff}}$. Masses and spins of the final black holes are shown in Fig. 4. Except for GW170729 with its sizable positive $\chi_{\mathrm{eff}} = 0.37^{+0.21}_{-0.25}$, the medians of all final spin distributions are around approximately 0.7. The remnant of GW170729 has a median final spin of $a_f = 0.81^{+0.07}_{-0.13}$ and is consistent with 0.7 at 90% confidence.

### D. Distance, inclination, and sky location

The luminosity distance $d_L$ of a GW source is inversely proportional to the signal's amplitude. Six BBH events (GW170104, GW170809, GW170818, GW151012, GW170823, and GW170729) have median distances of about a Gpc or beyond, the most distant of which is GW170729 at $d_L = 2840^{+1400}_{-1360}$ Mpc, corresponding to a redshift of $0.49^{+0.19}_{-0.21}$. The closest BBH is GW170608, at $d_L = 320^{+120}_{-110}$ Mpc, while the BNS GW170817 is found at $d_L = 42^{+6}_{-13}$ Mpc. The significant uncertainty in the luminosity distance stems from the degeneracy between the distance and the binary's inclination, inferred from the signal amplitude [128,186,187]. We show joint posteriors of the luminosity distance and inclination $\theta_{JN}$ in the left in Fig. 7. In general, the inclination angle is only weakly constrained, and for most events it has a bimodal distribution around $\theta_{JN} = 90°$ with greatest support for the source being either face on or face off (angular momentum pointed parallel or antiparallel, respectively, to the line of sight). These orientations produce the greatest gravitational-wave amplitude and so are consistent with the largest distance. For GW170817, the $\theta_{JN}$ distribution has a single mode. For GW170809, GW170818, and GW150914, the 90% interval contains only a single mode so that the fifth percentile lies above $\theta_{JN} = 90°$. Orientations of the total orbital angular momentum that are strongly misaligned with the line of sight are, in general, disfavored due to the weaker emitted GW signal compared to observing a binary face on ($\theta_{JN} = 0°$) or face off ($\theta_{JN} = 180°$). For GW170818, the misalignment is more likely, with the probability that $45° < \theta_{JN} < 135°$ being 0.38.

This probability is less than 0.36 for all other events. An inclination close to $\theta_{JN} = 90°$ would enhance subdominant modes in the GW signal but also result in a weaker emitted signal and, to compensate, a closer source. A more precise measurement of the inclination will be possible for strongly precessing binaries [162,188].

This analysis assumes that the emitted GW signal is not affected by gravitational lensing. Lensing would make GW mergers appear closer than they are and reduce their inferred, redshift-corrected source-frame masses, depending on the true distance and magnification factor of the lens. Motivated by the heavy BBHs observed by LIGO and Virgo, Ref. [189] claims that four of the published BBH observations are magnified by gravitational lensing. On the other hand, it has been pointed out that at LIGO's and Virgo's current sensitivities it is unlikely but not impossible that one of the GWs is multiply imaged. The analysis in Reference [190] concludes that lensing by massive galaxy clusters of one of our BBH GW detections can be rejected at the $4\sigma$ level.

In the right in Fig. 7, we show the joint posterior between the luminosity distance (or redshift) and source-frame chirp mass. We see that overall luminosity distance and chirp mass are positively correlated, as expected for unlensed BBHs observations.

An observed GW signal is registered with different arrival times at the detector sites. The observed time delays and amplitude and phase consistency of the signals at the sites allow us to localize the signal on the sky [191–193]. Two detectors can constrain the sky location to a broken annulus [194–197], and the presence of additional detectors in the network improves localization [19,198–200]. Figure 8 shows the sky localizations for all GW events. Both panels show posteriors in celestial coordinates which indicate the origin of the signal. In general, the credible regions of sky position are made up of a collection of disconnected components determined by the pattern of sensitivity of the individual detectors. The top shows localizations for confidently detected $O2$ events that were communicated to EM observers and are discussed further in Ref. [22]. The results for the credible regions and sky areas are different from those shown in Ref. [22] because of updates in the data calibration and choice of waveform models. The bottom shows localizations for $O1$ events, along with $O2$ events not previously released to EM observers. The sky area is expected to scale inversely with the square of the SNR [20,197]. This trend is followed for events detected by the two LIGO detectors. Several events (GW170729, GW170809, GW170814, GW170817, and GW170818) are observed with the two LIGO detectors and Virgo, which improves the sky localization [201]. The SNR contributed by Virgo can significantly shrink the area. We find the smallest 90% sky localization areas for GW170817: 16 deg$^2$ and GW170818: 39 deg$^2$.

### E. GW170817

We carry out a reanalysis of GW170817 using a set of waveform models including tidal effects described in detail in Appendix B 2. This analysis follows the one performed in Ref. [97], employs the same settings, but uses the recalibrated $O2$ data. We restrict the sky location to the known position of SSS17a/AT 2017gfo as determined by electromagnetic observations [21]. When computing the source-frame masses from the detector-frame masses, we use the redshift for NGC 4993 from Ref. [94] and its associated uncertainties. Updated posteriors for masses and the effective tidal deformability parameter $\tilde{\Lambda}$ are shown in Fig. 9. For the results presented here, we allow the tidal parameters to vary independently rather than being determined by a common equation of state [202]. Results are consistent with those presented previously in Ref. [97] with slight differences in the derived tidal deformability, discussed below. Posterior distributions for SEOBNRv4T and TEOBResumS are obtained from RapidPE. In contrast to the BBH events discussed above, GW170817 is completely dominated by the inspiral phase of the binary coalescence. The merger and postmerger happen at frequencies above 1 kHz, where LIGO and Virgo are less sensitive. The distributions of component masses are shown in the top in Fig. 9. With 90% probability, the mass of the larger NS $m_1$ for the IMRPhenomPv2NRT model is contained in the range $[1.36, 1.84]\ M_\odot$ ($[1.36, 1.58]\ M_\odot$) and the smaller NS $m_2$ in $[1.03, 1.36]\ M_\odot$ ($[1.18, 1.36]\ M_\odot$) for the high-spin (low-spin) prior. In Fig. 5, we show contours for the mass ratio and aligned effective spin posteriors for the IMRPhenomPv2NRT model assuming the high-spin prior. The results are consistent with those presented in Ref. [97]. The effective precession spin $\chi_p$ shown in the bottom right in Fig. 5 peaks at lower values than the prior, and the KL divergence $D_{\mathrm{KL}}^{\chi_p}$ between this prior and the posterior is $0.19^{+0.04}_{-0.03}$ bits. When conditioning the prior on the measured $\chi_{\mathrm{eff}}$, $D_{\mathrm{KL}}^{\chi_p}$ decreases to $0.07^{+0.01}_{-0.02}$ bits, providing very little evidence for precession. The strongly constrained $\chi_{\mathrm{eff}}$ restricts most of the spin d.o.f. into the orbital plane, and in-plane spins are large only when the binary's inclination angle approaches 180°, where they have the least impact on the waveform.

We show marginal posteriors for the effective tidal parameter $\tilde{\Lambda}$ in the bottom in Fig. 9. The prior and posterior for $\tilde{\Lambda}$ go to zero as $\tilde{\Lambda} \to 0$ because of the flat prior on the component deformability parameters $\Lambda_1$ and $\Lambda_2$. We reweight the posterior for $\tilde{\Lambda}$ by dividing by the prior used, effectively imposing a flat prior in $\tilde{\Lambda}$. The reweighted posterior has nonzero support at $\tilde{\Lambda} = 0$. We find bounds on the effective tidal parameter that are about 10% wider compared to the results presented in Ref. [97]. For the high-spin prior, the 90% upper limit on the tidal parameter is 686 for IMRPhenomPv2NRT, compared to the value of 630 found in Ref. [97]. The upper limit for SEOBNRv4NRT is very close, 664, and the value for TaylorF2 is higher at 816. For SEOBNRv4T and TEOBResumS, we find 843 and 841, respectively. For the low-spin prior, we quote the two-sided 90% highest posterior density (HPD) credible interval on $\tilde{\Lambda}$ that does not contain $\tilde{\Lambda} = 0$. This 90% HPD interval is the smallest interval that contains 90% of the probability. For IMRPhenomPv2NRT, we obtain $\tilde{\Lambda} = 330^{+438}_{-251}$, which is slightly higher than the interval $300^{+420}_{-230}$ found in Ref. [97]. For SEOBNRv4NRT, we find $\tilde{\Lambda} = 305^{+432}_{-241}$ and for TaylorF2 $394^{+557}_{-321}$. For SEOBNRv4T and TEOBResumS, we find $349^{+394}_{-349}$ and $405^{+545}_{-375}$, respectively. The posteriors produced by these two models agree better for the low-spin prior. This result is consistent with the very good agreement between the models for small spins $|\chi_i| \leq 0.15$ shown in Ref. [33]. For reference, we also show contours for a representative subset of theoretical EOS models given by piecewise-polytrope fits from Ref. [203]. These fits are evaluated using the IMRPhenomPv2NRT component mass posteriors, and the sharp cutoff to the right of each EOS posterior corresponds to the equal mass ratio boundary. As found in Ref. [97], the EOSs MS1, MS1b, and H4 lie outside the 90% credible upper limit and are therefore disfavored.

In Table III, we quote conservative estimates of key final-state parameters for GW170817 obtained from fits to NR simulations of quasicircular binary neutron star mergers [204–206]. We do not assume the type of final remnant and quote quantities at either the moment of merger or after the postmerger GW transient. Lower limits of radiated energy up to the merger and peak luminosity are given at 1% credible level. The final mass is computed from the radiated energy including the postmerger transient as an upper limit at 99% credible level. For the final angular momentum, we quote an upper bound computed from the radiated energy and using the phenomenological universal relation found in Ref. [204].

### F. Comparison against previously published results

We compare PE results between the original published $O1$ and $O2$ analyses for GW150914, GW151012, GW151226, GW170104, GW170608, and GW170814 and the reanalysis performed here. The values presented here supersede previously published results. For some events, we see differences in the overall posteriors that are due to a different choice of waveform models that have been combined. This difference is especially the case when comparing against previous results that combine samples between spin-aligned and effective precession models and mostly affects spin parameters. We first mention differences that are apparent when comparing results from the same waveform models in the original analysis and the reanalysis.

The source-frame total mass is consistent with the original analysis. For GW150914, we find an increase in the median of about 1 $M_{\odot}$ in this reanalysis when comparing between the same precessing waveform models because of the improved method for computing the power spectral density of the detector noise and the use of frequency-dependent calibration envelopes. For GW170104, we find the median of the total mass to be 0.3 $M_{\odot}$ higher because of the recalibration of the data and the noise subtraction. Similarly, we find an increase of about 0.2 $M_{\odot}$ in the total mass for GW151012 and a decrease in the total mass of about 0.3 $M_{\odot}$ for GW151226 and 0.2 $M_{\odot}$ for GW170608 in the reanalysis. The mass ratio and effective spin parameters are broadly consistent with the original analysis. GW170104 especially benefits from the noise subtraction. This subtraction increases the matched-filter SNR recovered by the parameter-estimation analysis from $13.3^{+0.2}_{-0.3}$ to $14.0^{+0.2}_{-0.3}$. The increase in SNR results in reduced parameter uncertainties [128]. For the effective spin parameter, the tightening of the posterior results in the loss of the tail at low values. The inferred value changes from $-0.12^{+0.21}_{-0.30}$ to $-0.04^{+0.17}_{-0.21}$; the upper limit remains about the same, and there is still little support for large aligned spins. For GW151226, we find from using the fully precessing model that the inferred effective aligned spin is $0.15^{+0.25}_{-0.11}$, and with the effective precession model it is $0.20^{+0.18}_{-0.08}$. The fully precessing model has some support at $\chi_{\rm eff} = 0$; the probability that $\chi_{\rm eff} < 0$ is, however, $< 0.01$. We find with 99% probability that at least one spin magnitude is greater than 0.28 compared to the value 0.2 in the $O1$ analysis. We discuss further differences between results obtained from the two BBH waveform models in Appendix B 2.

## VI. WAVEFORM RECONSTRUCTIONS

In the previous section, we present estimates of the source properties for each event based on different relativistic models of the emitted gravitational waveform. Such models, however, do not necessarily incorporate all physical effects. Here, we take an independent approach to determine the GW signal present in the data and assess the consistency with the waveform model-based analysis.

Figure 10 shows the time-frequency maps of the gravitational-wave strain data measured in the detector where the higher SNR is recorded [207], as well as three different types of waveform reconstructions for all GW events from BBHs. Two of those waveform reconstructions provide an independent estimate of the most probable signal: Instead of relying on waveform models, these algorithms exclusively use the coherent gravitational-wave energy measured by the detector network, requiring only weak assumptions on the form of the signal for the reconstruction.

The first method, BayesWave, represents the waveform as a sum of sine-Gaussian wavelets $h(\vec{\lambda}; t) = \sum_{j=1}^{N} \Psi(\vec{\lambda}_j; t)$, where the number of wavelets used in the reconstruction, $N$, and the parameters describing each wavelet, $\vec{\lambda}_j$, are explored by a transdimensional Markov chain Monte Carlo algorithm [53] (blue bands in the top panels). In comparison, we also show the waveforms obtained from the Bayesian inference results in Sec. V with the precessing waveform model PhenomP (red bands). The 90% credible regions are computed by selecting a discrete collection of times $t_k$, and computing the waveform at these times from a large number of fair draws from the posterior distribution of the model parameters; they indicate the range of waveforms that are consistent with the data for a particular model.

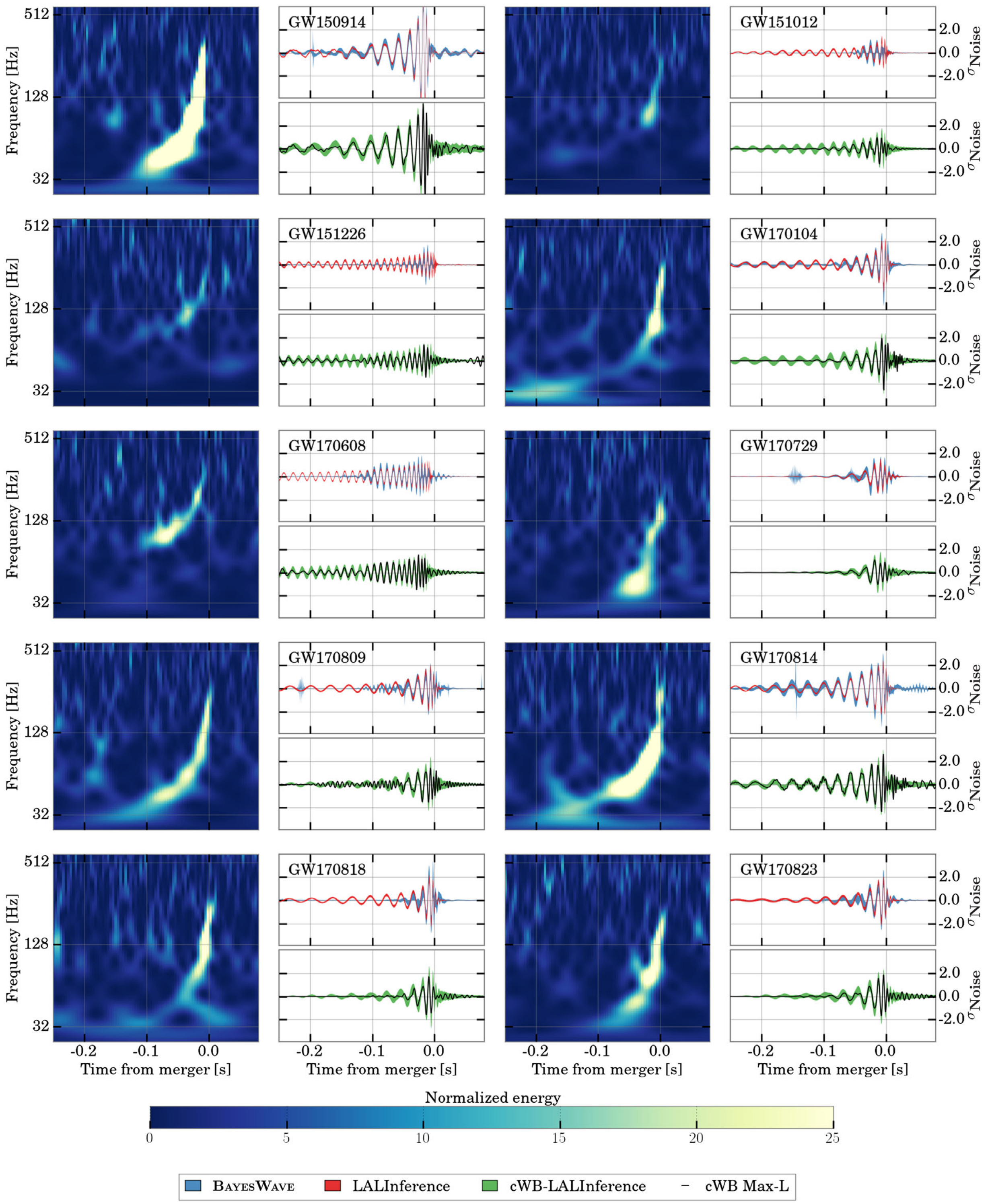

FIG. 10. Time-frequency maps and reconstructed signal waveforms for the ten BBH events. Each event is represented with three panels showing whitened data from the LIGO detector where the higher SNR is recorded. The first panel shows a normalized time-frequency power map of the GW strain. The remaining pair of panels shows time-domain reconstructions of the whitened signal, in units of the standard deviation of the noise. The upper panels show the 90% credible intervals from the posterior probability density functions of the waveform time series, inferred using CBC waveform templates from Bayesian inference (LALInference) with the PhenomP model (red band) and by the BayesWave wavelet model (blue band) [53]. The lower panels show the point estimates from the cWB search (solid lines), along with a 90% confidence interval (green band) derived from cWB analyses of simulated waveforms from the LALInference CBC parameter estimation injected into data near each event. Visible differences between the different reconstruction methods are verified to be consistent with a noise origin (see the text for details).

The wavelet-based method trades sensitivity for flexibility and, therefore, is unable to discern the early inspiral or ringdown part of the CBC waveforms where the signals are weaker and/or spread out over time. The CBC waveform models use strict assumptions about the shape of the waveforms and, based on those assumptions, predict the shape of the signal in places where it is weak compared to the detector noise. Comparing results from the different methods enables a visual assessment of the reconstructed signals, both with and without physical assumptions about the source. Regions where the 90% credible bands are not overlapping do not necessarily imply any physical discrepancy, instead arising from differences in the models for the GW signals. Furthermore, as can be seen in several panels in Fig. 10, the 90% credible intervals for the BayesWave reconstructions sometimes include features that are not present in the template-based reconstructions. We verify that these outliers are absent from the BayesWave 50% credible intervals (not shown here), indicating that they have low significance. Analogous features are seen when reconstructing simulated signals added to real data and should not be misinterpreted as evidence for disagreement with the template-based reconstructions. Their origin is more mundane: They are caused by small random coherent features in the noise that are seen by BayesWave; similar behavior is also seen for cWB, the second reconstruction method used (see below). Quantitative comparisons of the template- and wavelet-based methods include computing overlaps between the reconstructions and reanalyzing the data with the wavelet-based analysis after the best-fit CBC waveform has been subtracted. Both comparisons agree within expectations from analysis of simulations, and we conclude that there are no detectable discrepancies between the wavelet- and template-based reconstructions, i.e., that the template-based methods agree with the data within the uncertainties.

The bottom panels show the waveform reconstructions obtained from the second model-independent method, cWB, which reconstructs the maximum likelihood signal waveforms $\boldsymbol{h} = \{h_H(t), h_L(t)\}$ triggered in multiple detectors by a GW event. Since the reconstruction does not use any specific waveform model, the components of $\boldsymbol{h}$ are effectively the denoised detector responses normalized by the noise power spectral amplitude. To test the consistency between the cWB reconstructed signals $\boldsymbol{h}$ and the CBC parameter estimation results in Sec. V, the following method is used: (a) Waveforms from the CBC parameter estimation posterior samples are generated and injected at random times in the data around the GW event, (b) the cWB pipeline is run on these data and $\tilde{\boldsymbol{h}}$—the best-fit waveforms expected for a given waveform model—are reconstructed, and (c) the waveforms $\tilde{\boldsymbol{h}}$ are used to construct the confidence intervals. This method combines both the CBC and cWB reconstruction errors to produce confidence intervals for the denoised signal waveforms obtained with cWB. In addition, the confidence interval is robust to non-Gaussian detector noise, which may affect both the CBC and cWB reconstructions. Figure 10 shows the maximum likelihood waveforms reconstructed by cWB: A comparison with the 90% confidence interval indicates good agreement with the model-based parameter estimation results.

Further reconstructions are made available through the Gravitational Wave Open Science Center [57].

## VII. MERGER RATES OF COMPACT BINARY SYSTEMS

This section presents bounds on the astrophysical merger rates of BNS, NSBH, and BBH systems in the local Universe, derived from the search results presented in earlier sections. These bounds supersede earlier estimates and limits from previous LIGO-Virgo results [4,15,18,208]. Our merger rate estimates are derived by modeling the search results of each pipeline as a mixture of a set of astrophysical events and a set of background (noise) events of terrestrial origin [90]. Here, we describe how merger rates, as well as the probabilities that each candidate event is of astrophysical or terrestrial origin, are calculated.

Since we now have confident detections of different astrophysical event types—BNS and BBH—a more sophisticated treatment is necessary as compared to previous results. We define four categories: one terrestrial and three astrophysical categories (BNS, NSBH, and BBH). For each category, the distribution of the pipelines' ranking statistic values—generally denoted here as $x$—is empirically determined. Terrestrial quantities are denoted by the $T$ label—thus, the probability distribution of terrestrial event $x$ values is written $p(x|T)$—while the set of astrophysical categories is denoted by $A_i$, where $i$ labels the three types of binaries considered. For a given category $A_i$ and a population configuration $\{\theta\}$, i.e., the assumed spin and mass distribution, the ranking statistic distribution is given as $p(x|A_i, \{\theta\})$. In practice, differing $\{\theta\}$ do not significantly affect the shape of $p(x|A_i, \{\theta\})$ over the range of ranking statistics considered here [19].

All four categories are assumed to contribute events according to a Poisson process with mean $\Lambda$. In each astrophysical category, the mean can be further described as the product of the accessible spacetime volume $\langle VT \rangle$ for a source population $\{\theta\}$ and the astrophysical rate density $R_i$. The terrestrial $\Lambda_T$ and astrophysical $\Lambda_i$ count parameters are determined by fitting the mixture of $\Lambda_T p(x|T)$ and a given $\Lambda_i p(x|A_i, \{\theta\})$. Since each model is computed from the outputs of a given search, for the purposes of computing quantities such as the probability of astrophysical origin, each pipeline is treated separately.

Figure 11 shows the resulting astrophysical foreground and terrestrial background models, as well as the observed number of events above a ranking statistic threshold: on the left, PyCBC results, restricted to events with masses compatible with a BBH, with chirp mass $> 4.35\ M_\odot$ (so that BNS candidate events including GW170817 are not plotted); on the right, GstLAL results including all events, with the signal counts summed over the three astrophysical categories BNS, NSBH, and BBH. In both searches, the background model falls exponentially with the detection statistic, with no non-Gaussian tails. The different detection statistic used in the PyCBC and GstLAL searches leads to differently shaped signal models. However, both searches show agreement between the search results and the sum of the foreground and terrestrial background models. In both panels in Fig. 11, we see three regions: At the high-ranking statistic threshold, the signal model dominates and the observed events are inconsistent with terrestrial noise; at the low-ranking statistic threshold, the noise model dominates and the observed events are fully consistent with terrestrial noise; and a narrow intermediate region where both noise and signal models are comparable, and the observed events are consistent with the sum of the two models. The list of marginal events in Tables II and IV come from that narrow intermediate region.

The accessible spacetime volume $\langle VT\rangle$ is estimated by injecting synthesized signals with parameters drawn from $\{\theta\}$ and recovering them using the search pipeline. For all $\{\theta\}$, the injections are assumed to be uniformly distributed in the comoving volume. Then the detection efficiency over redshift $f(z|\{\theta\})$ derived from the recovery campaign measures the fraction of the differential volume $dV/dz$ which is accessible to the network:

$$\langle VT\rangle_{\{\theta\}} = T_{\rm obs}\int_0^\infty f(z|\{\theta\})\frac{dV}{dz}\frac{1}{1+z}dz. \qquad (8)$$

The total $\langle VT\rangle$ is then the product of the accessible volume for a given population with the observational time $T_{\rm obs}$. The angle brackets indicate that the volume is averaged over members of the population drawn from $\{\theta\}$. In the following, we suppress the $\{\theta\}$ dependence on $\langle VT\rangle$ and $p(x|A_i)$ and, instead, indicate specific populations where they are relevant. The factor of $1+z$ arises from the conversion of source-frame time to detector-frame time which is integrated to obtain $T_{\rm obs}$. $\langle VT\rangle$ is estimated over smaller periods of observation time to better account for time-varying detector sensitivity changes. When summed over analysis periods, the total observation time is 0.46 y. Additional details on how $\langle VT\rangle$ is measured can be found in Refs. [4,209]. A more generalized approach [210] to obtaining $\langle VT\rangle$ is used by PyCBC.

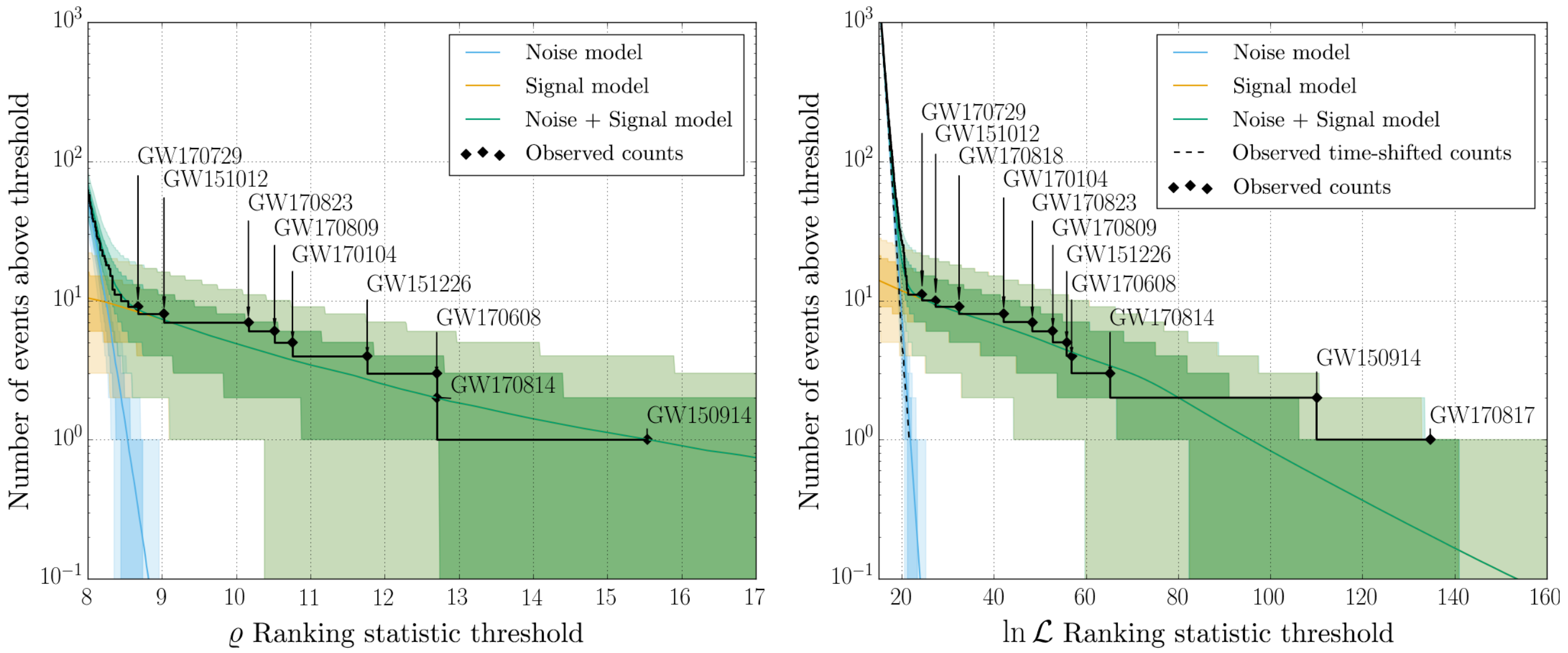

FIG. 11. Astrophysical signal and terrestrial noise event models compared with results for the matched-filter searches, PyCBC (left) and GstLAL (right), versus the respective search's ranking statistic: $\varrho$ for PyCBC [73] and $\ln\mathcal{L}$ for GstLAL [9,82]. These ranking statistics are not the same as the SNRs reported in Table I; see citations for details. For each panel, the solid colored lines show the median estimated rate ("model") of signal, noise, or signal plus noise events above a given ranking statistic threshold, while shaded regions show the estimated model uncertainties on the combined and individual models at 68% and 95% confidence. The observed number of events above the ranking statistic threshold is indicated by the black line, with confidently detected events (Sec. IV B) labeled. The PyCBC signal model and observed events are restricted to events with masses compatible with a BBH, with a chirp mass $> 4.35\ M_\odot$ (so that BNS candidate events including GW170817 are not plotted); the GstLAL signal model includes all events, with the signal counts summed over the three astrophysical categories BNS, NSBH, and BBH. The different ranking statistic used in the PyCBC and GstLAL searches lead to differently shaped signal models. The black dashed line in the GstLAL plot shows a realization of the cumulative counts in time-shifted data, reinforcing its consistency with the noise model.

Up to a normalization constant, the joint rate posterior has the form of

$$p(\{R\},\Lambda_T,\{\langle VT\rangle\}|\{x\}) = p(\{R\},\Lambda_T,\{\langle VT\rangle\})\prod_{\mu}\left\{\Lambda_T p(x_\mu|T) + \sum_i R_i\langle VT\rangle_i p(x_\mu|A_i)\right\}\exp\left(-\Lambda_T - \sum_i R_i\langle VT\rangle_i\right), \tag{9}$$

where $p(\{R\},\Lambda_T,\{\langle VT\rangle\})$ is the joint prior density and $\mu$ enumerates the event candidate ranking statistics for each search. We single out the terrestrial class, since we measure only its overall count $\Lambda_T$ and are otherwise uninterested in its properties as an event rate density. To account for statistical and calibration uncertainty in $\langle VT\rangle$, all searches marginalize over an 18% relative uncertainty incorporated into the prior on $\langle VT\rangle$. To quantify our uncertainty on the mass and spin distributions (encoded in $\{\theta\}$) of the source populations, we examine different populations. As mentioned before, $f(z|\{\theta\})$ is mostly unaffected by these choices, but the $\langle VT\rangle$ estimated is strongly influenced by the assumed population. Thus, a separate rate posterior is obtained for each population tested, and we also quote the union of population-specific credible intervals as the overall rate interval.

Previous GW BBH and BNS event rate distributions treat each source category independently—event candidates are assigned a category *a priori* based on the properties measured by a given search. In the following, PyCBC and cWB retain this approach. In addition to the rates derived in this way, we also present the results from an enhanced model jointly treating multiple astrophysical event categories [89]. This enhancement accounts for correlations between the source categories by measuring the response of the GstLAL search template banks to the astrophysical populations. We expect that the methods should agree for two reasons: negligible correlation between the BBH and BNS category and no significant candidates in the intervening NSBH category. Since an inspection of the corresponding rate posteriors confirms this expectation, we present a single search-combined posterior for both BBH populations. In the BNS category, the rates are also compatible between searches, but, because they are derived from differing methods, we present the BNS rate posteriors separately.

We update the event rates for the BBH and BNS categories with the additional events and observing time. In addition to the two categories with confirmed detections, we also revisit the NSBH category. In contrast to the BNS and BBH categories, there are no confident detections in the NSBH spaces (see Tables I and II and the absence of significant candidates in Table IV). Hence, we update the upper limits on the NSBH event rate from $O1$. Instead of using $O1$ or earlier detections as a prior on the $O2$ measurement, we reanalyze $O1$ and $O2$ as a whole and use an uninformative prior on the result. As in Ref. [4], we use the Jeffreys prior for a Poisson rate parameter, proportional to $R_i^{-1/2}$, for BNS and BBH, while for NSBH we use a prior uniform in $R_i$ which yields a conservative upper limit bound.

### A. Event classification

To determine the probability that a given candidate originated in one of the four categories, the models are marginalized over the counts with the ranking statistic distributions fixed at the value of the ranking statistic of the candidate. The distribution that is marginalized is the ratio of the category under consideration versus all categories (including terrestrial):

$$p_{A_i}(x_\mu|\{x\}) = \int p(\{R\},\Lambda_T,\{\langle VT\rangle\}|\{x\}) \times \frac{R_i\langle VT\rangle_i p(x_\mu|A_i)}{\Lambda_T p(x_\mu|T) + \sum_j R_j\langle VT\rangle_j p(x_\mu|A_j)} \times d\{R\}d\Lambda_T d\{\langle VT\rangle\}. \tag{10}$$

Thus, we obtain $p_{\rm terrestrial}$, $p_{\rm BBH}$, $p_{\rm BNS}$, and $p_{\rm NSBH}$, which are mutually exclusive categorizations. The *overall* probability of astrophysical origin sums the expression over all categories in $\{A\}$.

We expect different values of $p_{A_i}$ to be assigned to any given event by different search pipelines. This assignment is due to differences in the averaged efficiency of various methods to discriminate signal from noise events and also to the effects of random noise fluctuations on the ranking statistics assigned to a specific event. We also expect systematic uncertainties in the quoted probabilities due to our lack of knowledge of the true event populations, for instance, the mass distribution of BNS and NSBH mergers.

Parameter estimation is not performed on all candidates used to obtain rate estimates, so only the search masses and rankings are used to derive the astrophysical probabilities. Table IV shows the per-pipeline assigned probability values for each of the relevant categories. The cWB search does not have a specific event type corresponding to NSBH or BNS; thus, we treat all cWB search events as BBH candidates. The astrophysical probabilities from PyCBC are estimated by applying simple chirp mass cuts to the set of events with ranking statistic $\rho > 8$: Events with $\mathcal{M}<2.1$ are considered as candidate BNS, those with $\mathcal{M} > 4.35$ as candidate BBH, where the lower bound assumes two 5 $M_\odot$ BHs, and all remaining events as potential NSBH. We note that the value of the boundary between NSBH and BBH is chosen somewhat arbitrarily, given the uncertainty as to the exact value in our current understanding due to, for example, the formation and environment of the source. The astrophysical probabilities from GstLAL in Table IV are estimated using the pipeline response to injected synthetic signals, where neutron stars are assumed to have masses in the range 1–3 $M_\odot$ and black holes are assumed to have masses of 3 $M_\odot$ or larger. The details can be found in Ref. [89]. We note that the different definitions used by these three pipelines in classifying events as BNS, NSBH, or BBH reflect current astrophysical uncertainties in such classifications. Other, yet different definitions are used in order to compute event rates in the following subsections.

### B. Binary black hole event rates

After the detection of GW170104, the event rate of BBH mergers had been measured to lie between 12 and 213 $\mathrm{Gpc^{-3}\,y^{-1}}$ [15]. This measurement included the four events identified at that time. The $\langle VT \rangle$, and hence the rates, are derived from a set of assumed BBH populations. In $O1$, two distributions of the primary mass—one uniform in the log and one a power law $p(m_1) \propto m_1^{-\alpha}$ with an index of $\alpha = 2.3$—are used as representative extremes. In both populations shown here, the mass distribution cuts off at a lower mass of 5 $M_\odot$. The mass distributions cut off at a maximum mass of 50 $M_\odot$. The detector network is sensitive to binaries with a larger mass; however, the new cutoff is motivated by both more sophisticated modeling of the mass spectrum [55] preferring maximum BH masses much smaller than the previous limit of 100 $M_\odot$, as well as astrophysical processes which are expected to truncate the distribution [136]. The BH spin distribution has magnitude uniform in [0, 1]. The PyCBC search uses a spin tilt distribution which is isotropic over the unit sphere, and GstLAL uses a distribution that aligns BH spins to the orbital angular momentum.

The posteriors on the rate distributions are shown in Fig. 12. Including all events, the event rate is now measured to be $R = 56^{+44}_{-27}$ $\mathrm{Gpc^{-3}\,y^{-1}}$ (GstLAL) and $R = 57^{+47}_{-29}$ $\mathrm{Gpc^{-3}\,y^{-1}}$ (PyCBC) for the power-law distribution. For the uniform in log distribution, we obtain $R = 18.1^{+13.9}_{-8.7}\mathrm{Gpc^{-3}\,y^{-1}}$ (GstLAL) and $R = 19.5^{+15.2}_{-9.7}$ $\mathrm{Gpc^{-3}\,y^{-1}}$ (PyCBC). The difference in $\langle VT \rangle$ and rate distributions between the two spin populations is smaller than the uncertainty from calibration. Therefore, we present in Fig. 12 the rate distribution for both assumed mass distributions, combined over searches as an averaging over the spin configurations. The union of the intervals combined over both populations lies in 9.7–101 $\mathrm{Gpc^{-3}\,y^{-1}}$. GW170608 is included in the estimation of $\Lambda$ for BBH, but, given difficulties in characterizing the amount of time in which it could have occurred, its analysis period is not included in the overall $\langle VT \rangle$. We believe this exclusion introduces a bias that is no larger than the already accounted for calibration uncertainty.

A more detailed analysis [4] previously showed that both of the assumed populations used here are consistent with an inferred fit to the power-law index $\alpha$ as measured from the population of events known at the time. An update to this analysis using all current detections and examining a variety of plausible mass and spin distributions is explored in Ref. [55]. Allowing for a self-consistent fit to the event rate while varying a power-law model with a spectral index and maximum and minimum primary mass, the rate interval is found to be $53^{+56}_{-28}$ $\mathrm{Gpc^{-3}\,y^{-1}}$. This result is consistent with the intervals obtained from the fixed parameter populations used here. Within the same model, we obtain a 90% interval of the distribution for the power-law index of $\alpha = 1.3^{+1.4}_{-1.7}$. Compared with the earlier analysis [4], this result favors somewhat shallower power-law indices.

### C. Binary neutron star event rates

The discovery of GW170817 is the only unambiguous BNS candidate obtained in $O2$. Regardless, it provides a means to independently measure the rate of binary neutron star mergers. Previous estimates [211–213] from observations are derived from the properties of neutron star binaries with a pulsar component [214]. Earlier analyses [18,208] use a population model of binary neutron stars with uniform component masses in the 1–2 $M_\odot$ range and obtain an event rate interval of 320–4740 $\mathrm{Gpc}^{-3}\,\mathrm{y}^{-1}$. In addition to updating this rate to account for all available data from $O1$ and $O2$, we also introduce another fiducial population, serving two purposes. The first is to emulate a distribution assumed previously [208], which models both components as uncorrelated Gaussians. The overall mass distribution is centered at 1.33 $M_\odot$ with a standard deviation of 0.09 $M_\odot$. Second, this distribution can be considered as a bracket on the event rate from the upper end, since its $\langle VT\rangle$ over the population is smaller than the value obtained from the uniform set. To facilitate comparison and keep commensurate ranges of masses between the two distributions, we expand the uniform set to have component masses distributed between 0.8 and 2.3 $M_\odot$.

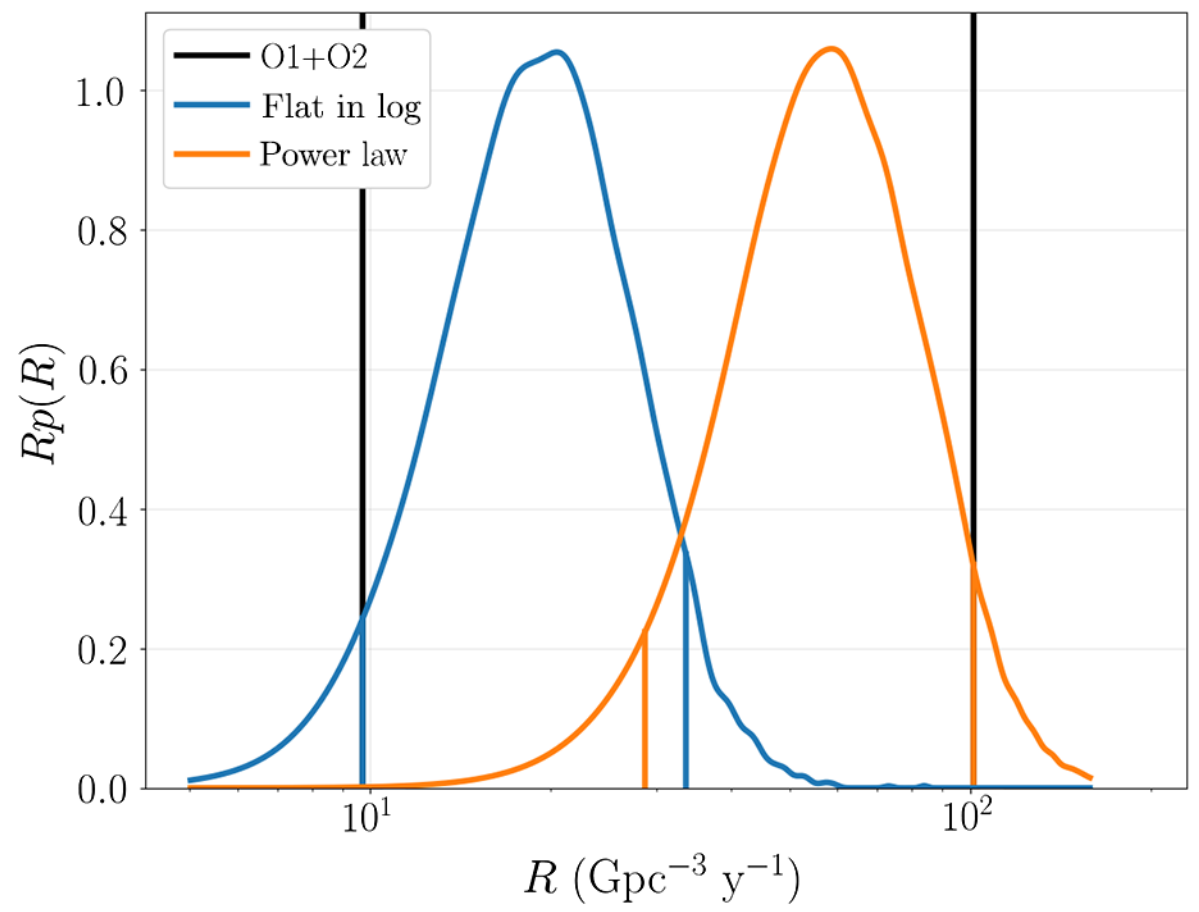

FIG. 12. This figure shows the posterior distribution—combined from the results of PyCBC and GstLAL—on the BBH event rate for the flat in log (blue) and power-law (orange) mass distributions. The symmetric 90% confidence intervals are indicated with vertical lines beneath the posterior distribution. The union of intervals is indicated in black.

TABLE IV. This table lists the probability of each event candidate belonging to a given source category. Each category is further delineated by the probabilities derived from each search pipeline's output. Where candidate values are not indicated, the search did not calculate a probability for that category or did not have a candidate with a ranking statistic sufficient for interest. $p_{A_i}$ values below $10^{-3}$ are shown as zero. The astrophysical category is the sum over the BNS, NSBH, and BBH categories (where available), and the sum over astrophysical and terrestrial is unity.

| | GstLAL | | | | | PyCBC | | | | | cWB | |
|---|---|---|---|---|---|---|---|---|---|---|---|---|
| | Terrestrial | BNS | NSBH | BBH | Astrophysical | Terrestrial | BNS | NSBH | BBH | Astrophysical | Terrestrial | BBH |
| GW150914 | 0 | 0 | 0.0064 | 0.99 | 1 | 0 | ... | ... | 1 | 1 | 0 | 1 |
| 151008[a] | ... | ... | ... | ... | ... | 0.73 | ... | ... | 0.27 | 0.27 | ... | ... |
| 151012.2 | 0.98 | 0.022 | 0.0012 | 0 | 0.023 | ... | ... | ... | ... | ... | ... | ... |
| GW151012 | 0.001 | 0 | 0.031 | 0.97 | 1 | 0.04 | ... | ... | 0.96 | 0.96 | ... | ... |
| 151116[b] | ... | ... | ... | ... | ... | $\sim$1 | $\ll$ 05 | ... | ... | $\ll$ 0.5 | ... | ... |
| GW151226 | 0 | 0 | 0.12 | 0.88 | 1 | 0 | ... | ... | 1 | 1 | 0.05 | 0.95 |
| 161202 | 0.97 | 0.034 | 0 | 0 | 0.034 | ... | ... | ... | ... | ... | ... | ... |
| 161217 | 0.98 | 0 | 0.011 | 0.0078 | 0.018 | ... | ... | ... | ... | ... | ... | ... |
| GW170104 | 0 | 0 | 0.0028 | 1 | 1 | 0 | ... | ... | 1 | 1 | 0 | 1 |
| 170208 | 0.98 | 0 | 0.011 | 0.0088 | 0.02 | ... | ... | ... | ... | ... | ... | ... |
| 170219 | 0.98 | 0.019 | 0 | 0 | 0.02 | ... | ... | ... | ... | ... | ... | ... |
| 170405 | 1 | 0.004 | 0 | 0 | 0.004 | ... | ... | ... | ... | ... | ... | ... |
| 170412 | 0.94 | 0 | 0.029 | 0.032 | 0.06 | ... | ... | ... | ... | ... | ... | ... |
| 170423 | 0.91 | 0.086 | 0 | 0 | 0.086 | ... | ... | ... | ... | ... | ... | ... |
| GW170608 | 0 | 0 | 0.084 | 0.92 | 1 | 0 | ... | ... | 1 | 1 | 0 | 1 |
| 170616[b] | ... | ... | ... | ... | ... | $\sim$1 | ... | $\ll$ 0.5 | ... | $\ll$ 0.5 | ... | ... |
| 170630 | 0.98 | 0.02 | 0 | 0 | 0.02 | ... | ... | ... | ... | ... | ... | ... |
| 170705 | 0.99 | 0 | 0.006 | 0.0061 | 0.012 | ... | ... | ... | ... | ... | ... | ... |
| 170720 | 0.99 | 0 | 0.0077 | 0.002 | 0.0097 | ... | ... | ... | ... | ... | ... | ... |
| GW170729 | 0.018 | 0 | 0 | 0.98 | 0.98 | 0.48 | ... | ... | 0.52 | 0.52 | 0.057 | 0.94 |
| GW170809 | 0 | 0 | 0.0064 | 0.99 | 1 | 0 | ... | ... | 1 | 1 | ... | ... |
| GW170814 | 0 | 0 | 0.0024 | 1 | 1 | 0 | ... | ... | 1 | 1 | 0 | 1 |
| GW170817 | 0 | 1 | 0 | 0 | 1 | 0 | 1 | ... | ... | 1 | ... | ... |
| GW170818 | 0 | 0 | 0.0053 | 0.99 | 1 | ... | ... | ... | ... | ... | ... | ... |
| GW170823 | 0 | 0 | 0.0059 | 0.99 | 1 | 0 | ... | ... | 1 | 1 | 0.0043 | 1 |

[a]Calculated assuming that this event is a member of the BBH population for PyCBC, though its NSBH probability could also be non-negligible.

[b]The astrophysical probability for categories with few or zero detected events can be strongly influenced by the assumed prior on rates and physical property distributions. As such, we provide only upper bounds on these values.

The event rate distribution for each search and mass distribution is shown in Fig. 13. The differences in the distribution between the searches are a consequence of the ranking statistic threshold applied to either. PyCBC measures a smaller $\langle VT\rangle$, because its fiducial threshold is higher than GstLAL. Despite the threshold difference, the two searches find similar values for $\Lambda_{\mathrm{BNS}}$, and hence the rate for GstLAL is lower than for PyCBC. For the uniform mass set, we obtain an interval at 90% confidence of $R = 800^{+1970}_{-680}\ \mathrm{Gpc}^{-3}\,\mathrm{y}^{-1}$ (PyCBC) and $R = 662^{+1609}_{-565}\ \mathrm{Gpc}^{-3}\,\mathrm{y}^{-1}$ (GstLAL), and for the Gaussian set we obtain $R = 1210^{+3230}_{-1040}\ \mathrm{Gpc}^{-3}\,\mathrm{y}^{-1}$ (PyCBC) and $R = 920^{+2220}_{-790}\ \mathrm{Gpc}^{-3}\,\mathrm{y}^{-1}$ (GstLAL). These values are consistent with previous observational values (both GW and radio pulsar) as well as more recent observationally driven investigations [215]. The union of the intervals combined over both populations lies in 110–3840 $\mathrm{Gpc}^{-3}\,\mathrm{y}^{-1}$.

### D. Neutron star black hole event rates

The NSBH space is a unique challenge both to model astrophysically and for which to produce accurate waveforms. Astrophysical models span a wide range of potential mass ratios and spin configurations, and there are no electromagnetic observational examples. Hence, we take an approach similar to previous analyses [208] and examine specific points in the mass space while considering two component spin configurations: isotropic and orbital angular momentum aligned as described in Sec. VII B.

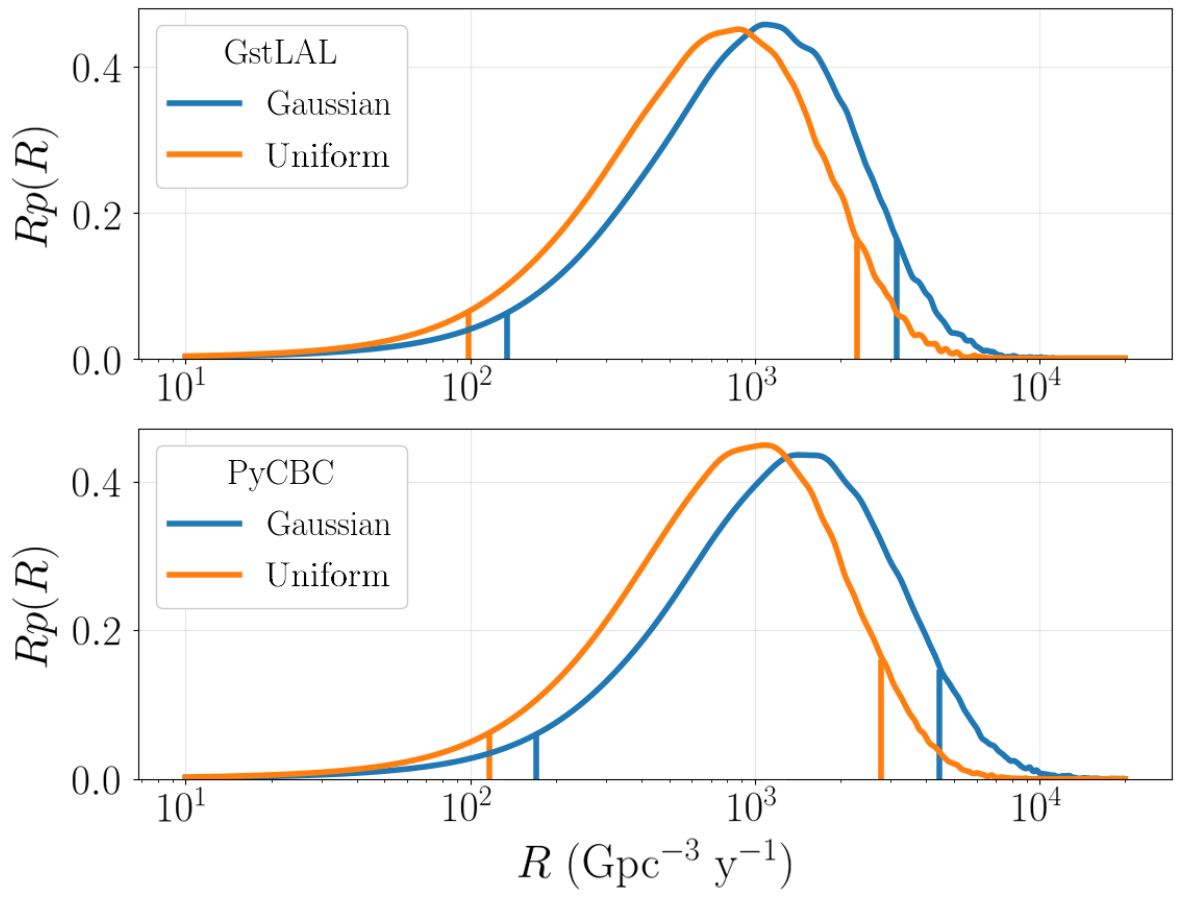

FIG. 13. This figure shows the posterior distributions of the BNS event rate for the GstLAL and PyCBC searches. The uniform mass distribution corresponds to the orange curves, and Gaussian mass distributions correspond to the blue curves. The symmetric 90% confidence intervals are indicated with vertical lines beneath the posterior distributions.

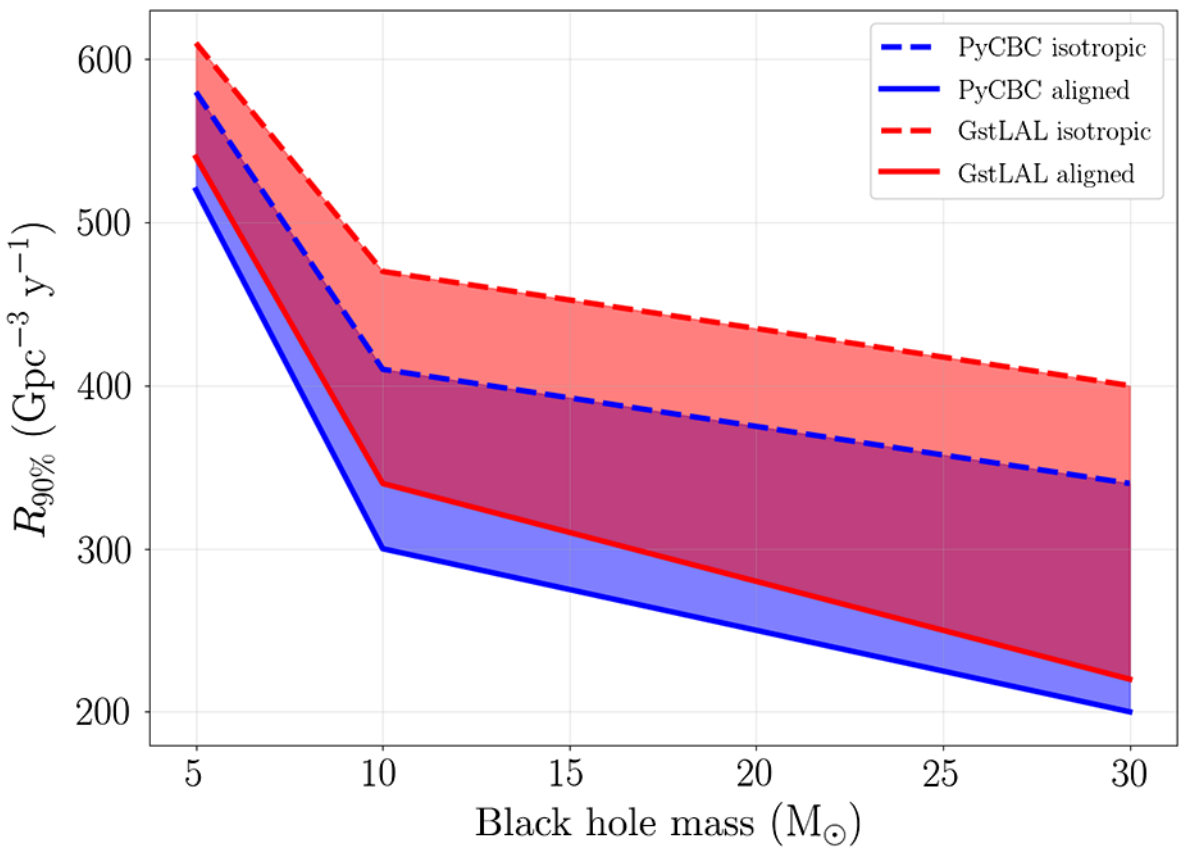

FIG. 14. This figure shows the 90% rate upper limit for the NSBH category, measured at a set of three discrete BH masses (5, 10, and 30 $M_\odot$) with the fiducial NS mass fixed to 1.4 $M_\odot$. The upper limit is evaluated for both matched-filter search pipelines, with GstLAL corresponding to red curves and PyCBC to blue. We also show two choices of spin distributions: isotropic (dashed lines) and aligned spin (solid lines).

Since there are no confident detection candidates in the NSBH category, we update the upper limit at 90% confidence in this category in Fig. 14. All upper limits are below 610 $\mathrm{Gpc^{-3}\,y^{-1}}$. Those results are obtained using a uniform prior over $R$. The Jeffreys prior (which also appeared in Ref. [208]) suppresses larger $R$ values. This prior choice would obtain a less conservative upper limit. This limit is now stronger at all masses than the "high" rate prediction [216] ($10^3$ $\mathrm{Gpc^{-3}\,y^{-1}}$) for NSBH sources.

## VIII. CONCLUSIONS

In GWTC-1, we have reported the results from three GW searches for compact mergers during the first and second observing runs by the advanced GW detector network. Advanced LIGO and Advanced Virgo have confidently detected gravitational waves from ten stellar-mass binary black hole mergers and one binary neutron star inspiral. The signals were discovered using three independent analyses: two matched-filter searches [8,9] and one weakly modeled burst search [11]. We have reported four previously unpublished BBH signals discovered during $O2$, as well as updated FARs and parameter estimates for all previously reported GW detections. The reanalysis of $O1$ data did not reveal any new GW events, but improvements to the various detection pipelines have resulted in an increase of the significance of GW151012.

Including these four new BBH mergers, the observed BBHs span a wide range of component masses, from $7.7^{+2.2}_{-2.5}$ $M_\odot$ to $50.2^{+16.2}_{-10.2}$ $M_\odot$. One of the new events, GW170729, is found to be the highest-mass BBH observed to date, with GW170608 still being the lightest BBH [17]. The three other new events GW170809, GW170818, and GW170823 are all identified as heavy stellar-mass BBH mergers, ranging in total mass from $59.0^{+5.4}_{-4.1}$ $M_\odot$ to $68.7^{+10.8}_{-8.1}$ $M_\odot$.

Similar to previous results, we find that the spins of the individual black holes are only weakly constrained, though for GW151226 and also for GW170729 we find that $\chi_{\mathrm{eff}}$ is positive and thus can rule out two nonspinning black holes as their constituents at greater than the 90% credible level. On the other hand, $\chi_p$ is only weakly constrained in our measurements. Furthermore, we find that the higher mode content of the observed GW signals is weak enough that waveform models including them are not strongly preferred given our data (see Appendix B 3).

The binary mergers observed during $O1$ and $O2$ range in distance between $40^{+8}_{-15}$ Mpc for the binary neutron star inspiral GW170817 to $2840^{+1400}_{-1360}$ Mpc for GW170729, making it not only the heaviest BBH but also the most distant one observed to date. For the BNS merger GW170817, we have presented conservative upper limits on the properties of the remnant.

GW170818 is the second triple-coincident LIGO-Virgo GW event and is localized to an area of 39 $\mathrm{deg}^2$, making it the best localized BBH to date. A similar impact of Virgo on the sky localization was already seen for GW170814 [16], reaffirming the importance of a global GW detector network for accurately localizing GW sources [200].

The properties of the observations reported in this catalog are based on general relativistic waveform models. Tests of the consistency of these observations with GR can be found in Refs. [15,217,218].

We have also presented a set of 14 marginal candidate events identified by the two matched-filter searches. The number of observed marginal events is consistent with our expectation given the chosen FAR threshold, but it is not possible to say whether or not a particular marginal trigger is a real GW signal.

Even with the set of ten BBH and one BNS, several outstanding questions remain regarding the origin and evolution of the detected binaries. To date, no binary components have been observed in either of the two putative mass gaps [139,140]—one between NSs and BHs and the other one due to pair instability supernovae [136,219]. Gravitational-wave measurement of BH spins favors either small magnitudes or large misalignment with the orbital angular momentum. The latter favors a formation scenario where no spin alignment process is present, e.g., assembly in globular clusters [172,174]. Several studies [166–169,220–225] indicate that, with a few hundred detections, more detailed formation scenarios and evolutionary details can be parsed from the population. The BBH sample from $O1$ and $O2$ allows for new constraints on the primary mass power-law index $\alpha = 1.3^{+1.4}_{-1.7}$ [55].

Data products, including strain data and posterior samples, and postprocessing tools can be obtained from the accompanying data release [56].

The third observing run ($O3$) of Advanced LIGO and Virgo began on April 1, 2019. The inferred rate of BBH mergers is 9.7–101 $\mathrm{Gpc^{-3}\,y^{-1}}$ and for BNS 110–3840 $\mathrm{Gpc^{-3}\,y^{-1}}$, and for NSBH binaries we obtain an improved 90% upper limit of the merger rate of 610 $\mathrm{Gpc^{-3}\,y^{-1}}$; in combination with further sensitivity upgrades to both LIGO and Virgo as well as the prospects of the Japanese GW detector KAGRA [226–228] joining the network possibly towards the end of $O3$ in 2019, many tens of binary observations are anticipated in the coming years [200], which will be presented in forthcoming catalogs.

## ACKNOWLEDGMENTS

The authors gratefully acknowledge the support of the United States National Science Foundation (NSF) for the construction and operation of the LIGO Laboratory and Advanced LIGO as well as the Science and Technology Facilities Council (STFC) of the United Kingdom, the Max-Planck-Society (MPS), and the State of Niedersachsen/Germany for support of the construction of Advanced LIGO and construction and operation of the GEO600 detector. Additional support for Advanced LIGO was provided by the Australian Research Council. The authors gratefully acknowledge the Italian Istituto Nazionale di Fisica Nucleare (INFN), the French Centre National de la Recherche Scientifique (CNRS), and the Foundation for Fundamental Research on Matter supported by the Netherlands Organisation for Scientific Research, for the construction and operation of the Virgo detector and the creation and support of the EGO consortium. The authors also gratefully acknowledge research support from these agencies as well as by the Council of Scientific and Industrial Research of India, the Department of Science and Technology, India, the Science and Engineering Research Board (SERB), India, the Ministry of Human Resource Development, India, the Spanish Agencia Estatal de Investigación, the Vicepresidència i Conselleria d'Innovació, Recerca i Turisme and the Conselleria d'Educació i Universitat del Govern de les Illes Balears, the Conselleria d'Educació, Investigació, Cultura i Esport de la Generalitat Valenciana, the National Science Centre of Poland, the Swiss National Science Foundation (SNSF), the Russian Foundation for Basic Research, the Russian Science Foundation, the European Commission, the European Regional Development Funds (ERDF), the Royal Society, the Scottish Funding Council, the Scottish Universities Physics Alliance, the Hungarian Scientific Research Fund (OTKA), the Lyon Institute of Origins (LIO), the Paris Île-de-France Region, the National Research, Development and Innovation Office Hungary (NKFIH), the National Research Foundation of Korea, Industry Canada and the Province of Ontario through the Ministry of Economic Development and Innovation, the Natural Science and Engineering Research Council Canada, the Canadian Institute for Advanced Research, the Brazilian Ministry of Science, Technology, Innovations, and Communications, the International Center for Theoretical Physics South American Institute for Fundamental Research (ICTP-SAIFR), the Research Grants Council of Hong Kong, the National Natural Science Foundation of China (NSFC), the Leverhulme Trust, the Research Corporation, the Ministry of Science and Technology (MOST), Taiwan, and the Kavli Foundation. The authors gratefully acknowledge the support of the NSF, STFC, MPS, INFN, CNRS, and the State of Niedersachsen/Germany for provision of computational resources.

## APPENDIX A: CHARACTERIZATION OF TRANSIENT NOISE RELEVANT TO CATALOG TRIGGERS

The instrumental artifacts identified in time coincidence with triggers in this catalog can be split into four groups: *scattered light*, *60–200 Hz nonstationarity*, *short-duration transients*, and *blips*. Time-frequency spectrograms of each glitch class can be seen in Fig. 15. We discuss the challenges of analyzing the times surrounding the instrumental artifacts and mitigation methods that can be used to address excess noise in the data.

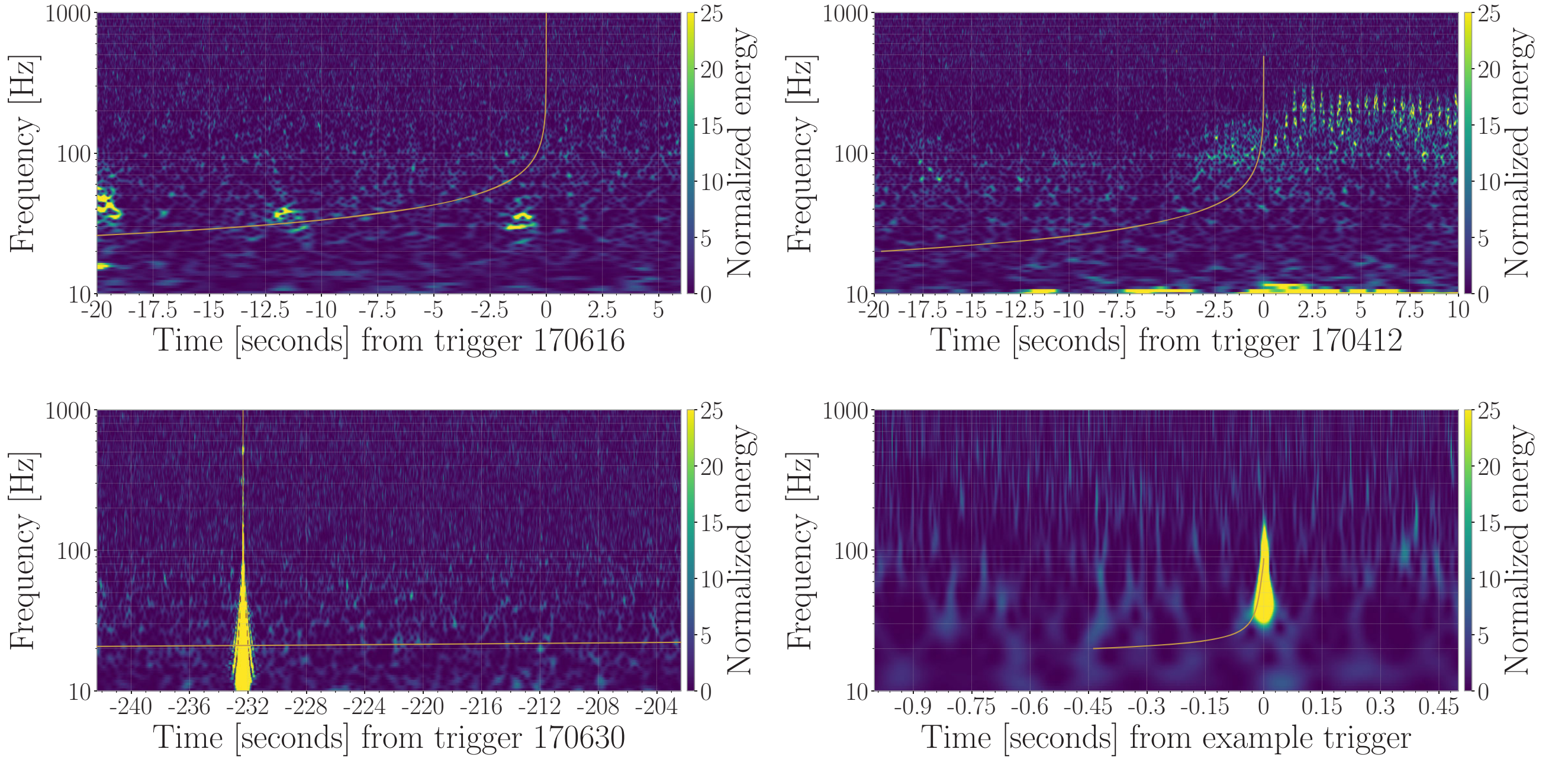

FIG. 15. Normalized spectrograms of the time around common noise artifacts with a time-frequency evolution of a related trigger template overlaid. Top left: Scattered light artifacts at Hanford with the template of trigger 170616 overlaid. Top right: A 60–200 Hz nonstationarity at Livingston with the template of trigger 170412 overlaid. Bottom left: A short-duration transient at Livingston with the template of trigger 170630 overlaid. Bottom right: A blip at Hanford with the template of a subthreshold high-mass trigger overlaid.

### 1. Scattered light

Scattered light is a common source of noise in the LIGO and Virgo detectors. Stray light is reflected back into the main interferometer path, resulting in excess power in the data [91,229,230]. Motion of the reflective surface, such as optic mounts, phase shift the reflected light. When this motion is smaller than the wavelength of the main laser, 1064 nm, the resultant artifacts are associated with stationary noise contributions to the interferometer spectrum. Larger motions result in archlike shapes in the time-frequency spectrograms. It is these larger motions that impact transient searches and which, therefore, concern us here. This motion is observed by auxiliary sensors and is used to help identify periods of scattering [231]. Scattering light can be present in the data for stretches of multiple hours during periods of increased ground motion [91]. Scattered light is one of the most common sources of background triggers in searches for compact binary coalescence signals [68,92,93]. Variability in the time-frequency morphology of scattered light leads to noise triggers for a wide variety of template parameters.

The strain noise amplitude of scattered light instrumental artifacts correlates with the intensity of the corresponding ground motion. Additionally, higher velocities of optic motion lead to higher-frequency content in the scattered light [229].

Since scattered light typically impacts low frequencies, it is possible to mitigate the impact by using only data from frequencies above the impacted region. Because it is often related to monitored optic motion, subtraction of the artifacts based on a nonlinear relationship with this optic motion may be possible.

### 2. 60–200 Hz nonstationarity

The 60–200 Hz nonstationarity appears in time-frequency spectrograms as excess power with slowly varying frequencies in clusters of multiple minutes [68]. The structure of the excess power appears similar to scattered light but at higher frequencies than predicted by available witnesses of optic motion. This type of nonstationarity occurs in both Hanford and Livingston at a rate of 1–2 times per day, making it unlikely for an astrophysical signal to be correlated by chance. While correlations between excess seismic noise and the nonstationarity exist, no clear witnesses have been identified. The structure of this class of artifact changed after mitigation of the motion of baffles used to block stray light [232,233]. This change suggests that baffles may be involved in the production of the 60–200 Hz nonstationarity.

In previous observing runs, the rate of these artifacts caused multiple hours of data to be vetoed. In time that is not vetoed, these transients can create significant triggers in the background of matched-filter searches [68]. This instrumental artifact particularly affects signals that are in the sensitive band of the detector above 30 Hz for longer than 3 sec.

The long-duration, variable frequencies and lack of a clear witness make this nonstationarity a difficult target for noise subtraction. Efforts to completely mitigate the periods of nonstationarity are ongoing.

#### 3. Short-duration transient

Short-duration transients last less than one second and have a high amplitude. The highest-amplitude transients can cause overflows of the digital-to-analog converters used to control the positions of the test masses. These transients occur in both LIGO detectors at a rate of approximately 1 per hour, with their cause largely unknown.

The large amount of excess power due to these artifacts produces a large impulse response during the whitening process, affecting the ability of searches to optimally search the surrounding data [8]. For this reason, these artifacts are gated by the searches before analysis, using the procedure described in Ref. [8]. Notably, the instrumental artifact present in LIGO-Livingston during GW170817 [18] is of this class.

For systems like GW170817, gating is shown to remove short-duration transients without a significant bias to astrophysical parameter estimation [234]; full glitch subtraction with BayesWave [53], as used for the GW170817 parameter estimation, produces more robust results.

#### 4. Blips

Blip transients [91] are short, band-limited transients that occur in both LIGO detectors at a rate of roughly once per hour. Because of their subsecond duration and limited bandwidth, these transients often have a significant overlap with the shortest templates used in matched-filter searches. Templates that terminate between 50 and 100 Hz and have high ratios of component mass parameters have a similar morphology to these artifacts. Blip transients are particularly problematic, as they typically do not couple into any witness sensors used to monitor the detector, which makes it difficult to systematically remove them from the analyses. As such, these transients are the limiting noise source to modeled searches for high-mass compact binary coalescences in $O1$ and $O2$ [68,72,91,235].

Investigations into blips have identified multiple causes [236], but the vast majority of blips remain unexplained. Although these transients cannot be removed from the analysis entirely, signal morphology tests in matched-filter searches are used to mitigate their effects [72].

## APPENDIX B: PARAMETER-ESTIMATION DESCRIPTION

We use coherent Bayesian inference methods to extract the posterior distribution $p(\vec{\vartheta}|\vec{d})$ for the parameters $\vec{\vartheta}$ that characterize a compact binary coalescence associated with a particular GW event. Following Bayes' theorem [237,238], the posterior is proportional to the product of the likelihood of the data given the parameters and the prior (assumed) distribution of the parameters. The likelihood function depends on a noise-weighted inner product between the detector data $\vec{d}$ and a parametrized waveform model for the two GW polarizations $h_{+,\times}(\vec{\vartheta};t)$ which is projected onto the response of each detector to obtain the strain [128]. By marginalizing the posterior distribution over all but one or two parameters, it is then possible to generate credible intervals or credible regions for those parameters.

We sample the posterior distribution with stochastic sampling algorithms using an implementation of Markov chain Monte Carlo [239,240] and nested sampling available in the LALInference package [241] as part of the LSC Algorithm Library (LAL) [242]. Additional posterior results for computationally expensive waveform models are obtained with an alternative parallelized parameter-estimation code RapidPE [243,244].

We estimate the power spectral density (PSD) that enters the inner product using BayesWave [53,54]. The PSD is modeled as a cubic spline for the broadband structure and a sum of Lorentzians for the line features. A median PSD is computed from the resulting posterior probability density function of PSDs, defined separately at each frequency. This PSD is expected to lead to more stable and reliable inference. The parameter estimation analyses in $O1$ assumed uniform priors for the calibration uncertainties in frequency. For the analyses in this catalog, frequency-dependent spline calibration envelopes [98] are incorporated into the measured GW strain to factor in potential deviations from the true GW strain due to uncertainties in the detector calibration [64,65]. We marginalize over the additional calibration parameters. See Sec. II B in Ref. [97] for details.

The low-frequency cutoff for likelihood integration used for parameter-estimation analyses is set to 20 Hz with the following exceptions: For GW170817 the analysis starts at 23 Hz, for GW170818 at 16 Hz, and for the GW170608 Hanford data at 30 Hz.

Because of the expansion of the Universe, we measure redshifted masses from GW observations. We assume a standard flat ΛCDM cosmology with Hubble parameter $H_0 = 67.9\ \mathrm{km\,s^{-1}\,Mpc^{-1}}$ and matter density parameter $\Omega_m = 0.306$ [245] and compute the redshift from the measured luminosity distance distribution. To obtain physical binary masses in the source frame, we divide the observed detector-frame masses by $(1+z)$ [12,246]. For the events in this catalog, changing from the above cosmology to the updated *Planck* (TT, TE, EE + lowE + lensing + BAO) [247] cosmology results in a relative change in the redshift of <0.3% and a relative change in the source-frame total mass of <0.03%.

#### 1. Priors used in individual event analysis

We assume a jointly uniform prior in the detector-frame component masses with bounds chosen such that the posterior has support only in the interior of the domain.

Spin vectors are assumed to be isotropic on the sphere and uniform in spin magnitude. This prior is also used for models that enforce spins to be aligned with the orbital angular momentum. We use an isotropic prior for the location of the source on the sky. The distance prior is proportional to the luminosity distance squared. The prior distribution for the inclination angle $\theta_{JN}$ is assumed to be uniform in its cosine. The priors for polarization angle time and phase of coalescence are uniform.

For the reanalysis of GW170817, we choose two different spin priors, consistent with previous analyses [18,97]: $a_i \leq 0.89$ and $a_i \leq 0.05$. In addition to the BBH binary parameters, we also sample in the dimensionless tidal deformabilities $\Lambda_i$ of each NS. They are assumed to be jointly uniform within $0 \leq \Lambda_i \leq 5000$.

### 2. Waveform models

For analyses of BBH systems in this catalog, we use two waveform models [248]: an effective precession model (IMRPhenomPv2) [25,26,49] using the effective precession parameter $\chi_p$ and a full precession model (SEOBNRv3) [27,28,30] which includes generic two-spin inspiral precession dynamics. For both models, only the nonprecessing spin sector is tuned to NR simulations. Analyses with the effective precession model are carried out with LALInference. Analyses with the full precession model also use LALInference, except for GW170814, where we use RapidPE. For most BBH events, results from the two waveform models are consistent, and the data give us little reason to prefer one model over the other. Posteriors generated with LALInference and RapidPE for the two models also agree well for most of the events presented here. We point out notable differences in results between the BBH waveform models below.

To quantify the agreement between the models, we compute the Jensen-Shannon divergence (JSD) [249] between posterior distributions obtained with the two BBH waveform models. The JSD is a symmetrized and smoothed measure of distance between two probability distributions $p(x)$ and $q(x)$ defined as

$$D_{\mathrm{JS}}(p|q) = \frac{1}{2}\left[D_{\mathrm{KL}}(p|s) + D_{\mathrm{KL}}(q|s)\right], \qquad \text{(B1)}$$

where $s = 1/2(p+q)$ and

$$D_{\mathrm{KL}}(p|q) = \int p(x)\log_2\left(\frac{p(x)}{q(x)}\right)dx \qquad \text{(B2)}$$

is the Kullback-Leibler divergence (KLD) between the distributions $p$ and $q$, measured in bits. The JSD fulfills the bound $0 \leq D_{\mathrm{JS}}(p|q) \leq 1$ when measured in bits. We compute results for the KLDs for the JSDs between BBH models and the KLDs between prior and posteriors shown in Tables V and VI from kernel density estimates of random draws from the prior and posterior distributions and quoting the median and 90% intervals.

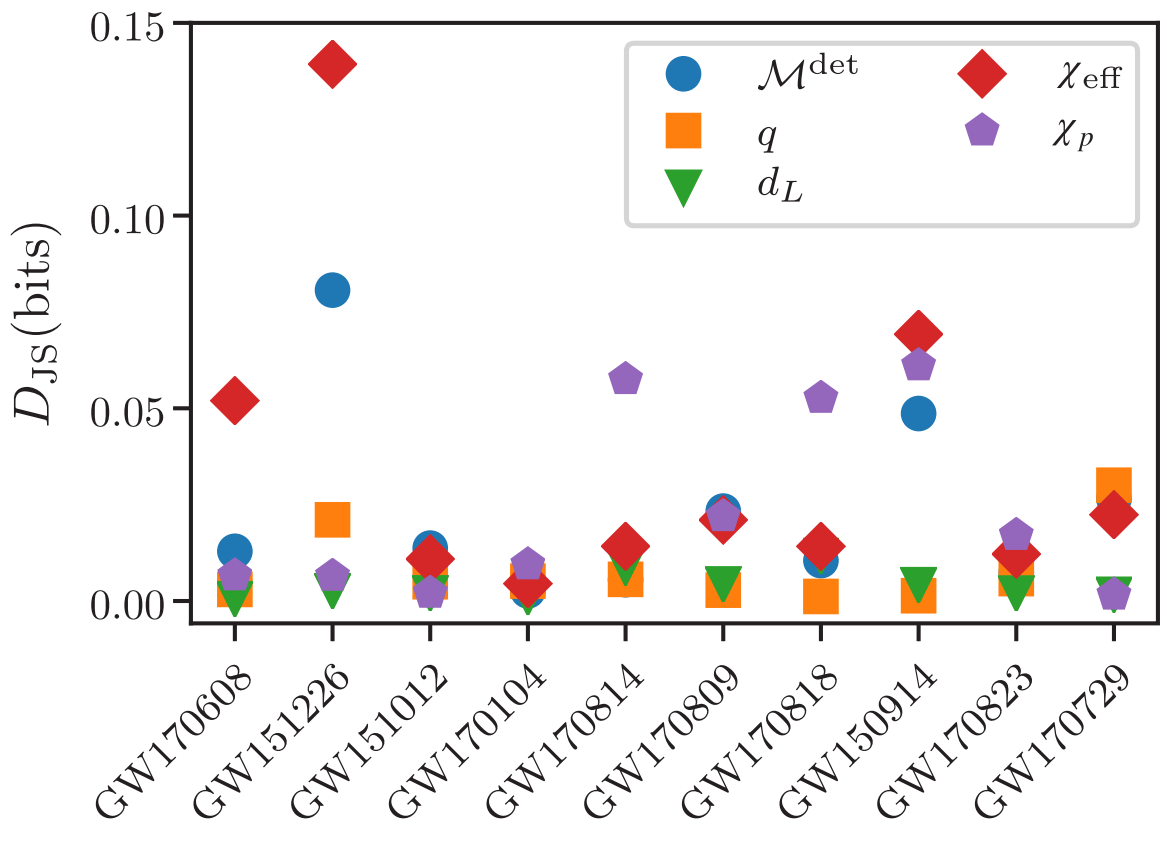

FIG. 16. Jensen-Shannon divergence between the two precessing BBH waveform models for key binary parameters, detector-frame chirp mass, mass ratio, luminosity distance, effective aligned spin, and effective precession spin.

We plot the JSD for all GW events for selected binary parameters in Fig. 16. The JSD values are, in general, smaller than approximately 0.05 bits, which indicates that the posteriors from the two BBH waveform models agree well. The largest value, $D_{\mathrm{JS}} = 0.14$ bits, is found for the $\chi_{\mathrm{eff}}$ distributions for GW151226. This result indicates the different $\chi_{\mathrm{eff}}$ probability density function (PDFs) measured by the two models for this event, as mentioned in Sec. V F. Further notable differences, quoted for the two BBH waveform models (SEOBNRv3 and IMRPhenomPv2) are the detector-frame chirp mass $\mathcal{M}^{\mathrm{det}}$ for GW151226: ($9.68^{+0.08}_{-0.08}\,M_{\odot}$, $9.70^{+0.07}_{-0.07}\,M_{\odot}$), the effective aligned spin $\chi_{\mathrm{eff}}$ for GW150914: ($0.01^{+0.12}_{-0.13}$, $-0.03^{+0.11}_{-0.12}$), and GW170608: ($0.02^{+0.19}_{-0.07}$, $0.04^{+0.19}_{-0.06}$), the effective precession spin $\chi_p$ for GW150914: ($0.30^{+0.34}_{-0.20}$, $0.39^{+0.47}_{-0.31}$), GW170814: ($0.37^{+0.47}_{-0.28}$, $0.34^{+0.43}_{-0.26}$), and GW170818: ($0.42^{+0.38}_{-0.28}$, $0.56^{+0.33}_{-0.39}$), and the mass ratio for GW170729: ($0.72^{+0.25}_{-0.28}$, $0.63^{+0.32}_{-0.26}$).

Because of the good overall agreement between waveforms, we present in our overall results posterior distributions for BBH coalescences that are averaged between the two models (using SEOBNR samples from LALInference and for GW170814 from RapidPE) and incorporate an equal number of samples from either model. These overall samples are used in the discussion of the source properties below. Waveforms from the full precession model need to be generated sufficiently far away from merger to enable cleanly attaching the merger-ringdown part to the inspiral-plunge part of the waveform. For several events, this procedure requires generating the full precession model from a lower starting frequency than for the effective precession model. To quote frequency-dependent quantities at a consistent reference frequency, which is a prerequisite for combining samples, we evolve the samples from the full precession model forward in time from 10 Hz to the fiducial reference frequency of 20 Hz.

We use marginalization over the arrival time and phase of the signal as an approximation for the fully precessing SEOBNRv3 model to make the analyses more computationally tractable. This approximation is valid if the $(\ell, m) = (2, \pm 1)$ observer-frame modes are subdominant compared to $(2, \pm 2)$ modes, as is the case for nearly face-on–face-off binaries. It is demonstrated that the impact of this approximation is negligible for GW150914 [96] and for GW170104 and GW151226 based on preliminary analyses.

*Waveform models for BNS.*—Unlike in the analysis of GW170817 in Ref. [97], we use three frequency-domain waveform models: the purely analytical TaylorF2 [35,36,38,100–112] and two point-particle models which add a fit to the phase evolution from tidal effects [32,99], SEOBNRv4NRT [29,32,77,99] and IMRPhenomPv2NRT [25,26,32,49,99,250]. These models are fast enough to be used as templates in LALInference. We supplement our results with two time-domain models (SEOBNRv4T [31] and TEOBResumS [33,113]), where the analysis is performed with RapidPE. In addition to tidal effects, IMRPhenomPv2NRT also includes the spin-induced quadrupole moment that enters in the phasing up to 3PN order [251,252]. In contrast to the analysis in Ref. [97], the terms up to 2PN order are now also included in SEOBNRv4NRT and SEOBNRv4T. The EOS dependence of each NS's spin-induced quadrupole moment is included by relating it to the tidal parameter of each NS using the quasiuniversal relations of Ref. [253]. For the spin-aligned models used for GW170817, the phase at coalescence is analytically marginalized out [241].

### 3. Impact of higher harmonics in the waveform

Waveform models including higher modes beyond the leading-order quadrupole contribution that cover the entire parameter space of our analyses were not available at the time of writing of this paper. Here, we systematically compare all $O2$ BBH observations with NR simulations supplemented by NR surrogate waveforms [254] that cover mass ratios up to $q = 0.2$ and aligned effective spins up to $|\chi_{\rm eff}| \sim 0.8$. These calculations focus on the impact of higher modes on the measurement of intrinsic parameters (i.e., masses and spins) using RapidPE [244,255,256] techniques. We find no compelling evidence that higher-order modes substantially affect our measurement of mass or spin parameters for any event. Instead, we find that they only modestly influence the interpretation of any observation, i.e., at a level smaller than our current statistical measurement uncertainty. For instance, we find for GW170729 a Bayes factor of approximately 1.4 for higher modes versus a pure quadrupole model. Assuming that GR is correct and these modes are present, however, we infer a modestly different mass ratio distribution with and without higher modes, with a mean (median) value of $q$ to be 0.61 (0.58) and 0.66(0.65), respectively, using the fiducial prior. Similarly, for 170809, we find a revised $\chi_{\rm eff}$ distribution which is symmetric about a median value of zero.

We conclude that the higher-mode content of the GW signals is weak enough that models including them are not strongly preferred given our data. This conclusion is consistent with the fact that the contribution from higher modes is highly suppressed for signals emitted by binaries with mass ratio $q \gtrsim 0.5$, total masses $\lesssim 100\ M_\odot$, and weak support for an edge on inclination $\theta_{JN} = 90°$, as is the case for the observed BBHs [257,258]. Our results agree with those in Refs. [144,259], which find that in these cases higher modes mostly affect the estimation of the inclination angle and luminosity distance.

## APPENDIX C: IMPACT OF PRIORS ON BAYESIAN PARAMETER ESTIMATION

In Bayesian inference, certain parameters inferred from measurements can be sensitive to the choice of priors and, thus, affect the interpretation of the observed events, with GW observations being no exception [55,179,260–262]. In this Appendix, we illustrate the impact of prior choices on the inference of source properties for GW events. Specifically, we choose the BBH GW170809, which lies in the bulk of GW observations to date (see Sec. V), as our explicit example.

### 1. Prior choices

Our default prior choice for the analysis of a GW event is uninformative. For GW170809, we choose the following *default* prior, henceforth referred to as P1:

(a) Component masses $m_i$ are distributed uniformly with the constraints that $m_1 \geq m_2$ and the total mass lies between $25\ M_\odot \leq M \leq 100\ M_\odot$. Priors on the chirp mass $\mathcal{M}$ and mass ratio $q \geq 0.125$ are determined by Jacobian transformations.
(b) The dimensionless spin magnitudes $0 \leq a_i \leq 0.99$ are distributed uniformly.
(c) The spin directions at a reference frequency $f_{\rm ref}$ are distributed isotropically (uniform in $\cos(\theta_i)$) on the unit sphere.
(d) The sources are distributed uniformly in volume with a maximum luminosity distance of $d_L \leq 4$ Gpc.
(e) The binary orientation is assumed to be isotropic.
(f) The coalescence time $t_c$ and phase $\phi_c$ are distributed uniformly.

Let us now consider two *alternative* prior choices:

P2: A *volumetric spin prior* in which the spin components are distributed uniformly inside the sphere $V = 4\pi(a_{\rm max}^3 - a_{\rm min}^3)/3$. This choice replaces assumptions (b) and (c) in the default prior P1.

TABLE V. KL divergences (in bits) between the prior and posterior for the effective aligned spin $\chi_{\rm eff}$ and the effective precession spin $\chi_p$. For the computation of the KL divergence for $\chi_p$, we quote the KL divergence with the prior conditioned on the $\chi_{\rm eff}$ posterior, $D_{\rm KL}^{\chi_p}(\chi_{\rm eff})$, and without conditioning, $D_{\rm KL}^{\chi_p}$. For GW170817, $D_{\rm KL}^{\chi_p}$ is given for the high spin prior. The median and 90% interval for the KL divergences is estimated by computing the statistic for repeated draws of a subset of the posterior and prior PDFs. Single-detector optimal SNRs from parameter-estimation analyses for Hanford (H), Livingston (L), and Virgo (V).

| Event | GW150914 | GW151012 | GW151226 | GW170104 | GW170608 | GW170729 | GW170809 | GW170814 | GW170817 | GW170818 | GW170823 |
|---|---|---|---|---|---|---|---|---|---|---|---|
| $D_{\rm KL}^{\chi_{\rm eff}}$ | $0.71^{+0.04}_{-0.03}$ | $0.23^{+0.03}_{-0.02}$ | $1.32^{+0.11}_{-0.06}$ | $0.54^{+0.03}_{-0.03}$ | $0.97^{+0.03}_{-0.05}$ | $1.83^{+0.07}_{-0.09}$ | $0.71^{+0.03}_{-0.03}$ | $0.99^{+0.05}_{-0.07}$ | $2.32^{+0.08}_{-0.10}$ | $0.50^{+0.04}_{-0.03}$ | $0.32^{+0.04}_{-0.03}$ |
| $D_{\rm KL}^{\chi_p}$ | $0.16^{+0.03}_{-0.02}$ | $0.09^{+0.03}_{-0.02}$ | $0.17^{+0.03}_{-0.04}$ | $0.05^{+0.01}_{-0.01}$ | $0.07^{+0.01}_{-0.02}$ | $0.09^{+0.02}_{-0.02}$ | $0.05^{+0.01}_{-0.01}$ | $0.02^{+0.01}_{-0.01}$ | $0.19^{+0.04}_{-0.03}$ | $0.06^{+0.02}_{-0.01}$ | $0.03^{+0.01}_{-0.01}$ |
| $D_{\rm KL}^{\chi_p}(\chi_{\rm eff})$ | $0.09^{+0.02}_{-0.02}$ | $0.08^{+0.02}_{-0.01}$ | $0.12^{+0.05}_{-0.02}$ | $0.07^{+0.02}_{-0.01}$ | $0.08^{+0.02}_{-0.02}$ | $0.03^{+0.01}_{-0.01}$ | $0.06^{+0.01}_{-0.01}$ | $0.13^{+0.03}_{-0.02}$ | $0.07^{+0.01}_{-0.02}$ | $0.09^{+0.02}_{-0.01}$ | $0.03^{+0.01}_{-0.01}$ |
| H SNR | $20.6^{+1.6}_{-1.6}$ | $6.4^{+1.3}_{-1.3}$ | $9.8^{+1.5}_{-1.4}$ | $9.5^{+1.3}_{-1.6}$ | $12.1^{+1.6}_{-1.6}$ | $5.9^{+1.1}_{-1.1}$ | $5.9^{+1.4}_{-1.4}$ | $9.3^{+1.0}_{-1.2}$ | $18.9^{+1.0}_{-1.0}$ | $4.6^{+0.9}_{-0.8}$ | $6.8^{+1.4}_{-1.2}$ |
| L SNR | $14.2^{+1.6}_{-1.4}$ | $5.8^{+1.2}_{-1.2}$ | $6.9^{+1.2}_{-1.1}$ | $9.9^{+1.5}_{-1.3}$ | $9.2^{+1.5}_{-1.2}$ | $8.3^{+1.4}_{-1.4}$ | $10.7^{+1.6}_{-1.8}$ | $14.3^{+1.5}_{-1.4}$ | $26.3^{+1.4}_{-1.3}$ | $9.7^{+1.5}_{-1.5}$ | $9.2^{+1.7}_{-1.5}$ |
| V SNR | … | … | … | … | … | $1.7^{+1.0}_{-1.1}$ | $1.1^{+1.2}_{-0.8}$ | $4.1^{+1.1}_{-1.1}$ | $3.0^{+0.2}_{-0.2}$ | $4.2^{+0.8}_{-0.7}$ | … |

P3: In addition to the volumetric spin prior, we also assume a uniform distribution in luminosity distance. This choice replaces assumptions (b)–(d) in P1.

Naturally, the priors for derived parameters, such as the effective aligned spin [23,121] given in Eq. (4) or the effective precession spin [125],

$$\chi_p = \frac{1}{B_1 m_1^2} \max\left(B_1 S_{1\perp}, B_2 S_{2\perp}\right) > 0, \tag{C1}$$

where $B_1 = 2 + 3q/2$ and $B_2 = 2 + 3/(2q)$, are coupled to the choice of priors for the mass ratio, spin magnitudes, and spin directions.

In current GW observations, small values of $\chi_{\rm eff}$ are preferred and $\chi_p$ is unconstrained, while spin measurements in x-ray binaries point to a range of spin magnitudes [263], including high spins. The volumetric spin prior P2 adds weight to higher spin values in comparison to the default spin prior, allowing us to test the robustness of the measurement, which can be important for understanding our inferences on the underlying binary population [55,262]. The additional assumption of a uniform in distance distribution in P3 may seem unnatural at first, but it provides a strong test on the robustness of the inference of the luminosity distance which is important for the cross-correlation with galaxy catalogs. Furthermore, it has computational advantages when low-significance events are considered in combination with nested sampling [264].

If the data are uninformative about a parameter, the choice of prior will play a dominant role in determining the shape of the posterior probability distribution. In the left column in Fig. 17, we show the different prior choices for a subset of physical parameters.

### 2. Comparison of posteriors under different prior assumptions

Here, we detail the *posterior* probability distributions obtained from a Bayesian parameter estimation on GW170809 with the different prior choices P1, P2, and P3. The results are obtained using the nested sampling algorithm implemented in LALINFERENCE [241] and the precessing waveform model IMRPhenomPv2 [25,26,49]. Marginalized one-dimensional PDFs for various parameters under the three different prior assumptions are shown in the right column in Fig. 17: The posterior PDFs for well-measured parameters have similar shapes irrespective of the assumed prior (e.g., the chirp mass), whereas they are very different, and hence prior dependent, for ill-measured parameters such as the effective precession spin.

TABLE VI. KL divergences (in bits) between the prior and posterior distribution for various parameters for GW170809 under the three different prior assumptions P1, P2, and P3. We note that the values for P1 are different from the ones in Table V, as we consider only the IMRPhenomPv2 model here.

| Prior | P1 | P2 | P3 |
|---|---|---|---|
| $D_{\rm KL}^{\chi_{\rm eff}}$ | $0.69^{+0.03}_{-0.04}$ | $1.04^{+0.04}_{-0.05}$ | $0.96^{+0.04}_{-0.03}$ |
| $D_{\rm KL}^{\chi_p}$ | $0.02^{+0.01}_{-0.01}$ | $0.03^{+0.01}_{-0.01}$ | $0.02^{+0.01}_{-0.01}$ |
| $D_{\rm KL}^{\chi_p}(\chi_{\rm eff})$ | $0.04^{+0.01}_{-0.01}$ | $0.05^{+0.01}_{-0.01}$ | $0.06^{+0.01}_{-0.01}$ |
| $D_{\rm KL}^{\mathcal{M}^{\rm det}}$ | $2.25^{+0.06}_{-0.06}$ | $2.07^{+0.07}_{-0.05}$ | $2.02^{+0.05}_{-0.07}$ |
| $D_{\rm KL}^{M_{\rm tot}^{\rm det}}$ | $2.20^{+0.06}_{-0.08}$ | $1.84^{+0.06}_{-0.05}$ | $1.77^{+0.05}_{-0.05}$ |
| $D_{\rm KL}^{q}$ | $0.73^{+0.03}_{-0.04}$ | $0.64^{+0.03}_{-0.03}$ | $0.54^{+0.03}_{-0.03}$ |
| $D_{\rm KL}^{d_L}$ | $4.59^{+0.12}_{-0.20}$ | $4.58^{+0.11}_{-0.17}$ | $2.04^{+0.07}_{-0.05}$ |

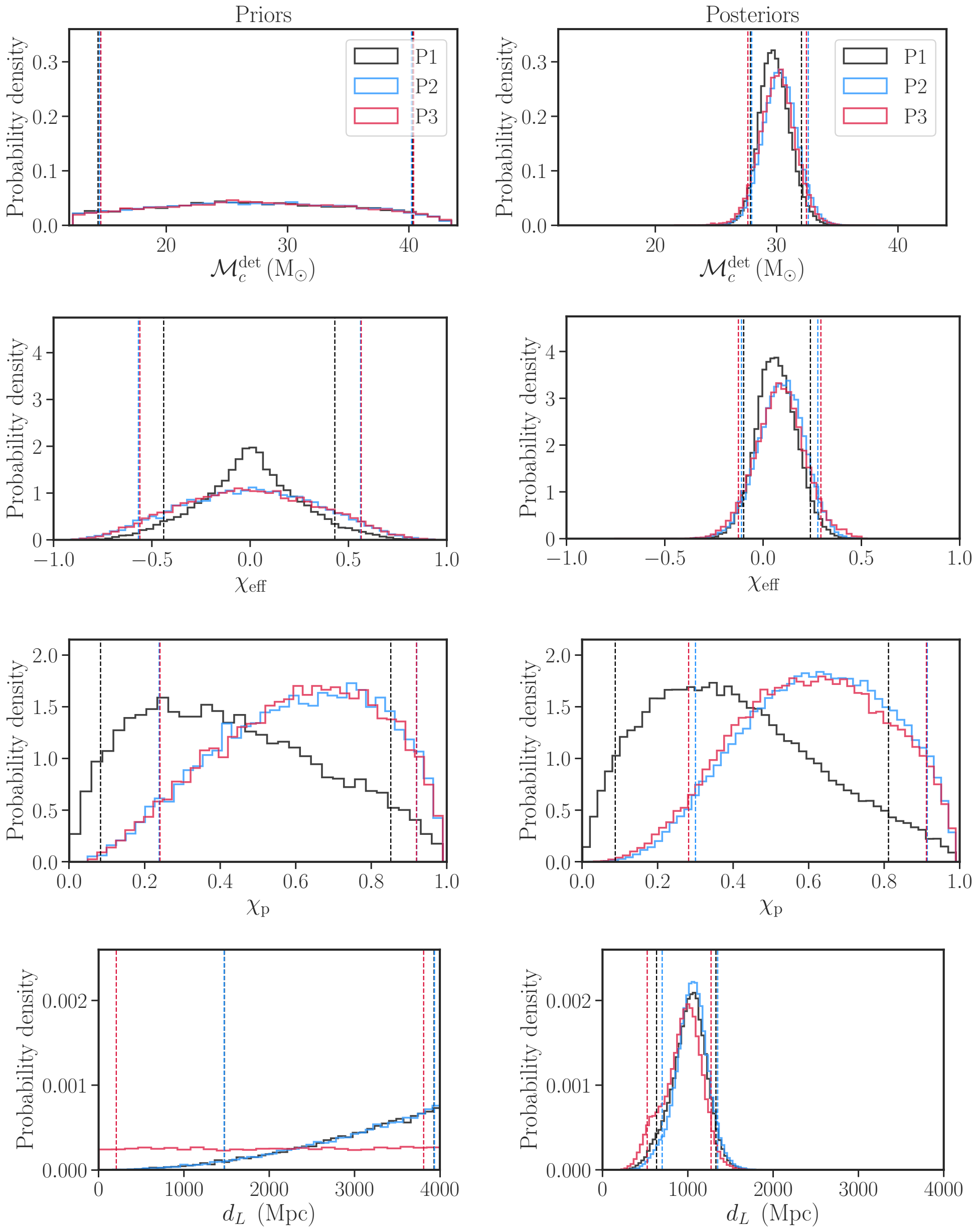

FIG. 17. Example prior and posterior distributions for GW170809. Left column: The four panels show the three different prior choices P1 (black), P2 (blue), and P3 (red) for four different physical parameters: the chirp mass, the effective aligned spin, the effective precession spin, and the luminosity distance. Right column: The four panels show the corresponding posterior probability distributions for the same four physical parameters obtained under the three different prior assumptions P1 (black), P2 (blue), and P3 (red). In all panels, the dashed vertical lines indicate the 90% credible intervals.

To quantify the impact of the choice of prior on parameter estimation and hence our observations from Fig. 17, we use the Kullback-Leibler (KL) divergence $D_{\rm KL}$ [159] as defined in Eq. (B2). This divergence allows us to determine the information gain between the prior and the posterior distributions. The results are summarized in Table VI. A similar spread on parameters, where applicable, is reported in Ref. [261], where $D_{\mathrm{KL}}^{\chi_{\mathrm{eff}}} \sim \mathcal{O}(1)$ and $D_{\mathrm{KL}}^{\chi_{\mathrm{p}}} \sim \mathcal{O}(10^{-2})$. As expected, parameters that have a dominant impact on the binary phasing, for example, the chirp mass $\mathcal{M}$ and the effective aligned spin $\chi_{\mathrm{eff}}$, are well measured and robust against different prior choices. Other parameters such as the effective precession spin $\chi_p$, however, are relatively poorly constrained in current observations, and the KL divergence approaches zero, implying that we predominantly recover the priors.

[1] B. P. Abbott *et al.* (LIGO Scientific Collaboration, Virgo Collaboration), *Observation of Gravitational Waves from a Binary Black Hole Merger*, Phys. Rev. Lett. **116**, 061102 (2016).

[2] B. P. Abbott *et al.* (LIGO Scientific Collaboration, Virgo Collaboration), *GW151226: Observation of Gravitational Waves from a 22-Solar-Mass Binary Black Hole Coalescence*, Phys. Rev. Lett. **116**, 241103 (2016).

[3] B. P. Abbott *et al.* (LIGO Scientific Collaboration, Virgo Collaboration), *GW150914: First Results from the Search for Binary Black Hole Coalescence with Advanced LIGO*, Phys. Rev. D **93**, 122003 (2016).

[4] B. P. Abbott *et al.* (LIGO Scientific Collaboration, Virgo Collaboration), *Binary Black Hole Mergers in the First Advanced LIGO Observing Run*, Phys. Rev. X **6**, 041015 (2016).

[5] J. Aasi *et al.* (LIGO Scientific Collaboration), *Advanced LIGO*, Classical Quantum Gravity **32**, 074001 (2015).

[6] F. Acernese *et al.* (Virgo Collaboration), *Advanced Virgo: A Second-Generation Interferometric Gravitational Wave Detector*, Classical Quantum Gravity **32**, 024001 (2015).

[7] A. H. Nitz, I. W. Harry, J. L. Willis, C. M. Biwer, D. A. Brown, L. P. Pekowsky, T. Dal Canton, A. R. Williamson, T. Dent, C. D. Capano, T. J. Massinger, A. K. Lenon, A. B. Nielsen, and M. Cabero, PyCBC Software, http://github.com/ligo-cbc/pycbc.

[8] S. A. Usman *et al.*, *The PyCBC Search for Gravitational Waves from Compact Binary Coalescence*, Classical Quantum Gravity **33**, 215004 (2016).

[9] S. Sachdev *et al.*, *The GstLAL Search Analysis Methods for Compact Binary Mergers in Advanced LIGO's Second and Advanced Virgo's First Observing Runs*, arXiv:1901.08580.

[10] C. Messick *et al.*, *Analysis Framework for the Prompt Discovery of Compact Binary Mergers in Gravitational-Wave Data*, Phys. Rev. D **95**, 042001 (2017).

[11] S. Klimenko *et al.*, *Method for Detection and Reconstruction of Gravitational Wave Transients with Networks of Advanced Detectors*, Phys. Rev. D **93**, 042004 (2016).

[12] A. Krolak and B. F. Schutz, *Coalescing Binaries—Probe of the Universe*, Gen. Relativ. Gravit. **19**, 1163 (1987).

[13] B. P. Abbott *et al.* (Virgo, LIGO Scientific Collaboration), *Search for Sub-Solar Mass Ultracompact Binaries in Advanced LIGO's First Observing Run*, arXiv:1808.04771.

[14] The event GW151012 was previously referred to as LVT151012. Here, we retire the LVT nomenclature; all candidate events with an estimated FAR of less than 1 per 30 days *and* a probability of > 0.5 of being of astrophysical origin [see Eq. (10) for the definition] are henceforth denoted with the GW prefix. All other candidates are referred to as marginal.

[15] B. P. Abbott *et al.* (VIRGO, LIGO Scientific Collaboration), *GW170104: Observation of a 50-Solar-Mass Binary Black Hole Coalescence at Redshift 0.2*, Phys. Rev. Lett. **118**, 221101 (2017).

[16] B. P. Abbott *et al.* (Virgo, LIGO Scientific Collaboration), *GW170814: A Three-Detector Observation of Gravitational Waves from a Binary Black Hole Coalescence*, Phys. Rev. Lett. **119**, 141101 (2017).

[17] B. P. Abbott *et al.* (Virgo, LIGO Scientific Collaboration), *GW170608: Observation of a 19-Solar-Mass Binary Black Hole Coalescence*, Astrophys. J. **851**, L35 (2017).

[18] B. P. Abbott *et al.* (Virgo, LIGO Scientific Collaboration), *GW170817: Observation of Gravitational Waves from a Binary Neutron Star Inspiral*, Phys. Rev. Lett. **119**, 161101 (2017).

[19] B. F. Schutz, *Networks of Gravitational Wave Detectors and Three Figures of Merit*, Classical Quantum Gravity **28**, 125023 (2011).

[20] S. Fairhurst, *Source Localization with an Advanced Gravitational Wave Detector Network*, Classical Quantum Gravity **28**, 105021 (2011).

[21] B. P. Abbott *et al.* [GROND, SALT Group, OzGrav, DFN, INTEGRAL, Virgo, Insight-Hxmt, MAXI Team, Fermi-LAT, J-GEM, RATIR, IceCube, CAASTRO, LWA, ePESSTO, GRAWITA, RIMAS, SKA South Africa/MeerKAT, H.E.S.S., 1M2H Team, IKI-GW Follow-up, Fermi GBM, Pi of Sky, DWF (Deeper Wider Faster Program), Dark Energy Survey, MASTER, AstroSat Cadmium Zinc Telluride Imager Team, Swift, Pierre Auger, ASKAP, VINROUGE, JAGWAR, Chandra Team at McGill University, TTU-NRAO, GROWTH, AGILE Team, MWA, ATCA, AST3, TOROS, Pan-STARRS, NuSTAR, ATLAS Telescopes, BOOTES, CaltechNRAO, LIGO Scientific, High Time Resolution Universe Survey, Nordic Optical Telescope, Las Cumbres Observatory Group, TZAC Consortium, LOFAR, IPN, DLT40, Texas Tech University, HAWC, ANTARES, KU, Dark Energy Camera GW-EM, CALET, Euro VLBI Team, ALMA Collaboration], *Multi-Messenger Observations of a Binary Neutron Star Merger*, Astrophys. J. **848**, L12 (2017).

[22] B. P. Abbott *et al.* (LIGO Scientific, Virgo Collaboration), *Low-Latency Gravitational Wave Alerts for Multi-Messenger Astronomy during the Second Advanced LIGO and Virgo Observing Run*, Astrophys. J. **875**, 161 (2019).

[23] P. Ajith *et al.*, *Inspiral-Merger-Ringdown Waveforms for Black-Hole Binaries with Non-Precessing Spins*, Phys. Rev. Lett. **106**, 241101 (2011).

[24] L. Santamaría *et al.*, *Matching Post-Newtonian and Numerical Relativity Waveforms: Systematic Errors and a New Phenomenological Model for Non-Precessing Black Hole Binaries*, Phys. Rev. D **82**, 064016 (2010).

[25] M. Hannam, P. Schmidt, A. Bohé, L. Haegel, S. Husa, F. Ohme, G. Pratten, and M. Pürrer, *Simple Model of Complete Precessing Black-Hole-Binary Gravitational Waveforms*, Phys. Rev. Lett. **113**, 151101 (2014).

[26] S. Khan, S. Husa, M. Hannam, F. Ohme, M. Pürrer, X. J. Forteza, and A. Bohé, *Frequency-Domain Gravitational Waves from Nonprecessing Black-Hole Binaries. II. A Phenomenological Model for the Advanced Detector Era*, Phys. Rev. D **93**, 044007 (2016).
[27] Y. Pan, A. Buonanno, A. Taracchini, L. E. Kidder, A. H. Mroué, H. P. Pfeiffer, M. A. Scheel, and B. Szilágyi, *Inspiral-Merger-Ringdown Waveforms of Spinning, Precessing Black-Hole Binaries in the Effective-One-Body Formalism*, Phys. Rev. D **89**, 084006 (2014).
[28] A. Taracchini *et al.*, *Effective-One-Body Model for Black-Hole Binaries with Generic Mass Ratios and Spins*, Phys. Rev. D **89**, 061502 (2014).
[29] A. Bohé *et al.*, *An Improved Effective-One-Body Model of Spinning, Nonprecessing Binary Black Holes for the Era of Gravitational-Wave Astrophysics with Advanced Detectors*, Phys. Rev. D **95**, 044028 (2017).
[30] S. Babak, A. Taracchini, and A. Buonanno, *Validating the Effective-One-Body Model of Spinning, Precessing Binary Black Holes against Numerical Relativity*, Phys. Rev. D **95**, 024010 (2017).
[31] T. Hinderer *et al.*, *Effects of Neutron-Star Dynamic Tides on Gravitational Waveforms within the Effective-One-Body Approach*, Phys. Rev. Lett. **116**, 181101 (2016).
[32] T. Dietrich *et al.*, *Matter Imprints in Waveform Models for Neutron Star Binaries: Tidal and Self-Spin Effects*, arXiv:1804.02235.
[33] A. Nagar *et al.*, *Time-Domain Effective-One-Body Gravitational Waveforms for Coalescing Compact Binaries with Nonprecessing Spins, Tides and Self-Spin Effects*, Phys. Rev. D **98**, 104052 (2018).
[34] L. Blanchet, B. R. Iyer, C. M. Will, and A. G. Wiseman, *Gravitational Wave Forms from Inspiralling Compact Binaries to Second Post-Newtonian Order*, Classical Quantum Gravity **13**, 575 (1996).
[35] T. Damour, P. Jaranowski, and G. Schaefer, *Dimensional Regularization of the Gravitational Interaction of Point Masses*, Phys. Lett. B **513**, 147 (2001).
[36] L. Blanchet, T. Damour, G. Esposito-Farese, and B. R. Iyer, *Dimensional Regularization of the Third Post-Newtonian Gravitational Wave Generation from Two Point Masses*, Phys. Rev. D **71**, 124004 (2005).
[37] A. Buonanno, B. Iyer, E. Ochsner, Y. Pan, and B. S. Sathyaprakash, *Comparison of Post-Newtonian Templates for Compact Binary Inspiral Signals in Gravitational-Wave Detectors*, Phys. Rev. D **80**, 084043 (2009).
[38] L. Blanchet, *Gravitational Radiation from Post-Newtonian Sources and Inspiralling Compact Binaries*, Living Rev. Relativity **17**, 2 (2014).
[39] A. Buonanno and T. Damour, *Effective One-Body Approach to General Relativistic Two-Body Dynamics*, Phys. Rev. D **59**, 084006 (1999).
[40] A. Buonanno and T. Damour, *Transition from Inspiral to Plunge in Binary Black Hole Coalescences*, Phys. Rev. D **62**, 064015 (2000).
[41] T. Damour, P. Jaranowski, and G. Schaefer, *Effective One Body Approach to the Dynamics of Two Spinning Black Holes with Next-to-Leading Order Spin-Orbit Coupling*, Phys. Rev. D **78**, 024009 (2008).
[42] T. Damour and A. Nagar, *An Improved Analytical Description of Inspiralling and Coalescing Black-Hole Binaries*, Phys. Rev. D **79**, 081503 (2009).
[43] E. Barausse and A. Buonanno, *An Improved Effective-One-Body Hamiltonian for Spinning Black-Hole Binaries*, Phys. Rev. D **81**, 084024 (2010).
[44] T. Damour and A. Nagar, *The Effective-One-Body Approach to the General Relativistic Two Body Problem*, Lect. Notes Phys. **905**, 273 (2016).
[45] F. Pretorius, *Evolution of Binary Black Hole Spacetimes*, Phys. Rev. Lett. **95**, 121101 (2005).
[46] M. Campanelli, C. O. Lousto, P. Marronetti, and Y. Zlochower, *Accurate Evolutions of Orbiting Black-Hole Binaries without Excision*, Phys. Rev. Lett. **96**, 111101 (2006).
[47] J. G. Baker, J. Centrella, D.-I. Choi, M. Koppitz, and J. van Meter, *Gravitational Wave Extraction from an Inspiraling Configuration of Merging Black Holes*, Phys. Rev. Lett. **96**, 111102 (2006).
[48] A. H. Mroue *et al.*, *Catalog of 174 Binary Black Hole Simulations for Gravitational Wave Astronomy*, Phys. Rev. Lett. **111**, 241104 (2013).
[49] S. Husa, S. Khan, M. Hannam, M. Pürrer, F. Ohme, X. J. Forteza, and A. Bohé, *Frequency-Domain Gravitational Waves from Nonprecessing Black-Hole Binaries. I. New Numerical Waveforms and Anatomy of the Signal*, Phys. Rev. D **93**, 044006 (2016).
[50] T. Chu, H. Fong, P. Kumar, H. P. Pfeiffer, M. Boyle, D. A. Hemberger, L. E. Kidder, M. A. Scheel, and B. Szilagyi, *On the Accuracy and Precision of Numerical Waveforms: Effect of Waveform Extraction Methodology*, Classical Quantum Gravity **33**, 165001 (2016).
[51] D. Davis, T. J. Massinger, A. P. Lundgren, J. C. Driggers, A. L. Urban, and L. K. Nuttall, *Improving the Sensitivity of Advanced LIGO Using Noise Subtraction*, Classical Quantum Gravity **36**, 055011 (2019).
[52] J. C. Driggers *et al.* (LIGO Scientific Collaboration), *Improving Astrophysical Parameter Estimation via Offline Noise Subtraction for Advanced LIGO*, Phys. Rev. D **99**, 042001 (2019).
[53] N. J. Cornish and T. B. Littenberg, *BayesWave: Bayesian Inference for Gravitational Wave Bursts and Instrument Glitches*, Classical Quantum Gravity **32**, 135012 (2015).
[54] T. B. Littenberg and N. J. Cornish, *Bayesian Inference for Spectral Estimation of Gravitational Wave Detector Noise*, Phys. Rev. D **91**, 084034 (2015).
[55] B. P. Abbott *et al.* (LIGO Scientific Collaboration, Virgo Collaboration), *Binary Black Hole Population Properties Inferred from the First and Second Observing Runs of Advanced LIGO and Advanced Virgo*, arXiv:1811.12940.
[56] LIGO Scientific Collaboration, Virgo Collaboration, GWTC-1, https://doi.org/10.7935/82H3-HH23.
[57] LIGO Scientific Collaboration, Virgo Collaboration, Gravitational Wave Open Science Center, https://www.gw-openscience.org.
[58] B. P. Abbott *et al.* (Virgo, LIGO Scientific Collaboration), *GW150914: The Advanced LIGO Detectors in the Era of First Discoveries*, Phys. Rev. Lett. **116**, 131103 (2016).

[59] B. P. Abbott *et al.*, *Sensitivity of the Advanced LIGO Detectors at the Beginning of Gravitational Wave Astronomy*, Phys. Rev. D **93**, 112004 (2016); **97**, 059901(A) (2018).
[60] M. Walker, A. F. Agnew, J. Bidler, A. Lundgren, A. Macedo, D. Macleod, T. J. Massinger, O. Patane, and J. R. Smith, *Identifying Correlations between LIGO's Astronomical Range and Auxiliary Sensors Using Lasso Regression*, Classical Quantum Gravity **35**, 225002 (2018).
[61] Interferometer performance is commonly reduced to a single scalar number that quantifies the distance to which a single instrument could detect a $1.4\,M_\odot + 1.4\,M_\odot$ BNS merger, averaged over the sky location and orientation, with an average SNR of 8 [62,63].
[62] H.-Y. Chen, D. E. Holz, J. Miller, M. Evans, S. Vitale, and J. Creighton, *Distance Measures in Gravitational-Wave Astrophysics and Cosmology*, arXiv:1709.08079.
[63] L. S. Finn and D. F. Chernoff, *Observing Binary Inspiral in Gravitational Radiation: One Interferometer*, Phys. Rev. D **47**, 2198 (1993).
[64] C. Cahillane *et al.* (LIGO Scientific Collaboration), *Calibration Uncertainty for Advanced LIGO's First and Second Observing Runs*, Phys. Rev. D **96**, 102001 (2017).
[65] A. Viets *et al.*, *Reconstructing the Calibrated Strain Signal in the Advanced LIGO Detectors*, Classical Quantum Gravity **35**, 095015 (2018).
[66] I. Bartos, R. Bork, M. Factourovich, J. Heefner, S. Marka, Z. Marka, Z. Raics, P. Schwinberg, and D. Sigg, *The Advanced LIGO Timing System*, Classical Quantum Gravity **27**, 084025 (2010).
[67] F. Acernese *et al.* (Virgo Collaboration), *Calibration of Advanced Virgo and Reconstruction of the Gravitational Wave Signal $h(t)$ during the Observing Run O2*, Classical Quantum Gravity **35**, 205004 (2018).
[68] B. P. Abbott *et al.* (Virgo, LIGO Scientific Collaboration), *Effects of Data Quality Vetoes on a Search for Compact Binary Coalescences in Advanced LIGO's First Observing Run*, Classical Quantum Gravity **35**, 065010 (2018).
[69] T. Dal Canton, A. P. Lundgren, and A. B. Nielsen, *Impact of Precession on Aligned-Spin Searches for Neutron-Star–Black-Hole Binaries*, Phys. Rev. D **91**, 062010 (2015).
[70] B. Allen, W. G. Anderson, P. R. Brady, D. A. Brown, and J. D. E. Creighton, *FINDCHIRP: An Algorithm for Detection of Gravitational Waves from Inspiraling Compact Binaries*, Phys. Rev. D **85**, 122006 (2012).
[71] B. Allen, *$\chi^2$ Time-Frequency Discriminator for Gravitational Wave Detection*, Phys. Rev. D **71**, 062001 (2005).
[72] A. H. Nitz, *Distinguishing Short Duration Noise Transients in LIGO Data to Improve the PyCBC Search for Gravitational Waves from High Mass Binary Black Hole Mergers*, Classical Quantum Gravity **35**, 035016 (2018).
[73] A. H. Nitz, T. Dent, T. Dal Canton, S. Fairhurst, and D. A. Brown, *Detecting Binary Compact-Object Mergers with Gravitational Waves: Understanding and Improving the Sensitivity of the PyCBC Search*, Astrophys. J. **849**, 118 (2017).
[74] This statement is true of almost all analysis periods, but the last analysis period in $O2$ has a slightly shorter length of approximately 3.6 days.
[75] T. Dal Canton and I. Harry, *Designing a Template Bank to Observe Compact Binary Coalescences in Advanced LIGO's Second Observing Run*, arXiv:1705.01845.
[76] M. Pürrer, *Frequency Domain Reduced Order Model of Aligned-Spin Effective-One-Body Waveforms with Generic Mass-Ratios and Spins*, Phys. Rev. D **93**, 064041 (2016).
[77] M. Pürrer, *Frequency Domain Reduced Order Models for Gravitational Waves from Aligned-Spin Compact Binaries*, Classical Quantum Gravity **31**, 195010 (2014).
[78] In Ref. [79], this model is called SEOBNRv4_ROM.
[79] LIGO Scientific Collaboration, Virgo Collaboration, LALSuite, https://git.ligo.org/lscsoft/lalsuite.
[80] G. Faye, S. Marsat, L. Blanchet, and B. R. Iyer, *The Third and a Half Post-Newtonian Gravitational Wave Quadrupole Mode for Quasi-Circular Inspiralling Compact Binaries*, Classical Quantum Gravity **29**, 175004 (2012).
[81] As a reminder, GstLAL analyzes data from three interferometers, so the light travel time depends on the specific interferometers involved in any particular candidate.
[82] K. Cannon, C. Hanna, and J. Peoples, *Likelihood-Ratio Ranking Statistic for Compact Binary Coalescence Candidates with Rate Estimation*, arXiv:1504.04632.
[83] C. Hanna *et al.*, *Fast Evaluation of Multi-Detector Consistency for Real-Time Gravitational Wave Searches*, arXiv:1901.02227.
[84] D. Mukherjee *et al.*, *The GstLAL Template Bank for Spinning Compact Binary Mergers in the Second Observation Run of Advanced LIGO and Virgo*, arXiv:1812.05121.
[85] S. Klimenko, I. Yakushin, A. Mercer, and G. Mitselmakher, *Coherent Method for Detection of Gravitational Wave Bursts*, Classical Quantum Gravity **25**, 114029 (2008).
[86] B. P. Abbott *et al.* (LIGO Scientific Collaboration, Virgo Collaboration), *The Rate of Binary Black Hole Mergers Inferred from Advanced LIGO Observations Surrounding GW150914*, Astrophys. J. Lett. **833**, L1 (2016).
[87] W. M. Farr and J. R. Gair, *The Sirens of August: Detection of Five Gravitational Wave Events in August 2017 Is Consistent with a Constant Rate, LIGO Scientific Collaboration and Virgo Collaboration* Technical Report No. LIGO-T1800529, https://dcc.ligo.org/LIGO-T1800529/public.
[88] A. H. Nitz, C. Capano, A. B. Nielsen, S. Reyes, R. White, D. A. Brown, and B. Krishnan, *1-OGC: The First Open Gravitational-Wave Catalog of Binary Mergers from Analysis of Public Advanced LIGO Data*, arXiv:1811.01921.
[89] S. J. Kapadia *et al.*, *A Self-Consistent Method to Estimate the Rate of Compact Binary Coalescences with a Poisson Mixture Model*, arXiv:1903.06881.
[90] W. M. Farr, J. R. Gair, I. Mandel, and C. Cutler, *Counting and Confusion: Bayesian Rate Estimation with Multiple Populations*, Phys. Rev. D **91**, 023005 (2015).
[91] B. P. Abbott *et al.* (LIGO Scientific Collaboration, Virgo Collaboration), *Characterization of Transient Noise in Advanced LIGO Relevant to Gravitational Wave Signal GW150914*, Classical Quantum Gravity **33**, 134001 (2016).

[92] L. K. Nuttall, *Characterizing Transient Noise in the LIGO Detectors*, Phil. Trans. R. Soc. A **376**, 20170286 (2018).

[93] J. McIver, *Data Quality Studies of Enhanced Interferometric Gravitational Wave Detectors*, Classical Quantum Gravity **29**, 124010 (2012).

[94] A. J. Levan *et al.*, *The Environment of the Binary Neutron Star Merger GW170817*, Astrophys. J. **848**, L28 (2017).

[95] E. Oelker, T. Isogai, J. Miller, M. Tse, L. Barsotti, N. Mavalvala, and M. Evans, *Properties of the Binary Black Hole Merger GW150914*, Phys. Rev. Lett. **116**, 041102 (2016).

[96] B. P. Abbott *et al.* (Virgo, LIGO Scientific Collaboration), *Improved Analysis of GW150914 Using a Fully Spin-Precessing Waveform Model*, Phys. Rev. X **6**, 041014 (2016).

[97] B. P. Abbott *et al.* (LIGO Scientific, Virgo Collaboration), *Properties of the Binary Neutron Star Merger GW170817*, Phys. Rev. X **9**, 011001 (2019).

[98] W. M. Farr, B. Farr, and T. Littenberg, *Modelling Calibration Errors in CBC Waveforms*, Collaboration Technical Report No. LIGO-T1400682, https://dcc.ligo.org/LIGO-T1400682/public.

[99] T. Dietrich, S. Bernuzzi, and W. Tichy, *Closed-Form Tidal Approximants for Binary Neutron Star Gravitational Waveforms Constructed from High-Resolution Numerical Relativity Simulations*, Phys. Rev. D **96**, 121501 (2017).

[100] B. S. Sathyaprakash and S. V. Dhurandhar, *Choice of Filters for the Detection of Gravitational Waves from Coalescing Binaries*, Phys. Rev. D **44**, 3819 (1991).

[101] L. Blanchet, T. Damour, B. R. Iyer, C. M. Will, and A. G. Wiseman, *Gravitational Radiation Damping of Compact Binary Systems to Second Post-Newtonian Order*, Phys. Rev. Lett. **74**, 3515 (1995).

[102] L. Blanchet, T. Damour, G. Esposito-Farèse, and B. R. Iyer, *Gravitational Radiation from Inspiralling Compact Binaries Completed at the Third Post-Newtonian Order*, Phys. Rev. Lett. **93**, 091101 (2004).

[103] W. D. Goldberger and I. Z. Rothstein, *An Effective Field Theory of Gravity for Extended Objects*, Phys. Rev. D **73**, 104029 (2006).

[104] T. Damour and A. Nagar, *Effective One Body Description of Tidal Effects in Inspiralling Compact Binaries*, Phys. Rev. D **81**, 084016 (2010).

[105] J. Vines, É. É. Flanagan, and T. Hinderer, *Post-1-Newtonian Tidal Effects in the Gravitational Waveform from Binary Inspirals*, Phys. Rev. D **83**, 084051 (2011).

[106] D. Bini, T. Damour, and G. Faye, *Effective Action Approach to Higher-Order Relativistic Tidal Interactions in Binary Systems and Their Effective One Body Description*, Phys. Rev. D **85**, 124034 (2012).

[107] T. Damour, A. Nagar, and L. Villain, *Measurability of the Tidal Polarizability of Neutron Stars in Late-Inspiral Gravitational-Wave Signals*, Phys. Rev. D **85**, 123007 (2012).

[108] A. Bohé, S. Marsat, and L. Blanchet, *Next-to-Next-to-Leading Order Spin-Orbit Effects in the Gravitational Wave Flux and Orbital Phasing of Compact Binaries*, Classical Quantum Gravity **30**, 135009 (2013).

[109] K. G. Arun, A. Buonanno, G. Faye, and E. Ochsner, *Higher-Order Spin Effects in the Amplitude and Phase of Gravitational Waveforms Emitted by Inspiraling Compact Binaries: Ready-to-Use Gravitational Waveforms*, Phys. Rev. D **79**, 104023 (2009); *Erratum*, Phys. Rev. D **84**, 049901(E) (2011).

[110] B. Mikoczi, M. Vasuth, and L. A. Gergely, *Self-Interaction Spin Effects in Inspiralling Compact Binaries*, Phys. Rev. D **71**, 124043 (2005).

[111] A. Bohé, G. Faye, S. Marsat, and E. K. Porter, *Quadratic-in-Spin Effects in the Orbital Dynamics and Gravitational-Wave Energy Flux of Compact Binaries at the 3PN Order*, Classical Quantum Gravity **32**, 195010 (2015).

[112] C. K. Mishra, A. Kela, K. G. Arun, and G. Faye, *Ready-to-Use Post-Newtonian Gravitational Waveforms for Binary Black Holes with Nonprecessing Spins: An Update*, Phys. Rev. D **93**, 084054 (2016).

[113] S. Bernuzzi, A. Nagar, T. Dietrich, and T. Damour, *Modeling the Dynamics of Tidally Interacting Binary Neutron Stars up to the Merger*, Phys. Rev. Lett. **114**, 161103 (2015).

[114] W. Israel, *Event Horizons in Static Vacuum Space-Times*, Phys. Rev. **164**, 1776 (1967).

[115] B. Carter, *Axisymmetric Black Hole Has Only Two Degrees of Freedom*, Phys. Rev. Lett. **26**, 331 (1971).

[116] D. C. Robinson, *Uniqueness of the Kerr Black Hole*, Phys. Rev. Lett. **34**, 905 (1975).

[117] R. M. Wald, *General Relativity* (Chicago University, Chicago, 1984).

[118] S. Chandrasekhar, *The Mathematical Theory of Black Holes*, Oxford Classic Texts in the Physical Sciences (Oxford University Press, New York, 1992).

[119] T. A. Apostolatos, C. Cutler, G. J. Sussman, and K. S. Thorne, *Spin Induced Orbital Precession and Its Modulation of the Gravitational Wave Forms from Merging Binaries*, Phys. Rev. D **49**, 6274 (1994).

[120] L. E. Kidder, *Coalescing Binary Systems of Compact Objects to Post-Newtonian* 5/2 *Order. 5. Spin Effects*, Phys. Rev. D **52**, 821 (1995).

[121] É. Racine, *Analysis of Spin Precession in Binary Black Hole Systems Including Quadrupole-Monopole Interaction*, Phys. Rev. D **78**, 044021 (2008).

[122] E. Poisson and C. M. Will, *Gravitational Waves from Inspiraling Compact Binaries: Parameter Estimation Using Second Post-Newtonian Waveforms*, Phys. Rev. D **52**, 848 (1995).

[123] T. Damour, *Coalescence of Two Spinning Black Holes: An Effective One-Body Approach*, Phys. Rev. D **64**, 124013 (2001).

[124] P. Ajith, *Addressing the Spin Question in Gravitational-Wave Searches: Waveform Templates for Inspiralling Compact Binaries with Nonprecessing Spins*, Phys. Rev. D **84**, 084037 (2011).

[125] P. Schmidt, F. Ohme, and M. Hannam, *Towards Models of Gravitational Waveforms from Generic Binaries II: Modelling Precession Effects with a Single Effective Precession Parameter*, Phys. Rev. D **91**, 024043 (2015).

[126] P. C. Peters and J. Mathews, *Gravitational Radiation from Point Masses in a Keplerian Orbit*, Phys. Rev. **131**, 435 (1963).

[127] P. C. Peters, *Gravitational Radiation and the Motion of Two Point Masses*, Phys. Rev. **136**, B1224 (1964).

[128] C. Cutler and É. E. Flanagan, *Gravitational Waves from Merging Compact Binaries: How Accurately Can One Extract the Binary's Parameters from the Inspiral Wave Form?*, Phys. Rev. D **49**, 2658 (1994).
[129] B. Farr, E. Ochsner, W. M. Farr, and R. O'Shaughnessy, *A More Effective Coordinate System for Parameter Estimation of Precessing Compact Binaries from Gravitational Waves*, Phys. Rev. D **90**, 024018 (2014).
[130] M. Favata, *Systematic Parameter Errors in Inspiraling Neutron Star Binaries*, Phys. Rev. Lett. **112**, 101101 (2014).
[131] L. Wade, J. D. E. Creighton, E. Ochsner, B. D. Lackey, B. F. Farr, T. B. Littenberg, and V. Raymond, *Systematic and Statistical Errors in a Bayesian Approach to the Estimation of the Neutron-Star Equation of State Using Advanced Gravitational Wave Detectors*, Phys. Rev. D **89**, 103012 (2014).
[132] K. Belczynski, T. Bulik, C. L. Fryer, A. Ruiter, F. Valsecchi, J. S. Vink, and J. R. Hurley, *On the Maximum Mass of Stellar Black Holes*, Astrophys. J. **714**, 1217 (2010).
[133] M. Mapelli, L. Zampieri, E. Ripamonti, and A. Bressan, *Dynamics of Stellar Black Holes in Young Star Clusters with Different Metallicities—I. Implications for X-Ray Binaries*, Mon. Not. R. Astron. Soc. **429**, 2298 (2013).
[134] M. Spera, M. Mapelli, and A. Bressan, *The Mass Spectrum of Compact Remnants from the PARSEC Stellar Evolution Tracks*, Mon. Not. R. Astron. Soc. **451**, 4086 (2015).
[135] M. Spera and M. Mapelli, *Very Massive Stars, Pair-Instability Supernovae and Intermediate-Mass Black Holes with the Sevn Code*, Mon. Not. R. Astron. Soc. **470**, 4739 (2017).
[136] S. E. Woosley, *Pulsational Pair-Instability Supernovae*, Astrophys. J. **836**, 244 (2017).
[137] N. Giacobbo, M. Mapelli, and M. Spera, *Merging Black Hole Binaries: The Effects of Progenitor's Metallicity, Mass-Loss Rate and Eddington Factor*, Mon. Not. R. Astron. Soc. **474**, 2959 (2018).
[138] P. Marchant, N. Langer, P. Podsiadlowski, T. M. Tauris, and T. J. Moriya, *A New Route towards Merging Massive Black Holes*, Astron. Astrophys. **588**, A50 (2016).
[139] F. Özel, D. Psaltis, R. Narayan, and J. E. McClintock, *The Black Hole Mass Distribution in the Galaxy*, Astrophys. J. **725**, 1918 (2010).
[140] W. M. Farr, N. Sravan, A. Cantrell, L. Kreidberg, C. D. Bailyn, I. Mandel, and V. Kalogera, *The Mass Distribution of Stellar-Mass Black Holes*, Astrophys. J. **741**, 103 (2011).
[141] L. Kreidberg, C D. Bailyn, W. M. Farr, and V. Kalogera, *Mass Measurements of Black Holes in X-Ray Transients: Is There a Mass Gap?*, Astrophys. J. **757**, 36 (2012).
[142] J. Antoniadis *et al.*, *A Massive Pulsar in a Compact Relativistic Binary*, Science **340**, 1233232 (2013).
[143] J. Veitch, M. Pürrer, and I. Mandel, *Measuring Intermediate Mass Black Hole Binaries with Advanced Gravitational Wave Detectors*, Phys. Rev. Lett. **115**, 141101 (2015).
[144] P. B. Graff, A. Buonanno, and B. S. Sathyaprakash, *Missing Link: Bayesian Detection and Measurement of Intermediate-Mass Black-Hole Binaries*, Phys. Rev. D **92**, 022002 (2015).
[145] C.-J. Haster, Z. Wang, C. P. L. Berry, S. Stevenson, J. Veitch, and I. Mandel, *Inference on Gravitational Waves from Coalescences of Stellar-Mass Compact Objects and Intermediate-Mass Black Holes*, Mon. Not. R. Astron. Soc. **457**, 4499 (2016).
[146] A. Ghosh, W. Del Pozzo, and P. Ajith, *Estimating Parameters of Binary Black Holes from Gravitational-Wave Observations of Their Inspiral, Merger and Ringdown*, Phys. Rev. D **94**, 104070 (2016).
[147] M. Pürrer, M. Hannam, and F. Ohme, *Can We Measure Individual Black-Hole Spins from Gravitational-Wave Observations?*, Phys. Rev. D **93**, 084042 (2016).
[148] The fits for the final mass and spin can be different from the fits used internally in the waveform models used in the analyses.
[149] F. Hofmann, E. Barausse, and L. Rezzolla, *The Final Spin from Binary Black Holes in Quasi-Circular Orbits*, Astrophys. J. Lett. **825**, L19 (2016).
[150] X. Jiménez-Forteza, D. Keitel, S. Husa, M. Hannam, S. Khan, and M. Pürrer, *Hierarchical Data-Driven Approach to Fitting Numerical Relativity Data for Nonprecessing Binary Black Holes with an Application to Final Spin and Radiated Energy*, Phys. Rev. D **95**, 064024 (2017).
[151] J. Healy and C. O. Lousto, *Remnant of Binary Black-Hole Mergers: New Simulations and Peak Luminosity Studies*, Phys. Rev. D **95**, 024037 (2017).
[152] N. K. Johnson-McDaniel *et al.*, *Determining the Final Spin of a Binary Black Hole System Including In-Plane Spins: Method and Checks of Accuracy*, LIGO Scientific Collaboration and Virgo Collaboration Technical Report No. LIGO-T1600168, https://dcc.ligo.org/LIGO-T1600168/public.
[153] D. Keitel *et al.*, *The Most Powerful Astrophysical Events: Gravitational-Wave Peak Luminosity of Binary Black Holes as Predicted by Numerical Relativity*, Phys. Rev. D **96**, 024006 (2017).
[154] M. Campanelli, C. O. Lousto, and Y. Zlochower, *Spinning-Black-Hole Binaries: The Orbital Hang Up*, Phys. Rev. D **74**, 041501 (2006).
[155] E. Baird, S. Fairhurst, M. Hannam, and P. Murphy, *Degeneracy between Mass and Spin in Black-Hole-Binary Waveforms*, Phys. Rev. D **87**, 024035 (2013).
[156] M. Pürrer, M. Hannam, P. Ajith, and S. Husa, *Testing the Validity of the Single-Spin Approximation in Inspiral-Merger-Ringdown Waveforms*, Phys. Rev. D **88**, 064007 (2013).
[157] S. Vitale, R. Lynch, V. Raymond, R. Sturani, J. Veitch, and P. Graff, *Parameter Estimation for Heavy Binary-Black Holes with Networks of Second-Generation Gravitational-Wave Detectors*, Phys. Rev. D **95**, 064053 (2017).
[158] B. P. Abbott *et al.* (LIGO Scientific Collaboration, Virgo Collaboration), *Effects of Waveform Model Systematics on the Interpretation of GW150914*, Classical Quantum Gravity **34**, 104002 (2017).
[159] S. Kullback and R. A. Leibler, *On Information and Sufficiency*, Ann. Math. Stat. **22**, 79 (1951).
[160] C. E. Shannon, *A Mathematical Theory of Communication*, Bell Syst. Tech. J. **27** (1948).
[161] A. Vecchio, *LISA Observations of Rapidly Spinning Massive Black Hole Binary Systems*, Phys. Rev. D **70**, 042001 (2004).

[162] S. Vitale, R. Lynch, J. Veitch, V. Raymond, and R. Sturani, *Measuring the Spin of Black Holes in Binary Systems Using Gravitational Waves*, Phys. Rev. Lett. **112**, 251101 (2014).

[163] K. Chatziioannou, N. Cornish, A. Klein, and N. Yunes, *Spin-Precession: Breaking the Black Hole–Neutron Star Degeneracy*, Astrophys. J. **798**, L17 (2015).

[164] D. Gerosa, M. Kesden, U. Sperhake, E. Berti, and R. O'Shaughnessy, *Multi-Timescale Analysis of Phase Transitions in Precessing Black-Hole Binaries*, Phys. Rev. D **92**, 064016 (2015).

[165] S. Vitale, R. Lynch, R. Sturani, and P. Graff, *Use of Gravitational Waves to Probe the Formation Channels of Compact Binaries*, Classical Quantum Gravity **34**, 03LT01 (2017).

[166] S. Stevenson, C. P. L. Berry, and I. Mandel, *Hierarchical Analysis of Gravitational-Wave Measurements of Binary Black Hole Spin-Orbit Misalignments*, Mon. Not. R. Astron. Soc. **471**, 2801 (2017).

[167] W. M. Farr, S. Stevenson, M. C. Miller, I. Mandel, B. Farr, and A. Vecchio, *Distinguishing Spin-Aligned and Isotropic Black Hole Populations with Gravitational Waves*, Nature (London) **548**, 426 (2017).

[168] C. Talbot and E. Thrane, *Determining the Population Properties of Spinning Black Holes*, Phys. Rev. D **96**, 023012 (2017).

[169] B. Farr, D. E. Holz, and W. M. Farr, *Using Spin to Understand the Formation of LIGO and Virgo's Black Holes*, Astrophys. J. **854**, L9 (2018).

[170] D. Wysocki, J. Lange, and R. O'Shaughnessy, *Reconstructing Phenomenological Distributions of Compact Binaries via Gravitational Wave Observations*, arXiv:1805.06442.

[171] V. Kalogera, *Spin-Orbit Misalignment in Close Binaries with Two Compact Objects*, Astrophys. J. **541**, 319 (2000).

[172] I. Mandel and R. O'Shaughnessy, *Compact Binary Coalescences in the Band of Ground-Based Gravitational-Wave Detectors*, Classical Quantum Gravity **27**, 114007 (2010).

[173] K. Belczynski, V. Kalogera, F. A. Rasio, R. E. Taam, A. Zezas, T. Bulik, T. J. Maccarone, and N. Ivanova, *Compact Object Modeling with the StarTrack Population Synthesis Code*, Astrophys. J. Suppl. Ser. **174**, 223 (2008).

[174] C. L. Rodriguez, M. Zevin, C. Pankow, V. Kalogera, and F. A. Rasio, *Illuminating Black Hole Binary Formation Channels with Spins in Advanced LIGO*, Astrophys. J. **832**, L2 (2016).

[175] B. Liu and D. Lai, *Spin-Orbit Misalignment of Merging Black Hole Binaries with Tertiary Companions*, Astrophys. J. **846**, L11 (2017).

[176] B. Liu and D. Lai, *Black Hole and Neutron Star Binary Mergers in Triple Systems: Merger Fraction and Spin-Orbit Misalignment*, Astrophys. J. **863**, 68 (2018).

[177] F. Antonini, C. L. Rodriguez, C. Petrovich, and C. L. Fischer, *Precessional Dynamics of Black Hole Triples: Binary Mergers with Near-Zero Effective Spin*, Mon. Not. R. Astron. Soc. **480**, L58 (2018).

[178] M. U. Kruckow, T. M. Tauris, N. Langer, M. Kramer, and R. G. Izzard, *Progenitors of Gravitational Wave Mergers: Binary Evolution with the Stellar Grid-Based Code COMBINE*, Mon. Not. R. Astron. Soc. **481**, 1908 (2018).

[179] N. Giacobbo and M. Mapelli, *The Progenitors of Compact-Object Binaries: Impact of Metallicity, Common Envelope and Natal Kicks*, Mon. Not. R. Astron. Soc. **480**, 2011 (2018).

[180] J. A. Gonzalez, U. Sperhake, B. Bruegmann, M. Hannam, and S. Husa, *Total Recoil: The Maximum Kick from Nonspinning Black-Hole Binary Inspiral*, Phys. Rev. Lett. **98**, 091101 (2007).

[181] E. Berti, V. Cardoso, J. A. Gonzalez, U. Sperhake, M. Hannam, S. Husa, and B. Bruegmann, *Inspiral, Merger and Ringdown of Unequal Mass Black Hole Binaries: A Multipolar Analysis*, Phys. Rev. D **76**, 064034 (2007).

[182] A. Buonanno, L. E. Kidder, and L. Lehner, *Estimating the Final Spin of a Binary Black Hole Coalescence*, Phys. Rev. D **77**, 026004 (2008).

[183] M. Campanelli, C. O. Lousto, and Y. Zlochower, *The Last Orbit of Binary Black Holes*, Phys. Rev. D **73**, 061501 (2006).

[184] J. G. Baker, M. Campanelli, C. O. Lousto, and R. Takahashi, *Coalescence Remnant of Spinning Binary Black Holes*, Phys. Rev. D **69**, 027505 (2004).

[185] LIGO Scientific Collaboration, Virgo Collaboration, *Low-Latency Skymaps for Transient GW Events in LIGO-Virgo O1 and O2*, https://dcc.ligo.org/LIGO-P1900170/public.

[186] S. Nissanke, D. E. Holz, S. A. Hughes, N. Dalal, and J. L. Sievers, *Exploring Short Gamma-Ray Bursts as Gravitational-Wave Standard Sirens*, Astrophys. J. **725**, 496 (2010).

[187] B. Farr *et al.*, *Parameter Estimation on Gravitational Waves from Neutron-Star Binaries with Spinning Components*, Astrophys. J. **825**, 116 (2016).

[188] M. V. van der Sluys, C. Roever, A. Stroeer, N. Christensen, V. Kalogera, R. Meyer, and A. Vecchio, *Gravitational-Wave Astronomy with Inspiral Signals of Spinning Compact-Object Binaries*, Astrophys. J. **688**, L61 (2008).

[189] T. Broadhurst, J. M. Diego, and G. Smoot, *Reinterpreting Low Frequency LIGO/Virgo Events as Magnified Stellar-Mass Black Holes at Cosmological Distances*, arXiv:1802.05273.

[190] G. P. Smith, M. Jauzac, J. Veitch, W. M. Farr, R. Massey, and J. Richard, *What If LIGO's Gravitational Wave Detections Are Strongly Lensed by Massive Galaxy Clusters?*, Mon. Not. R. Astron. Soc. **475**, 3823 (2018).

[191] S. Fairhurst, *Triangulation of Gravitational Wave Sources with a Network of Detectors*, New J. Phys. **11**, 123006 (2009); *Erratum*, New J. Phys. **13**, 069602(E) (2011).

[192] K. Grover, S. Fairhurst, B. F. Farr, I. Mandel, C. Rodriguez, T. Sidery, and A. Vecchio, *Comparison of Gravitational Wave Detector Network Sky Localization Approximations*, Phys. Rev. D **89**, 042004 (2014).

[193] L. P. Singer and L. R. Price, *Rapid Bayesian Position Reconstruction for Gravitational-Wave Transients*, Phys. Rev. D **93**, 024013 (2016).

[194] M. M. Kasliwal and S. Nissanke, *On Discovering Electromagnetic Emission from Neutron Star Mergers: The Early Years of Two Gravitational Wave Detectors*, Astrophys. J. **789**, L5 (2014).

[195] L. P. Singer *et al.*, *The First Two Years of Electromagnetic Follow-Up with Advanced LIGO and Virgo*, Astrophys. J. **795**, 105 (2014).

[196] R. Essick, S. Vitale, E. Katsavounidis, G. Vedovato, and S. Klimenko, *Localization of Short Duration Gravitational-Wave Transients with the Early Advanced LIGO and Virgo Detectors*, Astrophys. J. **800**, 81 (2015).

[197] C. P. L. Berry *et al.*, *Parameter Estimation for Binary Neutron-Star Coalescences with Realistic Noise during the Advanced LIGO Era*, Astrophys. J. **804**, 114 (2015).

[198] J. Veitch, I. Mandel, B. Aylott, B. Farr, V. Raymond, C. Rodriguez, M. van der Sluys, V. Kalogera, and A. Vecchio, *Estimating Parameters of Coalescing Compact Binaries with Proposed Advanced Detector Networks*, Phys. Rev. D **85**, 104045 (2012).

[199] C. L. Rodriguez, B. Farr, V. Raymond, W. M. Farr, T. B. Littenberg, D. Fazi, and V. Kalogera, *Basic Parameter Estimation of Binary Neutron Star Systems by the Advanced LIGO/Virgo Network*, Astrophys. J. **784**, 119 (2014).

[200] B. P. Abbott *et al.* (LIGO Scientific Collaboration, Virgo Collaboration, KAGRA), *Prospects for Observing and Localizing Gravitational-Wave Transients with Advanced LIGO and Advanced Virgo*, Living Rev. Relativity **21**, 3 (2018).

[201] Optimal SNRs for Virgo from PE are shown in Table V. For GW170729 and GW170809, these values are below the GstLAL single-detector SNR threshold for Virgo of 3.5.

[202] B. P. Abbott *et al.* (LIGO Scientific, Virgo Collaboration), *GW170817: Measurements of Neutron Star Radii and Equation of State*, Phys. Rev. Lett. **121**, 161101 (2018).

[203] J. S. Read, C. Markakis, M. Shibata, K. Uryu, J. D. E. Creighton, and J. L. Friedman, *Measuring the Neutron Star Equation of State with Gravitational Wave Observations*, Phys. Rev. D **79**, 124033 (2009).

[204] F. Zappa, S. Bernuzzi, D. Radice, A. Perego, and T. Dietrich, *Gravitational-Wave Luminosity of Binary Neutron Stars Mergers*, Phys. Rev. Lett. **120**, 111101 (2018).

[205] S. Bernuzzi, A. Nagar, S. Balmelli, T. Dietrich, and M. Ujevic, *Quasiuniversal Properties of Neutron Star Mergers*, Phys. Rev. Lett. **112**, 201101 (2014).

[206] F. Zappa, S. Bernuzzi, and A. Perego, *Gravitational-Wave Energy, Luminosity and Angular Momentum from Numerical Relativity Simulations of Binary Neutron Stars Mergers*, LIGO Scientific Collaboration and Virgo Collaboration, Technical Report No. LIGO-T1800417, https://dcc.ligo.org/T1800417/public https://tds.virgo-gw.eu/?content=3&r=15019.

[207] LHO for the events GW150914, GW151012, GW151226, GW170608, and GW170729 and LLO for GW170104, GW170809, GW170814, GW170818, and GW170823.

[208] B. P. Abbott *et al.* (LIGO Scientific Collaboration, Virgo Collaboration), *Upper Limits on the Rates of Binary Neutron Star and Neutron Star-Black Hole Mergers from Advanced Ligo's First Observing Run*, Astrophys. J. Lett. **832**, L21 (2016).

[209] B. P. Abbott *et al.* (LIGO Scientific Collaboration, Virgo Collaboration), *Supplement: The Rate of Binary Black Hole Mergers Inferred from Advanced LIGO Observations Surrounding GW150914*, Astrophys. J. Suppl. **227**, 14 (2016).

[210] V. Tiwari, *Estimation of the Sensitive Volume for Gravitational-Wave Source Populations Using Weighted Monte Carlo Integration*, Classical Quantum Gravity **35**, 145009 (2018).

[211] M. Burgay *et al.*, *An Increased Estimate of the Merger Rate of Double Neutron Stars from Observations of a Highly Relativistic System*, Nature (London) **426**, 531 (2003).

[212] V. Kalogera, C. Kim, D. R. Lorimer, M. Burgay, N. D'Amico, A. Possenti, R. N. Manchester, A. G. Lyne, B. C. Joshi, M. A. McLaughlin, M. Kramer, J. M. Sarkissian, and F. Camilo, *The Cosmic Coalescence Rates for Double Neutron Star Binaries*, Astrophys. J. Lett. **601**, L179 (2004).

[213] C. Kim, B. B. P. Perera, and M. A. McLaughlin, *Implications of PSR J0737-3039B for the Galactic NS-NS Binary Merger Rate*, Mon. Not. R. Astron. Soc. **448**, 928 (2015).

[214] F. Özel, D. Psaltis, R. Narayan, and A. S. Villarreal, *On the Mass Distribution and Birth Masses of Neutron Stars*, Astrophys. J. **757**, 55 (2012).

[215] N. Pol, M. McLaughlin, and D. R. Lorimer, *Future Prospects for Ground-Based Gravitational Wave Detectors—The Galactic Double Neutron Star Merger Rate Revisited*, Astrophys. J. **870**, 71 (2019); *Erratum*, Astrophys. J. **874**, 186(E) (2019).

[216] J. Abadie *et al.* (VIRGO, LIGO Scientific Collaboration), *Predictions for the Rates of Compact Binary Coalescences Observable by Ground-Based Gravitational-Wave Detectors*, Classical Quantum Gravity **27**, 173001 (2010).

[217] B. P. Abbott *et al.* (LIGO Scientific, Virgo Collaboration), *Tests of General Relativity with the Binary Black Hole Signals from the LIGO-Virgo Catalog GWTC-1*, arXiv: 1903.04467.

[218] B. P. Abbott *et al.* (Virgo, LIGO Scientific Collaboration), *Tests of General Relativity with GW170817*, arXiv: 1811.00364.

[219] P. Marchant, M. Renzo, R. Farmer, K. M. W. Pappas, R. E. Taam, S. de Mink, and V. Kalogera, *Pulsational Pair-Instability Supernovae in Very Close Binaries*, arXiv: 1810.13412.

[220] S. Stevenson, F. Ohme, and S. Fairhurst, *Distinguishing Compact Binary Population Synthesis Models Using Gravitational Wave Observations of Coalescing Binary Black Holes*, Astrophys. J. **810**, 58 (2015).

[221] C. Talbot and E. Thrane, *Measuring the Binary Black Hole Mass Spectrum with an Astrophysically Motivated Parameterization*, Astrophys. J. **856**, 173 (2018).

[222] M. Fishbach and D. E. Holz, *Where Are LIGO's Big Black Holes?*, Astrophys. J. **851**, L25 (2017).

[223] M. Zevin, C. Pankow, C. L. Rodriguez, L. Sampson, E. Chase, V. Kalogera, and F. A. Rasio, *Constraining Formation Models of Binary Black Holes with Gravitational-Wave Observations*, Astrophys. J. Lett. **846**, 82 (2017).

[224] M. Fishbach, D. E. Holz, and B. Farr, *Are LIGO's Black Holes Made from Smaller Black Holes?*, Astrophys. J. Lett. **840**, L24 (2017).

[225] J. W. Barrett, S. M. Gaebel, C. J. Neijssel, A. Vigna-Gómez, S. Stevenson, C. P. L. Berry, W. M. Farr, and I. Mandel, *Accuracy of Inference on the Physics of Binary*

*Evolution from Gravitational-Wave Observations*, Mon. Not. R. Astron. Soc. **477**, 4685 (2018).
[226] K. Somiya (KAGRA Collaboration), *Detector Configuration of KAGRA: The Japanese Cryogenic Gravitational-Wave Detector*, Classical Quantum Gravity **29**, 124007 (2012).
[227] Y. Aso, Y. Michimura, K. Somiya, M. Ando, O. Miyakawa, T. Sekiguchi, D. Tatsumi, and H. Yamamoto (KAGRA Collaboration), *Interferometer Design of the KAGRA Gravitational Wave Detector*, Phys. Rev. D **88**, 043007 (2013).
[228] T. Akutsu *et al.* (KAGRA Collaboration), *Construction of KAGRA: An Underground Gravitational Wave Observatory*, Prog. Theor. Exp. Phys. **2018**, 013F01 (2018).
[229] T. Accadia *et al.*, *Noise from Scattered Light in Virgo's Second Science Run Data*, Classical Quantum Gravity **27**, 194011 (2010).
[230] D. J. Ottaway, P. Fritschel, and S. J. Waldman, *Impact of Upconverted Scattered Light on Advanced Interferometric Gravitational Wave Detectors*, Opt. Express **20**, 8329 (2012).
[231] G. Valdes, B. O'Reilly, and M. Diaz, *A Hilbert-Huang Transform Method for Scattering Identification in LIGO*, Classical Quantum Gravity **34**, 235009 (2017).
[232] O. Patane *et al.*, aLIGO LHO Logbook, https://alog.ligo-wa.caltech.edu/aLOG/index.php?callRep=43177.
[233] M. Zevin *et al.*, *Gravity Spy: Integrating Advanced LIGO Detector Characterization, Machine Learning, and Citizen Science*, Classical Quantum Gravity **34**, 064003 (2017).
[234] C. Pankow *et al.*, *Mitigation of the Instrumental Noise Transient in Gravitational-Wave Data Surrounding GW170817*, Phys. Rev. D **98**, 084016 (2018).
[235] B. P. Abbott *et al.* (Virgo, LIGO Scientific Collaboration), *Observing Gravitational-Wave Transient GW150914 with Minimal Assumptions*, Phys. Rev. D **93**, 122004 (2016); **94**, 069903(A) (2016).
[236] M. Cabero *et al.*, *Blip Glitches in Advanced LIGO Data*, arXiv:1901.05093.
[237] T. Bayes, *An Essay toward Solving a Problem in the Doctrine of Chances*, Phil. Trans. R. Soc. London **53**, 370 (1764).
[238] E. T. Jaynes, in *Probability Theory: The Logic of Science*, edited by G. L. Bretthorst (Cambridge University Press, Cambridge, England, 2003).
[239] C. Röver, R. Meyer, and N. Christensen, *Bayesian Inference on Compact Binary Inspiral Gravitational Radiation Signals in Interferometric Data*, Classical Quantum Gravity **23**, 4895 (2006).
[240] M. van der Sluys, V. Raymond, I. Mandel, C. Röver, N. Christensen, V. Kalogera, R. Meyer, and A. Vecchio, *Parameter Estimation of Spinning Binary Inspirals Using Markov-Chain Monte Carlo*, Classical Quantum Gravity **25**, 184011 (2008).
[241] J. Veitch *et al.*, *Parameter Estimation for Compact Binaries with Ground-Based Gravitational-Wave Observations Using the LALInference Software Library*, Phys. Rev. D **91**, 042003 (2015).
[242] LIGO Scientific Collaboration, Virgo Collaboration, LALSuite branch lalinference_o2, https://git.ligo.org/lscsoft/lalsuite/tree/lalinference_o2.
[243] C. Pankow, P. Brady, E. Ochsner, and R. O'Shaughnessy, *Novel Scheme for Rapid Parallel Parameter Estimation of Gravitational Waves from Compact Binary Coalescences*, Phys. Rev. D **92**, 023002 (2015).
[244] J. Lange, R. O'Shaughnessy, and M. Rizzo, *Rapid and Accurate Parameter Inference for Coalescing, Precessing Compact Binaries*, arXiv:1805.10457.
[245] P. A. R. Ade *et al.* (Planck Collaboration), *Planck 2015 Results. XIII. Cosmological Parameters*, Astron. Astrophys. **594**, A13 (2016).
[246] B. F. Schutz, *Determining the Hubble Constant from Gravitational Wave Observations*, Nature (London) **323**, 310 (1986).
[247] N. Aghanim *et al.* (Planck Collaboration), *Planck 2018 Results. VI. Cosmological Parameters*, arXiv:1807.06209.
[248] For clarity, we use the names of the waveform models as defined in the LIGO Algorithm Library (LAL) [79], as well as in technical publications.
[249] J. Lin, *Divergence Measures Based on the Shannon Entropy*, IEEE Trans. Inf. Theory **37**, 145 (1991).
[250] In the LIGO Algorithm Library (LAL) [79], as well as in technical publications, these models are referred to as SEOBNRv4_ROM_NRTidal and IMRPhenomPv2_NRTidal.
[251] E. Poisson, *Gravitational Waves from Inspiraling Compact Binaries: The Quadrupole Moment Term*, Phys. Rev. D **57**, 5287 (1998).
[252] I. Harry and T. Hinderer, *Observing and Measuring the Neutron-Star Equation-of-State in Spinning Binary Neutron Star Systems*, Classical Quantum Gravity **35**, 145010 (2018).
[253] K. Yagi and N. Yunes, *Approximate Universal Relations for Neutron Stars and Quark Stars*, Phys. Rep. **681**, 1 (2017).
[254] J. Blackman, S. E. Field, M. A. Scheel, C. R. Galley, C. D. Ott, M. Boyle, L. E. Kidder, H. P. Pfeiffer, and B. Szilágyi, *Numerical Relativity Waveform Surrogate Model for Generically Precessing Binary Black Hole Mergers*, Phys. Rev. D **96**, 024058 (2017).
[255] B. P. Abbott *et al.* (LIGO Scientific Collaboration, Virgo Collaboration), *Directly Comparing GW150914 with Numerical Solutions of Einstein's Equations for Binary Black Hole Coalescence*, Phys. Rev. D **94**, 064035 (2016).
[256] J. Lange *et al.*, *Parameter Estimation Method That Directly Compares Gravitational Wave Observations to Numerical Relativity*, Phys. Rev. D **96**, 104041 (2017).
[257] V. Varma and P. Ajith, *Effects of Nonquadrupole Modes in the Detection and Parameter Estimation of Black Hole Binaries with Nonprecessing Spins*, Phys. Rev. D **96**, 124024 (2017).
[258] J. C. Bustillo, S. Husa, A. M. Sintes, and M. Pürrer, *Impact of Gravitational Radiation Higher Order Modes on Single Aligned-Spin Gravitational Wave Searches for Binary Black Holes*, Phys. Rev. D **93**, 084019 (2016).
[259] P. Kumar, J. Blackman, S. E. Field, M. Scheel, C. R. Galley, M. Boyle, L. E. Kidder, H. P. Pfeiffer, B. Szilagyi, and S. A. Teukolsky, *Constraining the Parameters of GW150914 & GW170104 with Numerical Relativity Surrogates*, arXiv:1808.08004.

[260] A. R. Williamson, J. Lange, R. O'Shaughnessy, J. A. Clark, P. Kumar, J. C. Bustillo, and J. Veitch, *Systematic Challenges for Future Gravitational Wave Measurements of Precessing Binary Black Holes*, Phys. Rev. D **96**, 124041 (2017).

[261] S. Vitale, D. Gerosa, C.-J. Haster, K. Chatziioannou, and A. Zimmerman, *Impact of Bayesian Priors on the Characterization of Binary Black Hole Coalescences*, Phys. Rev. Lett. **119**, 251103 (2017).

[262] V. Tiwari, S. Fairhurst, and M. Hannam, *Constraining Black-Hole Spins with Gravitational Wave Observations*, Astrophys. J. **868**, 140 (2018).

[263] M. C. Miller and J. M. Miller, *The Masses and Spins of Neutron Stars and Stellar-Mass Black Holes*, Phys. Rep. **548**, 1 (2015).

[264] Y. Huang, H. Middleton, K. K. Y. Ng, S. Vitale, and J. Veitch, *Characterization of Low-Significance Gravitational-Wave Compact Binary Sources*, Phys. Rev. D **98**, 123021 (2018).

B. P. Abbott,[1] R. Abbott,[1] T. D. Abbott,[2] S. Abraham,[3] F. Acernese,[4,5] K. Ackley,[6] C. Adams,[7] R. X. Adhikari,[1] V. B. Adya,[8,9] C. Affeldt,[8,9] M. Agathos,[10] K. Agatsuma,[11] N. Aggarwal,[12] O. D. Aguiar,[13] L. Aiello,[14,15] A. Ain,[3] P. Ajith,[16] G. Allen,[17] A. Allocca,[18,19] M. A. Aloy,[20] P. A. Altin,[21] A. Amato,[22] A. Ananyeva,[1] S. B. Anderson,[1] W. G. Anderson,[23] S. V. Angelova,[24] S. Antier,[25] S. Appert,[1] K. Arai,[1] M. C. Araya,[1] J. S. Areeda,[26] M. Arène,[27] N. Arnaud,[25,28] K. G. Arun,[29] S. Ascenzi,[30,31] G. Ashton,[6] S. M. Aston,[7] P. Astone,[32] F. Aubin,[33] P. Aufmuth,[9] K. AultONeal,[34] C. Austin,[2] V. Avendano,[35] A. Avila-Alvarez,[26] S. Babak,[36,27] P. Bacon,[27] F. Badaracco,[14,15] M. K. M. Bader,[37] S. Bae,[38] P. T. Baker,[39] F. Baldaccini,[40,41] G. Ballardin,[28] S. W. Ballmer,[42] S. Banagiri,[43] J. C. Barayoga,[1] S. E. Barclay,[44] B. C. Barish,[1] D. Barker,[45] K. Barkett,[46] S. Barnum,[12] F. Barone,[4,5] B. Barr,[44] L. Barsotti,[12] M. Barsuglia,[27] D. Barta,[47] J. Bartlett,[45] I. Bartos,[48] R. Bassiri,[49] A. Basti,[18,19] M. Bawaj,[50,41] J. C. Bayley,[44] M. Bazzan,[51,52] B. Bécsy,[53] M. Bejger,[27,54] I. Belahcene,[25] A. S. Bell,[44] D. Beniwal,[55] B. K. Berger,[49] G. Bergmann,[8,9] S. Bernuzzi,[56,57] J. J. Bero,[58] C. P. L. Berry,[59] D. Bersanetti,[60] A. Bertolini,[37] J. Betzwieser,[7] R. Bhandare,[61] J. Bidler,[26] I. A. Bilenko,[62] S. A. Bilgili,[39] G. Billingsley,[1] J. Birch,[7] R. Birney,[24] O. Birnholtz,[58] S. Biscans,[1,12] S. Biscoveanu,[6] A. Bisht,[9] M. Bitossi,[28,19] M. A. Bizouard,[25] J. K. Blackburn,[1] J. Blackman,[46] C. D. Blair,[7] D. G. Blair,[63] R. M. Blair,[45] S. Bloemen,[64] N. Bode,[8,9] M. Boer,[65] Y. Boetzel,[66] G. Bogaert,[65] F. Bondu,[67] E. Bonilla,[49] R. Bonnand,[33] P. Booker,[8,9] B. A. Boom,[37] C. D. Booth,[68] R. Bork,[1] V. Boschi,[28] S. Bose,[69,3] K. Bossie,[7] V. Bossilkov,[63] J. Bosveld,[63] Y. Bouffanais,[27] A. Bozzi,[28] C. Bradaschia,[19] P. R. Brady,[23] A. Bramley,[7] M. Branchesi,[14,15] J. E. Brau,[70] T. Briant,[71] J. H. Briggs,[44] F. Brighenti,[72,73] A. Brillet,[65] M. Brinkmann,[8,9] V. Brisson,[25,†] P. Brockill,[23] A. F. Brooks,[1] D. D. Brown,[55] S. Brunett,[1] A. Buikema,[12] T. Bulik,[74] H. J. Bulten,[75,37] A. Buonanno,[36,76] D. Buskulic,[33] M. J. Bustamante Rosell,[77] C. Buy,[27] R. L. Byer,[49] M. Cabero,[8,9] L. Cadonati,[78] G. Cagnoli,[22,79] C. Cahillane,[1] J. Calderón Bustillo,[6] T. A. Callister,[1] E. Calloni,[80,5] J. B. Camp,[81] W. A. Campbell,[6] M. Canepa,[82,60] K. C. Cannon,[83] H. Cao,[55] J. Cao,[84] E. Capocasa,[27] F. Carbognani,[28] S. Caride,[85] M. F. Carney,[59] G. Carullo,[18] J. Casanueva Diaz,[19] C. Casentini,[30,31] S. Caudill,[37] M. Cavaglià,[86] F. Cavalier,[25] R. Cavalieri,[28] G. Cella,[19] P. Cerdá-Durán,[20] G. Cerretani,[18,19] E. Cesarini,[87,31] O. Chaibi,[65] K. Chakravarti,[3] S. J. Chamberlin,[88] M. Chan,[44] S. Chao,[89] P. Charlton,[90] E. A. Chase,[59] E. Chassande-Mottin,[27] D. Chatterjee,[23] M. Chaturvedi,[61] K. Chatziioannou,[91] B. D. Cheeseboro,[39] H. Y. Chen,[92] X. Chen,[63] Y. Chen,[46] H.-P. Cheng,[48] C. K. Cheong,[93] H. Y. Chia,[48] A. Chincarini,[60] A. Chiummo,[28] G. Cho,[94] H. S. Cho,[95] M. Cho,[76] N. Christensen,[65,96] Q. Chu,[63] S. Chua,[71] K. W. Chung,[93] S. Chung,[63] G. Ciani,[51,52] A. A. Ciobanu,[55] R. Ciolfi,[97,98] F. Cipriano,[65] A. Cirone,[82,60] F. Clara,[45] J. A. Clark,[78] P. Clearwater,[99] F. Cleva,[65] C. Cocchieri,[86] E. Coccia,[14,15] P.-F. Cohadon,[71] D. Cohen,[25] R. Colgan,[100] M. Colleoni,[101] C. G. Collette,[102] C. Collins,[11] L. R. Cominsky,[103] M. Constancio Jr.,[13] L. Conti,[52] S. J. Cooper,[11] P. Corban,[7] T. R. Corbitt,[2] I. Cordero-Carrión,[104] K. R. Corley,[100] N. Cornish,[53] A. Corsi,[85] S. Cortese,[28] C. A. Costa,[13] R. Cotesta,[36] M. W. Coughlin,[1] S. B. Coughlin,[68,59] J.-P. Coulon,[65] S. T. Countryman,[100] P. Couvares,[1] P. B. Covas,[101] E. E. Cowan,[78] D. M. Coward,[63] M. J. Cowart,[7] D. C. Coyne,[1] R. Coyne,[105] J. D. E. Creighton,[23] T. D. Creighton,[106] J. Cripe,[2] M. Croquette,[71] S. G. Crowder,[107] T. J. Cullen,[2] A. Cumming,[44] L. Cunningham,[44] E. Cuoco,[28] T. Dal Canton,[81] G. Dálya,[108] S. L. Danilishin,[8,9] S. D'Antonio,[31] K. Danzmann,[9,8] A. Dasgupta,[109] C. F. Da Silva Costa,[48] L. E. H. Datrier,[44] V. Dattilo,[28] I. Dave,[61] M. Davier,[25] D. Davis,[42] E. J. Daw,[110] D. DeBra,[49] M. Deenadayalan,[3] J. Degallaix,[22] M. De Laurentis,[80,5] S. Deléglise,[71] W. Del Pozzo,[18,19] L. M. DeMarchi,[59] N. Demos,[12] T. Dent,[8,9,111] R. De Pietri,[112,57] J. Derby,[26] R. De Rosa,[80,5] C. De Rossi,[22,28] R. DeSalvo,[113] O. de Varona,[8,9] S. Dhurandhar,[3] M. C. Díaz,[106] T. Dietrich,[37] L. Di Fiore,[5] M. Di Giovanni,[114,98] T. Di Girolamo,[80,5] A. Di Lieto,[18,19] B. Ding,[102] S. Di Pace,[115,32] I. Di Palma,[115,32] F. Di Renzo,[18,19] A. Dmitriev,[11] Z. Doctor,[92] F. Donovan,[12] K. L. Dooley,[68,86] S. Doravari,[8,9] I. Dorrington,[68] T. P. Downes,[23] M. Drago,[14,15] J. C. Driggers,[45] Z. Du,[84] J.-G. Ducoin,[25]

P. Dupej,[44] S. E. Dwyer,[45] P. J. Easter,[6] T. B. Edo,[110] M. C. Edwards,[96] A. Effler,[7] P. Ehrens,[1] J. Eichholz,[1] S. S. Eikenberry,[48] M. Eisenmann,[33] R. A. Eisenstein,[12] R. C. Essick,[92] H. Estelles,[101] D. Estevez,[33] Z. B. Etienne,[39] T. Etzel,[1] M. Evans,[12] T. M. Evans,[7] V. Fafone,[30,31,14] H. Fair,[42] S. Fairhurst,[68] X. Fan,[84] S. Farinon,[60] B. Farr,[70] W. M. Farr,[11] E. J. Fauchon-Jones,[68] M. Favata,[35] M. Fays,[110] M. Fazio,[116] C. Fee,[117] J. Feicht,[1] M. M. Fejer,[49] F. Feng,[27] A. Fernandez-Galiana,[12] I. Ferrante,[18,19] E. C. Ferreira,[13] T. A. Ferreira,[13] F. Ferrini,[28] F. Fidecaro,[18,19] I. Fiori,[28] D. Fiorucci,[27] M. Fishbach,[92] R. P. Fisher,[42,118] J. M. Fishner,[12] M. Fitz-Axen,[43] R. Flaminio,[33,119] M. Fletcher,[44] E. Flynn,[26] H. Fong,[91] J. A. Font,[20,120] P. W. F. Forsyth,[21] J.-D. Fournier,[65] S. Frasca,[115,32] F. Frasconi,[19] Z. Frei,[108] A. Freise,[11] R. Frey,[70] V. Frey,[25] P. Fritschel,[12] V. V. Frolov,[7] P. Fulda,[48] M. Fyffe,[7] H. A. Gabbard,[44] B. U. Gadre,[3] S. M. Gaebel,[11] J. R. Gair,[121] L. Gammaitoni,[40] M. R. Ganija,[55] S. G. Gaonkar,[3] A. Garcia,[26] C. García-Quirós,[101] F. Garufi,[80,5] B. Gateley,[45] S. Gaudio,[34] G. Gaur,[122] V. Gayathri,[123] G. Gemme,[60] E. Genin,[28] A. Gennai,[19] D. George,[17] J. George,[61] L. Gergely,[124] V. Germain,[33] S. Ghonge,[78] Abhirup Ghosh,[16] Archisman Ghosh,[37] S. Ghosh,[23] B. Giacomazzo,[114,98] J. A. Giaime,[2,7] K. D. Giardina,[7] A. Giazotto,[19,†] K. Gill,[34] G. Giordano,[4,5] L. Glover,[113] P. Godwin,[88] E. Goetz,[45] R. Goetz,[48] B. Goncharov,[6] G. González,[2] J. M. Gonzalez Castro,[18,19] A. Gopakumar,[125] M. L. Gorodetsky,[62] S. E. Gossan,[1] M. Gosselin,[28] R. Gouaty,[33] A. Grado,[126,5] C. Graef,[44] M. Granata,[22] A. Grant,[44] S. Gras,[12] P. Grassia,[1] C. Gray,[45] R. Gray,[44] G. Greco,[72,73] A. C. Green,[11,48] R. Green,[68] E. M. Gretarsson,[34] P. Groot,[64] H. Grote,[68] S. Grunewald,[36] P. Gruning,[25] G. M. Guidi,[72,73] H. K. Gulati,[109] Y. Guo,[37] A. Gupta,[88] M. K. Gupta,[109] E. K. Gustafson,[1] R. Gustafson,[127] L. Haegel,[101] O. Halim,[15,14] B. R. Hall,[69] E. D. Hall,[12] E. Z. Hamilton,[68] G. Hammond,[44] M. Haney,[66] M. M. Hanke,[8,9] J. Hanks,[45] C. Hanna,[88] M. D. Hannam,[68] O. A. Hannuksela,[93] J. Hanson,[7] T. Hardwick,[2] K. Haris,[16] J. Harms,[14,15] G. M. Harry,[128] I. W. Harry,[36] C.-J. Haster,[91] K. Haughian,[44] F. J. Hayes,[44] J. Healy,[58] A. Heidmann,[71] M. C. Heintze,[7] H. Heitmann,[65] P. Hello,[25] G. Hemming,[28] M. Hendry,[44] I. S. Heng,[44] J. Hennig,[8,9] A. W. Heptonstall,[1] Francisco Hernandez Vivanco,[6] M. Heurs,[8,9] S. Hild,[44] T. Hinderer,[129,37,130] D. Hoak,[28] S. Hochheim,[8,9] D. Hofman,[22] A. M. Holgado,[17] N. A. Holland,[21] K. Holt,[7] D. E. Holz,[92] P. Hopkins,[68] C. Horst,[23] J. Hough,[44] E. J. Howell,[63] C. G. Hoy,[68] A. Hreibi,[65] Y. Huang,[12,131] E. A. Huerta,[17] D. Huet,[25] B. Hughey,[34] M. Hulko,[1] S. Husa,[101] S. H. Huttner,[44] T. Huynh-Dinh,[7] B. Idzkowski,[74] A. Iess,[30,31] C. Ingram,[55] R. Inta,[85] G. Intini,[115,32] B. Irwin,[117] H. N. Isa,[44] J.-M. Isac,[71] M. Isi,[1] B. R. Iyer,[16] K. Izumi,[45] T. Jacqmin,[71] S. J. Jadhav,[132] K. Jani,[78] N. N. Janthalur,[132] P. Jaranowski,[133] A. C. Jenkins,[134] J. Jiang,[48] D. S. Johnson,[17] N. K. Johnson-McDaniel,[10] A. W. Jones,[11] D. I. Jones,[135] R. Jones,[44] R. J. G. Jonker,[37] L. Ju,[63] J. Junker,[8,9] C. V. Kalaghatgi,[68] V. Kalogera,[59] B. Kamai,[1] S. Kandhasamy,[86] G. Kang,[38] J. B. Kanner,[1] S. J. Kapadia,[23] S. Karki,[70] K. S. Karvinen,[8,9] R. Kashyap,[16] M. Kasprzack,[1] S. Katsanevas,[28] E. Katsavounidis,[12] W. Katzman,[7] S. Kaufer,[9] K. Kawabe,[45] N. V. Keerthana,[3] F. Kéfélian,[65] D. Keitel,[44] R. Kennedy,[110] J. S. Key,[136] F. Y. Khalili,[62] H. Khan,[26] I. Khan,[14,31] S. Khan,[8,9] Z. Khan,[109] E. A. Khazanov,[137] M. Khursheed,[61] N. Kijbunchoo,[21] Chunglee Kim,[138] J. C. Kim,[139] K. Kim,[93] W. Kim,[55] W. S. Kim,[140] Y.-M. Kim,[141] C. Kimball,[59] E. J. King,[55] P. J. King,[45] M. Kinley-Hanlon,[128] R. Kirchhoff,[8,9] J. S. Kissel,[45] L. Kleybolte,[142] J. H. Klika,[23] S. Klimenko,[48] T. D. Knowles,[39] P. Koch,[8,9] S. M. Koehlenbeck,[8,9] G. Koekoek,[37,143] S. Koley,[37] V. Kondrashov,[1] A. Kontos,[12] N. Koper,[8,9] M. Korobko,[142] W. Z. Korth,[1] I. Kowalska,[74] D. B. Kozak,[1] V. Kringel,[8,9] N. Krishnendu,[29] A. Królak,[144,145] G. Kuehn,[8,9] A. Kumar,[132] P. Kumar,[146] R. Kumar,[109] S. Kumar,[16] L. Kuo,[89] A. Kutynia,[144] S. Kwang,[23] B. D. Lackey,[36] K. H. Lai,[93] T. L. Lam,[93] M. Landry,[45] B. B. Lane,[12] R. N. Lang,[147] J. Lange,[58] B. Lantz,[49] R. K. Lanza,[12] A. Lartaux-Vollard,[25] P. D. Lasky,[6] M. Laxen,[7] A. Lazzarini,[1] C. Lazzaro,[52] P. Leaci,[115,32] S. Leavey,[8,9] Y. K. Lecoeuche,[45] C. H. Lee,[95] H. K. Lee,[148] H. M. Lee,[149] H. W. Lee,[139] J. Lee,[94] K. Lee,[44] J. Lehmann,[8,9] A. Lenon,[39] N. Leroy,[25] N. Letendre,[33] Y. Levin,[6,100] J. Li,[84] K. J. L. Li,[93] T. G. F. Li,[93] X. Li,[46] F. Lin,[6] F. Linde,[37] S. D. Linker,[113] T. B. Littenberg,[150] J. Liu,[63] X. Liu,[23] R. K. L. Lo,[93,1] N. A. Lockerbie,[24] L. T. London,[68] A. Longo,[151,152] M. Lorenzini,[14,15] V. Loriette,[153] M. Lormand,[7] G. Losurdo,[19] J. D. Lough,[8,9] C. O. Lousto,[58] G. Lovelace,[26] M. E. Lower,[154] H. Lück,[9,8] D. Lumaca,[30,31] A. P. Lundgren,[155] R. Lynch,[12] Y. Ma,[46] R. Macas,[68] S. Macfoy,[24] M. MacInnis,[12] D. M. Macleod,[68] A. Macquet,[65] F. Magaña-Sandoval,[42] L. Magaña Zertuche,[86] R. M. Magee,[88] E. Majorana,[32] I. Maksimovic,[153] A. Malik,[61] N. Man,[65] V. Mandic,[43] V. Mangano,[44] G. L. Mansell,[45,12] M. Manske,[23,21] M. Mantovani,[28] F. Marchesoni,[50,41] F. Marion,[33] S. Márka,[100] Z. Márka,[100] C. Markakis,[10,17] A. S. Markosyan,[49] A. Markowitz,[1] E. Maros,[1] A. Marquina,[104] S. Marsat,[36] F. Martelli,[72,73] I. W. Martin,[44] R. M. Martin,[35] D. V. Martynov,[11] K. Mason,[12] E. Massera,[110] A. Masserot,[33] T. J. Massinger,[1] M. Masso-Reid,[44] S. Mastrogiovanni,[115,32] A. Matas,[43,36] F. Matichard,[1,12] L. Matone,[100] N. Mavalvala,[12] N. Mazumder,[69] J. J. McCann,[63] R. McCarthy,[45] D. E. McClelland,[21] S. McCormick,[7] L. McCuller,[12] S. C. McGuire,[156] J. McIver,[1] D. J. McManus,[21] T. McRae,[21] S. T. McWilliams,[39] D. Meacher,[88] G. D. Meadors,[6] M. Mehmet,[8,9] A. K. Mehta,[16] J. Meidam,[37] A. Melatos,[99] G. Mendell,[45] R. A. Mercer,[23] L. Mereni,[22] E. L. Merilh,[45] M. Merzougui,[65]

S. Meshkov,[1] C. Messenger,[44] C. Messick,[88] R. Metzdorff,[71] P. M. Meyers,[99] H. Miao,[11] C. Michel,[22] H. Middleton,[99] E. E. Mikhailov,[157] L. Milano,[80,5] A. L. Miller,[48] A. Miller,[115,32] M. Millhouse,[53] J. C. Mills,[68] M. C. Milovich-Goff,[113] O. Minazzoli,[65,158] Y. Minenkov,[31] A. Mishkin,[48] C. Mishra,[159] T. Mistry,[110] S. Mitra,[3] V. P. Mitrofanov,[62] G. Mitselmakher,[48] R. Mittleman,[12] G. Mo,[96] D. Moffa,[117] K. Mogushi,[86] S. R. P. Mohapatra,[12] M. Montani,[72,73] C. J. Moore,[10] D. Moraru,[45] G. Moreno,[45] S. Morisaki,[83] B. Mours,[33] C. M. Mow-Lowry,[11] Arunava Mukherjee,[8,9] D. Mukherjee,[23] S. Mukherjee,[106] N. Mukund,[3] A. Mullavey,[7] J. Munch,[55] E. A. Muñiz,[42] M. Muratore,[34] P. G. Murray,[44] A. Nagar,[87,160,161] I. Nardecchia,[30,31] L. Naticchioni,[115,32] R. K. Nayak,[162] J. Neilson,[113] G. Nelemans,[64,37] T. J. N. Nelson,[7] M. Nery,[8,9] A. Neunzert,[127] K. Y. Ng,[12] S. Ng,[55] P. Nguyen,[70] D. Nichols,[129,37] A. B. Nielsen,[8] S. Nissanke,[129,37] A. Nitz,[8] F. Nocera,[28] C. North,[68] L. K. Nuttall,[155] M. Obergaulinger,[20] J. Oberling,[45] B. D. O'Brien,[48] G. D. O'Dea,[113] G. H. Ogin,[163] J. J. Oh,[140] S. H. Oh,[140] F. Ohme,[8,9] H. Ohta,[83] M. A. Okada,[13] M. Oliver,[101] P. Oppermann,[8,9] Richard J. Oram,[7] B. O'Reilly,[7] R. G. Ormiston,[43] L. F. Ortega,[48] R. O'Shaughnessy,[58] S. Ossokine,[36] D. J. Ottaway,[55] H. Overmier,[7] B. J. Owen,[85] A. E. Pace,[88] G. Pagano,[18,19] M. A. Page,[63] A. Pai,[123] S. A. Pai,[61] J. R. Palamos,[70] O. Palashov,[137] C. Palomba,[32] A. Pal-Singh,[142] Huang-Wei Pan,[89] B. Pang,[46] P. T. H. Pang,[93] C. Pankow,[59] F. Pannarale,[115,32] B. C. Pant,[61] F. Paoletti,[19] A. Paoli,[28] M. A. Papa,[8,23,9] A. Parida,[3] W. Parker,[7,156] D. Pascucci,[44] A. Pasqualetti,[28] R. Passaquieti,[18,19] D. Passuello,[19] M. Patil,[145] B. Patricelli,[18,19] B. L. Pearlstone,[44] C. Pedersen,[68] M. Pedraza,[1] R. Pedurand,[22,164] A. Pele,[7] S. Penn,[165] A. Perego,[57,166] C. J. Perez,[45] A. Perreca,[114,98] H. P. Pfeiffer,[36,91] M. Phelps,[8,9] K. S. Phukon,[3] O. J. Piccinni,[115,32] M. Pichot,[65] F. Piergiovanni,[72,73] G. Pillant,[28] L. Pinard,[22] M. Pirello,[45] M. Pitkin,[44] R. Poggiani,[18,19] D. Y. T. Pong,[93] S. Ponrathnam,[3] P. Popolizio,[28] E. K. Porter,[27] J. Powell,[154] A. K. Prajapati,[109] J. Prasad,[3] K. Prasai,[49] R. Prasanna,[132] G. Pratten,[101] T. Prestegard,[23] S. Privitera,[36] G. A. Prodi,[114,98] L. G. Prokhorov,[62] O. Puncken,[8,9] M. Punturo,[41] P. Puppo,[32] M. Pürrer,[36] H. Qi,[23] V. Quetschke,[106] P. J. Quinonez,[34] E. A. Quintero,[1] R. Quitzow-James,[70] F. J. Raab,[45] H. Radkins,[45] N. Radulescu,[65] P. Raffai,[108] S. Raja,[61] C. Rajan,[61] B. Rajbhandari,[85] M. Rakhmanov,[106] K. E. Ramirez,[106] A. Ramos-Buades,[101] Javed Rana,[3] K. Rao,[59] P. Rapagnani,[115,32] V. Raymond,[68] M. Razzano,[18,19] J. Read,[26] T. Regimbau,[33] L. Rei,[60] S. Reid,[24] D. H. Reitze,[1,48] W. Ren,[17] F. Ricci,[115,32] C. J. Richardson,[34] J. W. Richardson,[1] P. M. Ricker,[17] G. M. Riemenschneider,[160,167] K. Riles,[127] M. Rizzo,[59] N. A. Robertson,[1,44] R. Robie,[44] F. Robinet,[25] A. Rocchi,[31] L. Rolland,[33] J. G. Rollins,[1] V. J. Roma,[70] M. Romanelli,[67] R. Romano,[4,5] C. L. Romel,[45] J. H. Romie,[7] K. Rose,[117] D. Rosińska,[168,54] S. G. Rosofsky,[17] M. P. Ross,[169] S. Rowan,[44] A. Rüdiger,[8,9,†] P. Ruggi,[28] G. Rutins,[170] K. Ryan,[45] S. Sachdev,[1] T. Sadecki,[45] M. Sakellariadou,[134] O. Salafia,[57,166] L. Salconi,[28] M. Saleem,[29] F. Salemi,[8] A. Samajdar,[37] L. Sammut,[6] E. J. Sanchez,[1] L. E. Sanchez,[1] N. Sanchis-Gual,[20] V. Sandberg,[45] J. R. Sanders,[42] K. A. Santiago,[35] N. Sarin,[6] B. Sassolas,[22] B. S. Sathyaprakash,[88,68] P. R. Saulson,[42] O. Sauter,[127] R. L. Savage,[45] P. Schale,[70] M. Scheel,[46] J. Scheuer,[59] P. Schmidt,[64] R. Schnabel,[142] R. M. S. Schofield,[70] A. Schönbeck,[142] E. Schreiber,[8,9] B. W. Schulte,[8,9] B. F. Schutz,[68] S. G. Schwalbe,[34] J. Scott,[44] S. M. Scott,[21] E. Seidel,[17] D. Sellers,[7] A. S. Sengupta,[171] N. Sennett,[36] D. Sentenac,[28] V. Sequino,[30,31,14] A. Sergeev,[137] Y. Setyawati,[8,9] D. A. Shaddock,[21] T. Shaffer,[45] M. S. Shahriar,[59] M. B. Shaner,[113] L. Shao,[36] P. Sharma,[61] P. Shawhan,[76] H. Shen,[17] R. Shink,[172] D. H. Shoemaker,[12] D. M. Shoemaker,[78] S. ShyamSundar,[61] K. Siellez,[78] M. Sieniawska,[54] D. Sigg,[45] A. D. Silva,[13] L. P. Singer,[81] N. Singh,[74] A. Singhal,[14,32] A. M. Sintes,[101] S. Sitmukhambetov,[106] V. Skliris,[68] B. J. J. Slagmolen,[21] T. J. Slaven-Blair,[63] J. R. Smith,[26] R. J. E. Smith,[6] S. Somala,[173] E. J. Son,[140] B. Sorazu,[44] F. Sorrentino,[60] T. Souradeep,[3] E. Sowell,[85] A. P. Spencer,[44] A. K. Srivastava,[109] V. Srivastava,[42] K. Staats,[59] C. Stachie,[65] M. Standke,[8,9] D. A. Steer,[27] M. Steinke,[8,9] J. Steinlechner,[142,44] S. Steinlechner,[142] D. Steinmeyer,[8,9] S. P. Stevenson,[154] D. Stocks,[49] R. Stone,[106] D. J. Stops,[11] K. A. Strain,[44] G. Stratta,[72,73] S. E. Strigin,[62] A. Strunk,[45] R. Sturani,[174] A. L. Stuver,[175] V. Sudhir,[12] T. Z. Summerscales,[176] L. Sun,[1] S. Sunil,[109] J. Suresh,[3] P. J. Sutton,[68] B. L. Swinkels,[37] M. J. Szczepańczyk,[34] M. Tacca,[37] S. C. Tait,[44] C. Talbot,[6] D. Talukder,[70] D. B. Tanner,[48] M. Tápai,[124] A. Taracchini,[36] J. D. Tasson,[96] R. Taylor,[1] F. Thies,[8,9] M. Thomas,[7] P. Thomas,[45] S. R. Thondapu,[61] K. A. Thorne,[7] E. Thrane,[6] Shubhanshu Tiwari,[114,98] Srishti Tiwari,[125] V. Tiwari,[68] K. Toland,[44] M. Tonelli,[18,19] Z. Tornasi,[44] A. Torres-Forné,[177] C. I. Torrie,[1] D. Töyrä,[11] F. Travasso,[28,41] G. Traylor,[7] M. C. Tringali,[74] A. Trovato,[27] L. Trozzo,[178,19] R. Trudeau,[1] K. W. Tsang,[37] M. Tse,[12] R. Tso,[46] L. Tsukada,[83] D. Tsuna,[83] D. Tuyenbayev,[106] K. Ueno,[83] D. Ugolini,[179] C. S. Unnikrishnan,[125] A. L. Urban,[2] S. A. Usman,[68] H. Vahlbruch,[9] G. Vajente,[1] G. Valdes,[2] N. van Bakel,[37] M. van Beuzekom,[37] J. F. J. van den Brand,[75,37] C. Van Den Broeck,[37,180] D. C. Vander-Hyde,[42] J. V. van Heijningen,[63] L. van der Schaaf,[37] A. A. van Veggel,[44] M. Vardaro,[51,52] V. Varma,[46] S. Vass,[1] M. Vasúth,[47] A. Vecchio,[11] G. Vedovato,[52] J. Veitch,[44] P. J. Veitch,[55] K. Venkateswara,[169] G. Venugopalan,[1] D. Verkindt,[33] F. Vetrano,[72,73] A. Viceré,[72,73] A. D. Viets,[23] D. J. Vine,[170] J.-Y. Vinet,[65] S. Vitale,[12] T. Vo,[42] H. Vocca,[40,41] C. Vorvick,[45] S. P. Vyatchanin,[62] A. R. Wade,[1] L. E. Wade,[117]

M. Wade,[117] R. Walet,[37] M. Walker,[26] L. Wallace,[1] S. Walsh,[23] G. Wang,[14,19] H. Wang,[11] J. Z. Wang,[127] W. H. Wang,[106] Y. F. Wang,[93] R. L. Ward,[21] Z. A. Warden,[34] J. Warner,[45] M. Was,[33] J. Watchi,[102] B. Weaver,[45] L.-W. Wei,[8,9] M. Weinert,[8,9] A. J. Weinstein,[1] R. Weiss,[12] F. Wellmann,[8,9] L. Wen,[63] E. K. Wessel,[17] P. Weßels,[8,9] J. W. Westhouse,[34] K. Wette,[21] J. T. Whelan,[58] L. V. White,[42] B. F. Whiting,[48] C. Whittle,[12] D. M. Wilken,[8,9] D. Williams,[44] A. R. Williamson,[129,37] J. L. Willis,[1] B. Willke,[8,9] M. H. Wimmer,[8,9] W. Winkler,[8,9] C. C. Wipf,[1] H. Wittel,[8,9] G. Woan,[44] J. Woehler,[8,9] J. K. Wofford,[58] J. Worden,[45] J. L. Wright,[44] D. S. Wu,[8,9] D. M. Wysocki,[58] L. Xiao,[1] H. Yamamoto,[1] C. C. Yancey,[76] L. Yang,[116] M. J. Yap,[21] M. Yazback,[48] D. W. Yeeles,[68] Hang Yu,[12] Haocun Yu,[12] S. H. R. Yuen,[93] M. Yvert,[33] A. K. Zadrożny,[106,144] M. Zanolin,[34] F. Zappa,[56] T. Zelenova,[28] J.-P. Zendri,[52] M. Zevin,[59] J. Zhang,[63] L. Zhang,[1] T. Zhang,[44] C. Zhao,[63] M. Zhou,[59] Z. Zhou,[59] X. J. Zhu,[6] A. B. Zimmerman,[77,91] Y. Zlochower,[58] M. E. Zucker,[1,12] and J. Zweizig[1]

(LIGO Scientific Collaboration and Virgo Collaboration)

[1]*LIGO, California Institute of Technology, Pasadena, California 91125, USA*
[2]*Louisiana State University, Baton Rouge, Louisiana 70803, USA*
[3]*Inter-University Centre for Astronomy and Astrophysics, Pune 411007, India*
[4]*Università di Salerno, Fisciano, I-84084 Salerno, Italy*
[5]*INFN, Sezione di Napoli, Complesso Universitario di Monte S. Angelo, I-80126 Napoli, Italy*
[6]*OzGrav, School of Physics and Astronomy, Monash University, Clayton 3800, Victoria, Australia*
[7]*LIGO Livingston Observatory, Livingston, Louisiana 70754, USA*
[8]*Max Planck Institute for Gravitational Physics (Albert Einstein Institute), D-30167 Hannover, Germany*
[9]*Leibniz Universität Hannover, D-30167 Hannover, Germany*
[10]*University of Cambridge, Cambridge CB2 1TN, United Kingdom*
[11]*University of Birmingham, Birmingham B15 2TT, United Kingdom*
[12]*LIGO, Massachusetts Institute of Technology, Cambridge, Massachusetts 02139, USA*
[13]*Instituto Nacional de Pesquisas Espaciais, 12227-010 São José dos Campos, São Paulo, Brazil*
[14]*Gran Sasso Science Institute (GSSI), I-67100 L'Aquila, Italy*
[15]*INFN, Laboratori Nazionali del Gran Sasso, I-67100 Assergi, Italy*
[16]*International Centre for Theoretical Sciences, Tata Institute of Fundamental Research, Bengaluru 560089, India*
[17]*NCSA, University of Illinois at Urbana-Champaign, Urbana, Illinois 61801, USA*
[18]*Università di Pisa, I-56127 Pisa, Italy*
[19]*INFN, Sezione di Pisa, I-56127 Pisa, Italy*
[20]*Departamento de Astronomía y Astrofísica, Universitat de València, E-46100 Burjassot, València, Spain*
[21]*OzGrav, Australian National University, Canberra, Australian Capital Territory 0200, Australia*
[22]*Laboratoire des Matériaux Avancés (LMA), CNRS/IN2P3, F-69622 Villeurbanne, France*
[23]*University of Wisconsin–Milwaukee, Milwaukee, Wisconsin 53201, USA*
[24]*SUPA, University of Strathclyde, Glasgow G1 1XQ, United Kingdom*
[25]*LAL, Université Paris-Sud, CNRS/IN2P3, Université Paris-Saclay, F-91898 Orsay, France*
[26]*California State University Fullerton, Fullerton, California 92831, USA*
[27]*APC, AstroParticule et Cosmologie, Université Paris Diderot, CNRS/IN2P3, CEA/Irfu, Observatoire de Paris, Sorbonne Paris Cité, F-75205 Paris Cedex 13, France*
[28]*European Gravitational Observatory (EGO), I-56021 Cascina, Pisa, Italy*
[29]*Chennai Mathematical Institute, Chennai 603103, India*
[30]*Università di Roma Tor Vergata, I-00133 Roma, Italy*
[31]*INFN, Sezione di Roma Tor Vergata, I-00133 Roma, Italy*
[32]*INFN, Sezione di Roma, I-00185 Roma, Italy*
[33]*Laboratoire d'Annecy de Physique des Particules (LAPP), Université Grenoble Alpes, Université Savoie Mont Blanc, CNRS/IN2P3, F-74941 Annecy, France*
[34]*Embry-Riddle Aeronautical University, Prescott, Arizona 86301, USA*
[35]*Montclair State University, Montclair, New Jersey 07043, USA*
[36]*Max Planck Institute for Gravitational Physics (Albert Einstein Institute), D-14476 Potsdam-Golm, Germany*
[37]*Nikhef, Science Park 105, 1098 XG Amsterdam, Netherlands*
[38]*Korea Institute of Science and Technology Information, Daejeon 34141, South Korea*
[39]*West Virginia University, Morgantown, West Virginia 26506, USA*
[40]*Università di Perugia, I-06123 Perugia, Italy*
[41]*INFN, Sezione di Perugia, I-06123 Perugia, Italy*
[42]*Syracuse University, Syracuse, New York 13244, USA*

[43]*University of Minnesota, Minneapolis, Minnesota 55455, USA*
[44]*SUPA, University of Glasgow, Glasgow G12 8QQ, United Kingdom*
[45]*LIGO Hanford Observatory, Richland, Washington 99352, USA*
[46]*Caltech CaRT, Pasadena, California 91125, USA*
[47]*Wigner RCP, RMKI, H-1121 Budapest, Konkoly Thege Miklós út 29-33, Hungary*
[48]*University of Florida, Gainesville, Florida 32611, USA*
[49]*Stanford University, Stanford, California 94305, USA*
[50]*Università di Camerino, Dipartimento di Fisica, I-62032 Camerino, Italy*
[51]*Università di Padova, Dipartimento di Fisica e Astronomia, I-35131 Padova, Italy*
[52]*INFN, Sezione di Padova, I-35131 Padova, Italy*
[53]*Montana State University, Bozeman, Montana 59717, USA*
[54]*Nicolaus Copernicus Astronomical Center, Polish Academy of Sciences, 00-716, Warsaw, Poland*
[55]*OzGrav, University of Adelaide, Adelaide, South Australia 5005, Australia*
[56]*Theoretisch-Physikalisches Institut, Friedrich-Schiller-Universität Jena, D-07743 Jena, Germany*
[57]*INFN, Sezione di Milano Bicocca, Gruppo Collegato di Parma, I-43124 Parma, Italy*
[58]*Rochester Institute of Technology, Rochester, New York 14623, USA*
[59]*Center for Interdisciplinary Exploration and Research in Astrophysics (CIERA), Northwestern University, Evanston, Illinois 60208, USA*
[60]*INFN, Sezione di Genova, I-16146 Genova, Italy*
[61]*RRCAT, Indore, Madhya Pradesh 452013, India*
[62]*Faculty of Physics, Lomonosov Moscow State University, Moscow 119991, Russia*
[63]*OzGrav, University of Western Australia, Crawley, Western Australia 6009, Australia*
[64]*Department of Astrophysics/IMAPP, Radboud University Nijmegen, P.O. Box 9010, 6500 GL Nijmegen, Netherlands*
[65]*Artemis, Université Côte d'Azur, Observatoire Côte d'Azur, CNRS, CS 34229, F-06304 Nice Cedex 4, France*
[66]*Physik-Institut, University of Zurich, Winterthurerstrasse 190, 8057 Zurich, Switzerland*
[67]*Université Rennes, CNRS, Institut FOTON—UMR6082, F-3500 Rennes, France*
[68]*Cardiff University, Cardiff CF24 3AA, United Kingdom*
[69]*Washington State University, Pullman, Washington 99164, USA*
[70]*University of Oregon, Eugene, Oregon 97403, USA*
[71]*Laboratoire Kastler Brossel, Sorbonne Université, CNRS, ENS-Université PSL, Collège de France, F-75005 Paris, France*
[72]*Università degli Studi di Urbino "Carlo Bo," I-61029 Urbino, Italy*
[73]*INFN, Sezione di Firenze, I-50019 Sesto Fiorentino, Firenze, Italy*
[74]*Astronomical Observatory Warsaw University, 00-478 Warsaw, Poland*
[75]*VU University Amsterdam, 1081 HV Amsterdam, Netherlands*
[76]*University of Maryland, College Park, Maryland 20742, USA*
[77]*Department of Physics, University of Texas, Austin, Texas 78712, USA*
[78]*School of Physics, Georgia Institute of Technology, Atlanta, Georgia 30332, USA*
[79]*Université Claude Bernard Lyon 1, F-69622 Villeurbanne, France*
[80]*Università di Napoli "Federico II," Complesso Universitario di Monte S. Angelo, I-80126 Napoli, Italy*
[81]*NASA Goddard Space Flight Center, Greenbelt, Maryland 20771, USA*
[82]*Dipartimento di Fisica, Università degli Studi di Genova, I-16146 Genova, Italy*
[83]*RESCEU, University of Tokyo, Tokyo, 113-0033, Japan*
[84]*Tsinghua University, Beijing 100084, China*
[85]*Texas Tech University, Lubbock, Texas 79409, USA*
[86]*The University of Mississippi, University, Mississippi 38677, USA*
[87]*Museo Storico della Fisica e Centro Studi e Ricerche "Enrico Fermi," I-00184 Roma, Italyrico Fermi, I-00184 Roma, Italy*
[88]*The Pennsylvania State University, University Park, Pennsylvania 16802, USA*
[89]*National Tsing Hua University, Hsinchu City, 30013 Taiwan, Republic of China*
[90]*Charles Sturt University, Wagga Wagga, New South Wales 2678, Australia*
[91]*Canadian Institute for Theoretical Astrophysics, University of Toronto, Toronto, Ontario M5S 3H8, Canada*
[92]*University of Chicago, Chicago, Illinois 60637, USA*
[93]*The Chinese University of Hong Kong, Shatin, NT, Hong Kong*
[94]*Seoul National University, Seoul 08826, South Korea*
[95]*Pusan National University, Busan 46241, South Korea*
[96]*Carleton College, Northfield, Minnesota 55057, USA*

[97]INAF, Osservatorio Astronomico di Padova, I-35122 Padova, Italy
[98]INFN, Trento Institute for Fundamental Physics and Applications, I-38123 Povo, Trento, Italy
[99]OzGrav, University of Melbourne, Parkville, Victoria 3010, Australia
[100]Columbia University, New York, New York 10027, USA
[101]Universitat de les Illes Balears, IAC3—IEEC, E-07122 Palma de Mallorca, Spain
[102]Université Libre de Bruxelles, Brussels 1050, Belgium
[103]Sonoma State University, Rohnert Park, California 94928, USA
[104]Departamento de Matemáticas, Universitat de València, E-46100 Burjassot, València, Spain
[105]University of Rhode Island, Kingston, Rhode Island 02881, USA
[106]The University of Texas Rio Grande Valley, Brownsville, Texas 78520, USA
[107]Bellevue College, Bellevue, Washington 98007, USA
[108]MTA-ELTE Astrophysics Research Group, Institute of Physics, Eötvös University, Budapest 1117, Hungary
[109]Institute for Plasma Research, Bhat, Gandhinagar 382428, India
[110]The University of Sheffield, Sheffield S10 2TN, United Kingdom
[111]IGFAE, Campus Sur, Universidade de Santiago de Compostela, 15782 Spain
[112]Dipartimento di Scienze Matematiche, Fisiche e Informatiche, Università di Parma, I-43124 Parma, Italy
[113]California State University, Los Angeles, 5151 State University Drive, Los Angeles, California 90032, USA
[114]Università di Trento, Dipartimento di Fisica, I-38123 Povo, Trento, Italy
[115]Università di Roma "La Sapienza," I-00185 Roma, Italy
[116]Colorado State University, Fort Collins, Colorado 80523, USA
[117]Kenyon College, Gambier, Ohio 43022, USA
[118]Christopher Newport University, Newport News, Virginia 23606, USA
[119]National Astronomical Observatory of Japan, 2-21-1 Osawa, Mitaka, Tokyo 181-8588, Japan
[120]Observatori Astronòmic, Universitat de València, E-46980 Paterna, València, Spain
[121]School of Mathematics, University of Edinburgh, Edinburgh EH9 3FD, United Kingdom
[122]Institute of Advanced Research, Gandhinagar 382426, India
[123]Indian Institute of Technology Bombay, Powai, Mumbai 400 076, India
[124]University of Szeged, Dóm tér 9, Szeged 6720, Hungary
[125]Tata Institute of Fundamental Research, Mumbai 400005, India
[126]INAF, Osservatorio Astronomico di Capodimonte, I-80131, Napoli, Italy
[127]University of Michigan, Ann Arbor, Michigan 48109, USA
[128]American University, Washington, D.C. 20016, USA
[129]GRAPPA, Anton Pannekoek Institute for Astronomy and Institute of High-Energy Physics, University of Amsterdam, Science Park 904, 1098 XH Amsterdam, Netherlands
[130]Delta Institute for Theoretical Physics, Science Park 904, 1090 GL Amsterdam, Netherlands
[131]Department of Physics and Kavli Institute for Astrophysics and Space Research, Massachusetts Institute of Technology, Cambridge, Massachusetts 02139, USA
[132]Directorate of Construction, Services and Estate Management, Mumbai 400094 India
[133]University of Białystok, 15-424 Białystok, Poland
[134]King's College London, University of London, London WC2R 2LS, United Kingdom
[135]University of Southampton, Southampton SO17 1BJ, United Kingdom
[136]University of Washington Bothell, Bothell, Washington 98011, USA
[137]Institute of Applied Physics, Nizhny Novgorod, 603950, Russia
[138]Ewha Womans University, Seoul 03760, South Korea
[139]Inje University Gimhae, South Gyeongsang 50834, South Korea
[140]National Institute for Mathematical Sciences, Daejeon 34047, South Korea
[141]Ulsan National Institute of Science and Technology, Ulsan 44919, South Korea
[142]Universität Hamburg, D-22761 Hamburg, Germany
[143]Maastricht University, P.O. Box 616, 6200 MD Maastricht, Netherlands
[144]NCBJ, 05-400 Świerk-Otwock, Poland
[145]Institute of Mathematics, Polish Academy of Sciences, 00656 Warsaw, Poland
[146]Cornell University, Ithaca, New York 14850, USA
[147]Hillsdale College, Hillsdale, Michigan 49242, USA
[148]Hanyang University, Seoul 04763, South Korea
[149]Korea Astronomy and Space Science Institute, Daejeon 34055, South Korea
[150]NASA Marshall Space Flight Center, Huntsville, Alabama 35811, USA
[151]Dipartimento di Matematica e Fisica, Università degli Studi Roma Tre, I-00146 Roma, Italy

[152]*INFN, Sezione di Roma Tre, I-00146 Roma, Italy*
[153]*ESPCI, CNRS, F-75005 Paris, France*
[154]*OzGrav, Swinburne University of Technology, Hawthorn, Victoria 3122, Australia*
[155]*University of Portsmouth, Portsmouth, PO1 3FX, United Kingdom*
[156]*Southern University and A&M College, Baton Rouge, Louisiana 70813, USA*
[157]*College of William and Mary, Williamsburg, Virginia 23187, USA*
[158]*Centre Scientifique de Monaco, 8 quai Antoine Ier, MC-98000, Monaco*
[159]*Indian Institute of Technology Madras, Chennai 600036, India*
[160]*INFN Sezione di Torino, Via P. Giuria 1, I-10125 Torino, Italy*
[161]*Institut des Hautes Etudes Scientifiques, F-91440 Bures-sur-Yvette, France*
[162]*IISER-Kolkata, Mohanpur, West Bengal 741252, India*
[163]*Whitman College, 345 Boyer Avenue, Walla Walla, Washington 99362, USA*
[164]*Université de Lyon, F-69361 Lyon, France*
[165]*Hobart and William Smith Colleges, Geneva, New York 14456, USA*
[166]*Dipartimento di Fisica, Università degli Studi di Milano Bicocca, Piazza della Scienza 3, 20126 Milano, Italy*
[167]*Dipartimento di Fisica, Università degli Studi di Torino, I-10125 Torino, Italy*
[168]*Janusz Gil Institute of Astronomy, University of Zielona Góra, 65-265 Zielona Góra, Poland*
[169]*University of Washington, Seattle, Washington 98195, USA*
[170]*SUPA, University of the West of Scotland, Paisley PA1 2BE, United Kingdom*
[171]*Indian Institute of Technology, Gandhinagar Ahmedabad Gujarat 382424, India*
[172]*Université de Montréal/Polytechnique, Montreal, Quebec H3T 1J4, Canada*
[173]*Indian Institute of Technology Hyderabad, Sangareddy, Khandi, Telangana 502285, India*
[174]*International Institute of Physics, Universidade Federal do Rio Grande do Norte, Natal RN 59078-970, Brazil*
[175]*Villanova University, 800 Lancaster Avenue, Villanova, Pennsylvania 19085, USA*
[176]*Andrews University, Berrien Springs, Michigan 49104, USA*
[177]*Max Planck Institute for Gravitationalphysik (Albert Einstein Institute), D-14476 Potsdam-Golm, Germany*
[178]*Università di Siena, I-53100 Siena, Italy*
[179]*Trinity University, San Antonio, Texas 78212, USA*
[180]*Van Swinderen Institute for Particle Physics and Gravity, University of Groningen, Nijenborgh 4, 9747 AG Groningen, Netherlands*

[†]Deceased.